\documentclass{aa} 
\usepackage[varg]{txfonts}
\usepackage{graphicx}
\usepackage{bm}
\usepackage{color}
\usepackage{natbib}
\usepackage{hyperref}
\usepackage{pifont}
\newcommand{\xmark}{\ding{53}}
\hypersetup{colorlinks=true,linkcolor=blue,citecolor=blue,urlcolor=blue}
\usepackage[toc,page]{appendix}
\usepackage{ulem}
\usepackage[dvipsnames]{xcolor}
\usepackage{ulem}

\begin{document}


\title{General relativistic effects on the orbit of the S2 star with GRAVITY} 
\author{M. Grould\inst{1,2} \and F. H. Vincent\inst{1} \and T. Paumard\inst{1} \and G. Perrin\inst{1}}
\institute{\inst{1} LESIA, Observatoire de Paris, PSL Research University, CNRS UMR 8109, Universit\'e Pierre et Marie Curie, Universit\'e Paris Diderot, 5 place Jules Janssen, 92190 Meudon, France \\ \inst{2} LUTh, Observatoire de Paris, PSL Research University, CNRS UMR 8102, Universit\'e Paris Diderot, 5 place Jules Janssen, 92190 Meudon Cedex, France}
\abstract
{The first observations of the GRAVITY instrument obtained in 2016, have shown that it should become possible to probe the spacetime close to the supermassive black hole Sagittarius A* (Sgr~A*) at the Galactic center by using accurate astrometric positions of the S2 star.}
{The goal of this paper is to investigate the detection by GRAVITY of different relativistic effects affecting the astrometric and/or spectroscopic observations of S2 such as the transverse Doppler shift, the gravitational redshift, the pericenter advance and higher-order general relativistic (GR) effects, in particular the Lense-Thirring effect due to the angular momentum of the black hole.} 
{We implement seven stellar-orbit models to simulate both astrometric and spectroscopic observations of S2 beginning near its next pericenter passage in 2018. Each model takes into account a certain number of relativistic effects. The most accurate one is a fully GR model and is used to generate the mock observations of the star. For each of the six other models, we determine the minimal observation times above which it fails to fit the observations, showing the effects that should be detected. These threshold times are obtained for different astrometric accuracies as well as for different spectroscopic errors.}
{Transverse Doppler shift and gravitational redshift can be detected within a few months by using S2 observations obtained with pairs of accuracies $(\sigma_\mathrm{A}, \sigma_\mathrm{V})~=~(10-100~\mu$as, $1-10$~km/s) where $\sigma_\mathrm{A}$ and $\sigma_\mathrm{V}$ are the astrometric and spectroscopic accuracies, respectively. Gravitational lensing can be detected within a few years with $(\sigma_\mathrm{A}, \sigma_\mathrm{V})=(10~\mu$as, 10~km/s). Pericenter advance should be detected within a few years with $(\sigma_\mathrm{A}, \sigma_\mathrm{V})~=~(10~\mu$as, $1-10$~km/s). Cumulative high-order photon curvature contributions, including the Shapiro time delay, affecting spectroscopic measurements can be observed within a few months with $(\sigma_\mathrm{A}, \sigma_\mathrm{V})~=~(10~\mu$as, 1~km/s). By using a stellar-orbit model neglecting relativistic effects on the photon path except the major contribution of gravitational lensing, S2 observations obtained with accuracies $(\sigma_\mathrm{A}, \sigma_\mathrm{V})~=~(10~\mu$as, 10~km/s), and a black hole angular momentum $(a, i', \Omega')~=~(0.99, 45^\circ, 160^\circ)$, the $1\sigma$ error on the spin parameter $a$ is of about 0.4, 0.2, and 0.1 for a total observing run of 16, 30, and 47 years, respectively. The $1\sigma$ errors on the direction of the angular momentum reach $\sigma_{i'}~\approx~25^\circ$ and $\sigma_{\Omega'}~\approx~40^\circ$ when considering the three orbital periods run. We found that the uncertainties obtained with a less spinning black hole ($a=0.7$) are similar to those evaluated with $a=0.99$.}
{The combination of S2 observations obtained with the GRAVITY instrument and the spectrograph SINFONI (Spectrograph for INtegral Field Observations in the Near Infrared) also installed at the VLT (Very Large Telescope) will lead to the detection of various relativistic effects. Such detections will be possible with S2 monitorings obtained within a few months or years, depending on the effect. Strong constraints on the angular momentum of Sgr~A* (e.g., at $1\sigma~=~0.1$) with the S2 star will be possible with a simple stellar-orbit model without using a ray-tracing code but with approximating the gravitational lensing effect. However, long monitorings are necessary, and we thus must rely on the discovery of closer-in stars near Sgr~A* if we want to efficiently constrain the black hole parameters with stellar orbits in a short time, or monitor the flares if they orbit around the black hole.}

\titlerunning{Relativistic effects on the S2 orbit}
\maketitle



 \section{Introduction}

Decades of studies have demonstrated the presence of a compact object of several million solar masses at the center of the Galaxy \citep{1977ApJ...218L.103W,1996ApJ...472..153G, 1997MNRAS.284..576E, 1998ApJ...509..678G, 2008ApJ...689.1044G, 2009ApJ...692.1075G,2017ApJ...837...30G}. One of the finest pieces of evidence supporting the existence of this compact source was obtained with the monitoring of S stars in the central parsec over a dozen years by \cite{2008ApJ...689.1044G} and \cite{2009ApJ...692.1075G}. In particular, a complete orbit of the closest star to the Galactic center, named S2, has been obtained. Such observations combined with monitorings of other S stars led to a confident constraint of the mass of the compact object of $\approx (4.31 \pm 0.42) \times 10^{6} M_{\odot}$ (\citealt{2009ApJ...692.1075G}, see \citealt{2016ApJ...830...17B} or \citealt{2017ApJ...837...30G} for a recent improvement on the estimation of this mass). Nowadays, the assumption is that this object is probably a supermassive black hole described by general relativity (GR) \citep[see e.g.,][and references therein]{2009ApJ...701.1357B,2011ApJ...735..110B}. Several methods for proving the existence of this GR black hole  are investigated, such as observing the accretion disk around the object, the flares occurring near it, or the stellar orbits of stars close to Sagittarius A* (Sgr~A*) (\citealt{2008ApJ...674L..25W}; \citealt{2009ApJ...695...59D}; \citealt{2010PhRvD..81f2002M}; \citealt{2014PhRvD..90b4068G}; \citealt{2014ApJ...784....7B}; \citealt{2016CQGra..33j5015V}; \citealt{2016CQGra..33k3001J}). The second generation instrument at the VLT (Very Large Telescope), GRAVITY, is expected to better constrain the nature of this object \citep{2003SPIE.4841.1548E}. By using its astrometric accuracy of about 10~$\mu$as, it will probe spacetime in strong gravitational fields by observing stars and gas located near Sgr~A*.\\

Different theoretical studies have been performed in order to determine whether it will be possible to detect GR effects with stellar orbits. The main purpose is to prove that observations of stars orbiting Sgr~A* are affected by GR effects induced by the presence of a Kerr black hole. Several authors have shown that low-order GR effects should be detectable using astrometric and/or spectroscopic measurements, such as the pericenter advance, the transversal Doppler shift or the gravitational redshift \citep{1998AcA....48..653J, 2000ApJ...542..328F, 2005ApJ...622..878W, 2006ApJ...639L..21Z,2017ApJ...845...22P,2017arXiv170901598N}. In particular, \cite{1998AcA....48..653J} showed that it will be easy to reject a stellar-orbit model neglecting the pericenter advance when considering astrometric measurements obtained on stars whose semi-major axis can reach 2~mpc (the semi-major axis of the S2 star is 5~mpc), and whether the astrometric accuracy on data is $\lesssim0.1$~mas. If we consider an accuracy of about 20~$\mu$as, it should be possible to detect this effect with a star whose semi-major axis can reach 5~mpc. Furthermore,  \cite{2006ApJ...639L..21Z} considered both astrometric and spectroscopic measurements of several S stars and showed that low-order relativistic effects affecting spectroscopy could be detected when considering monitorings of 10 years with instruments reaching accuracies of 1.5~mas for astrometry and 25~km/s for spectroscopy. Other theoretical investigations have been done in order to constrain high-order effects such as the Lense-Thirring effect and the quadrupole moment of the black hole \citep{2007CQGra..24.1775K,2008ApJ...674L..25W,2009ApJ...690.1553K,2010PhRvD..81f2002M,2010ApJ...711..157A,2010ApJ...720.1303A}. More precisely, \cite{2008ApJ...674L..25W} demonstrated the possibility of constraining the quadrupole moment of the black hole by observing astrometric positions of at least two stars whose eccentricity and orbital period satisfy $e~\geqslant0.9$ and $T~\geqslant0.1$~years, respectively. \cite{2010ApJ...720.1303A} estimated the spectroscopic accuracies necessary to detect various relativistic effects with measurements obtained during one orbital period of S2 ($\approx~15.8$~years). For instance, these authors showed that an accuracy of 10~m/s is needed to constrain the Lense-Thirring effect in one orbital period. More recently, \cite{2015ApJ...809..127Z} and \cite{2016ApJ...827..114Y} have shown it possible to constrain the angular momentum of the black hole by using astrometric and spectroscopic observations of the S2 star. Both authors used a sophisticated stellar-orbit model including both the computation of null and time-like geodesics. In addition, works on the detection of gravitational lensing have been performed by \cite{2012ApJ...753...56B} and have shown that this effect is sufficiently important to be detected by GRAVITY; in particular, with the S17 star whose gravitational lensing induces an astrometric shift of about 30~$\mu$as in 2018. Besides, studies performed by \cite{2004ApJ...611.1045B, 2005ApJ...627..790B, 2012ApJ...753...56B} and \cite{2016MNRAS.458.3614J} showed that gravitational lensing is affected by the angular momentum of the black hole and can thus lead to a constraint on the Lense-Thirring effect. However, we will need very accurate instruments to detect such deviations since the astrometric shift reaches only a few microarcseconds.

The aim of this paper is to further develop the investigations performed by these various authors but only focusing on S2. The choice to only work on the S2 star is motivated by the fact that we do not know whether closer-in stars will be observed with GRAVITY. It is thus important to extract the maximum information from this star.
Considering different astrometric and spectroscopic accuracies for the S2 observations, we estimate the different minimal observation times above which it is possible to detect different relativistic effects. We are thus capable of determining the threshold times needed for the GRAVITY instrument to detect different relativistic effects. The astrometric accuracy of this instrument will improve the detection of relativistic effects and thus better constrain the nature of Sgr~A*. This paper can be considered as a first step in the development of the numerical tools necessary to interpret the forthcoming accurate GRAVITY data. The study performed in this paper is done by implementing different models allowing us to describe the future S2 data with different degrees of refinement in the implementation of the various relativistic effects. With such models we will be able to determine whether we can detect the different effects for a given pair of astrometric and spectroscopic accuracies, and thus determine which model can be used to interpret the forthcoming S2 data with minimal computing time.
A part of this paper is exclusively devoted to the angular momentum of the central black hole candidate, where we discuss the constraint on its norm and direction obtained by using simulated GRAVITY observations of the S2 star. Contrary to \cite{2015ApJ...809..127Z} and \cite{2016ApJ...827..114Y}, we use various stellar-orbit models and determine which one allows us to investigate the constraint on the black hole angular momentum parameters with minimal computing time whilst still obtaining good-quality fits. 

The paper is organized as follows: Sect.~\ref{obs} is devoted to the explanation on how S2 astrometric and spectroscopic observations are simulated. Sect.~\ref{models} defines the models fitted to the simulated observations and used to detect relativistic effects, and Sect.~\ref{fitting} explains the procedures used to estimate both the threshold times and the angular momentum of the black hole; the different results are also given in this section. Finally, conclusions and discussions are given in Sect.~\ref{conclusion}.


 \section{Mock observations of the S2 star}
 \label{obs}

In our study, we consider two observables: the astrometric positions and the radial velocities (spectroscopy) of the S2 star. To generate mock observations of the S2 star, we consider a fully GR model by using the ray-tracing code \textsc{Gyoto}\footnote{\url{http://gyoto.obspm.fr/}}~\citep{2011CQGra..28v5011V,2016A&A...591A.116G}. The Kerr metric is considered in this model, and thus all relativistic effects are taken into account, such as pericenter advance, transverse Doppler shift, gravitational redshift, gravitational lensing, Lense-Thirring effect and the Shapiro time delay (see Appendix~\ref{app_E_1} for a brief definition of these effects). The Roemer effect is also naturally taken into account in \textsc{Gyoto} (see Appendix~\ref{app_E_1}). In this full-GR model, gravitational lensing is obtained by making two assumptions: we neglect multiple images of the star (e.g., the secondary image) and we consider only one photon of the primary image. The first assumption is valid since the influence of multiple images on the astrometric positions of the S2 star is negligible ($\lesssim 0.5 \,\mu$as if only considering the secondary image). The second assumption means that we do not compute the flux of the primary image. It is valid since the star is far enough from the black hole to neglect the amplification effect due to lensing. One photon is thus sufficient to recover the astrometric position of the star. For more technical explanations about how we compute astrometric positions and radial velocities of S2 in the full-GR model with \textsc{Gyoto}, see Appendix~\ref{app_A}. \\

\begin{figure}[!t]
\centering
      \includegraphics[scale=0.45]{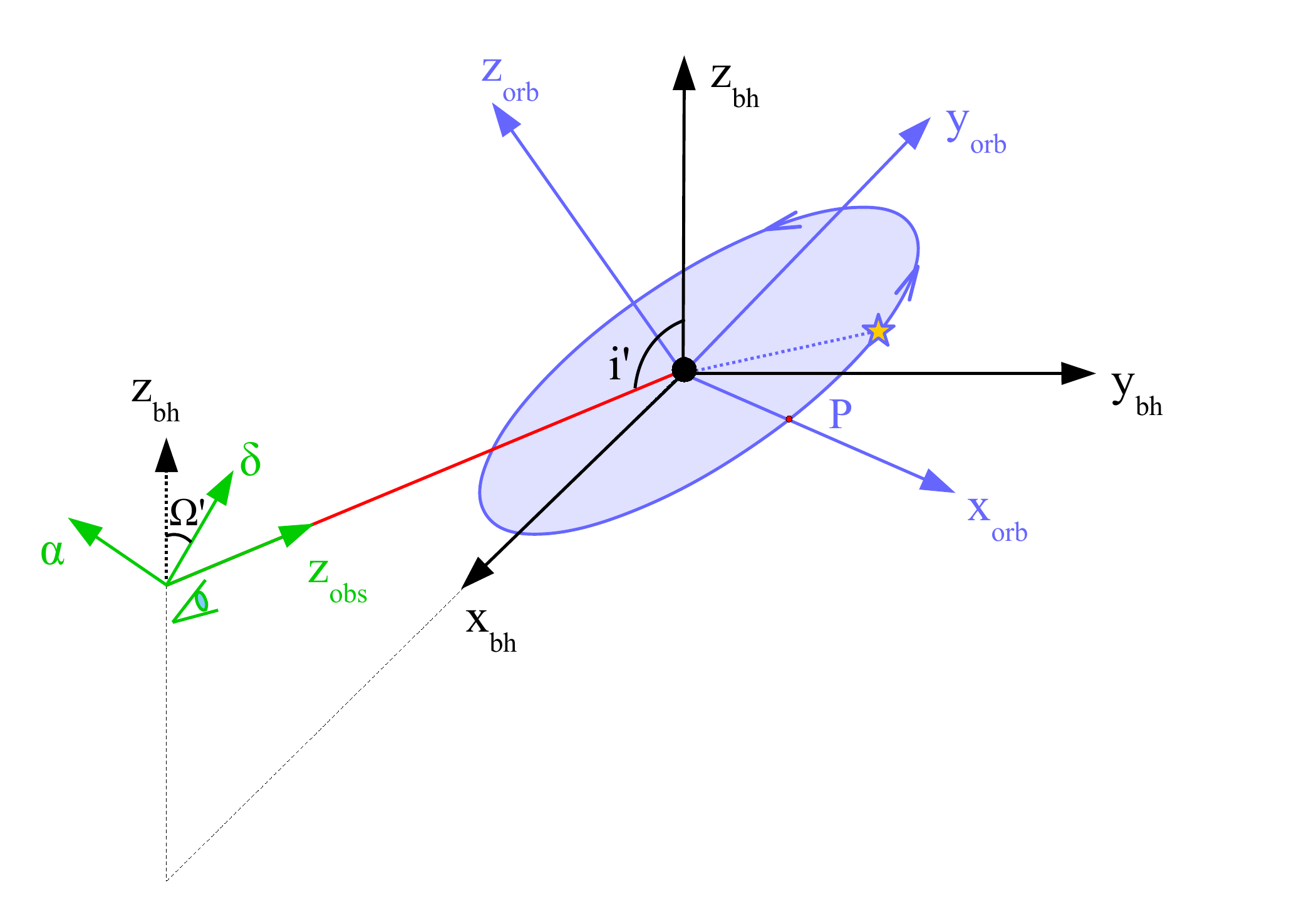}
   \caption{Illustration of the different reference frames: the \textit{black-hole frame} (in black) $(x_{\mathrm{bh}},y_{\mathrm{bh}},z_{\mathrm{bh}})$ labeled in Kerr-Schild coordinates, where the $z_{\mathrm{bh}}$-axis corresponds to the angular momentum-axis; the \textit{orbit frame} (in blue) $(x_{\mathrm{orb}},y_{\mathrm{orb}},z_{\mathrm{orb}})$; and finally, the \textit{observer frame} (in green) $(\alpha,\delta,z_{\mathrm{obs}})$. The angles $i'$ and $\Omega'$ allow to recover the direction of the angular momentum of the black hole; they give the position of the black-hole frame relative to the observer frame. The point P on the orbit of the star denotes the pericenter.}
   \label{frames}
\end{figure}

In this paper, we consider three different reference frames in Fig.~\ref{frames}: The \textit{black-hole frame} $(x_{\mathrm{bh}},y_{\mathrm{bh}},z_{\mathrm{bh}})$ centered on the black hole and labeled in Kerr-Schild coordinates \citep{2007arXiv0706.0622V}; the $z_{\mathrm{bh}}$-axis is taken along the angular momentum axis of the black hole; The \textit{orbit frame} $(x_{\mathrm{orb}},y_{\mathrm{orb}},z_{\mathrm{orb}})$ centered on the black hole and such that $(x_{\mathrm{orb}},y_{\mathrm{orb}})$ spans the plane of the orbit and $z_{\mathrm{orb}}$ is along the angular momentum of the orbit; and the \textit{observer frame} $(\alpha,\delta,z_{\mathrm{obs}})$ located at the observer position with $(\alpha,\delta)$ spanning the observer screen whose origin is located at the center of the screen and which corresponds to the apparent position of the black hole;  $z_{\mathrm{obs}}$ being directed towards the black hole.

\begin{figure*}[htb]
\centering
      \includegraphics[scale=0.45]{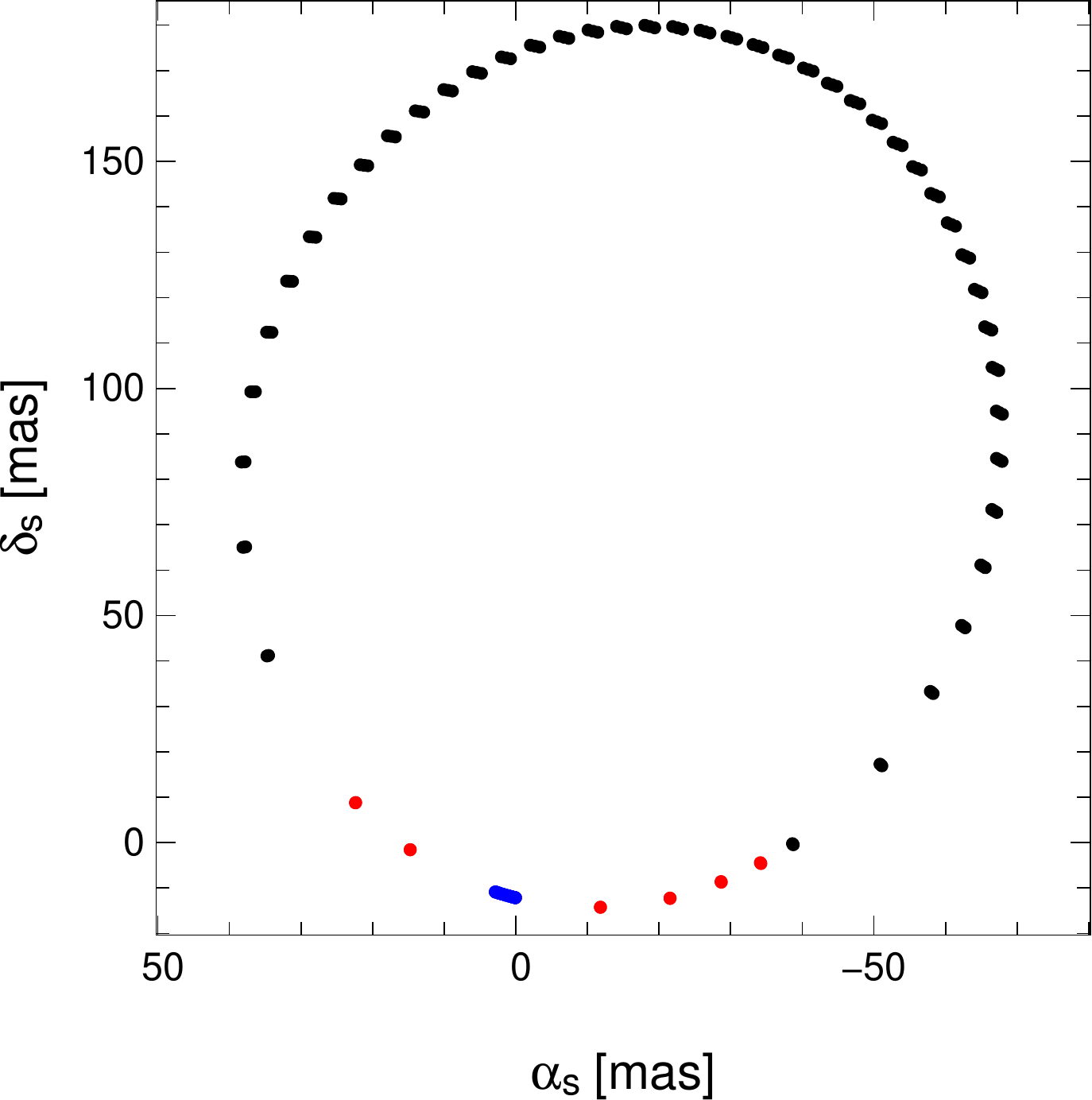}
      \quad
      \includegraphics[scale=0.45]{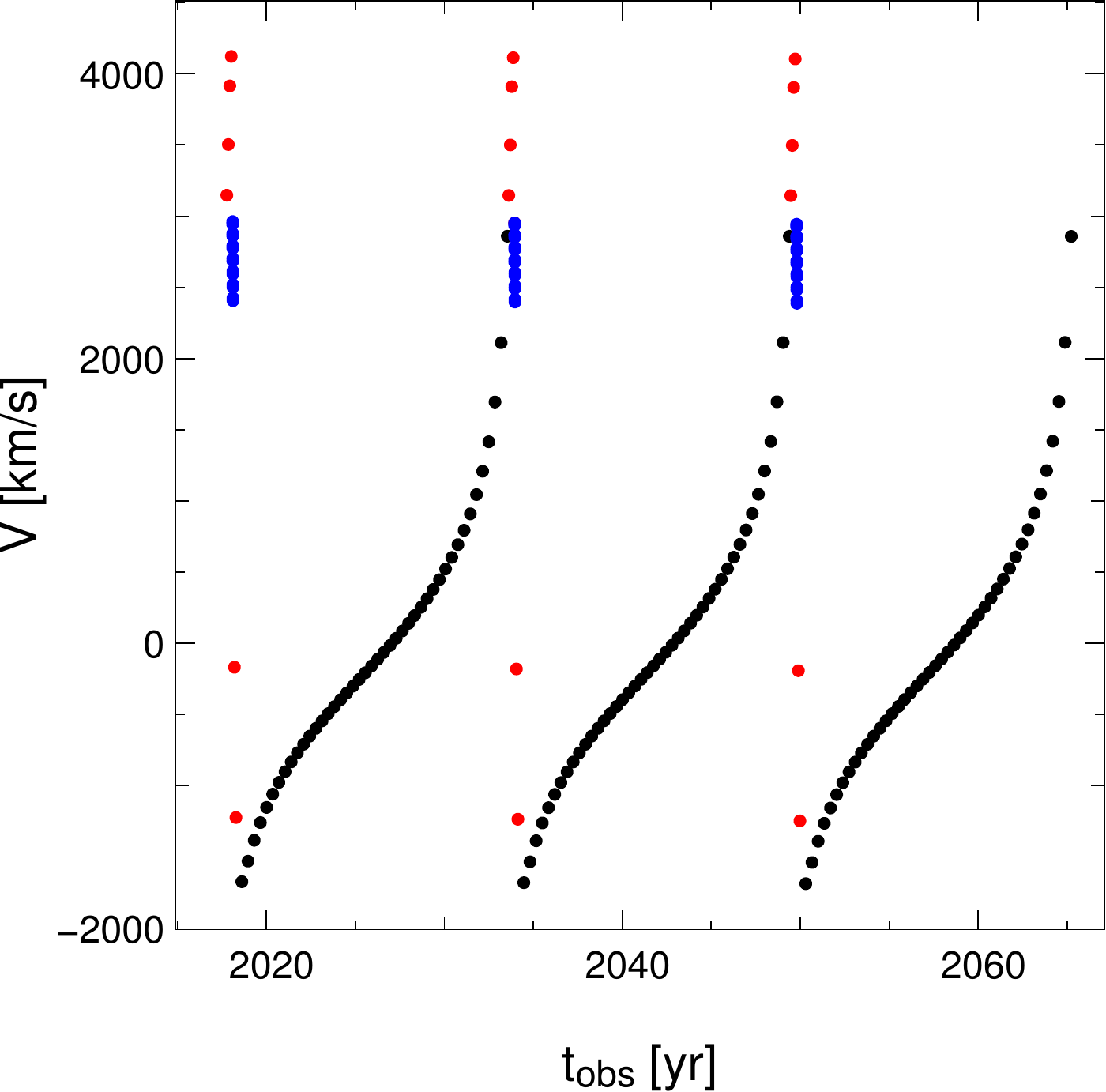} \\
        \quad
        
      \includegraphics[scale=0.45]{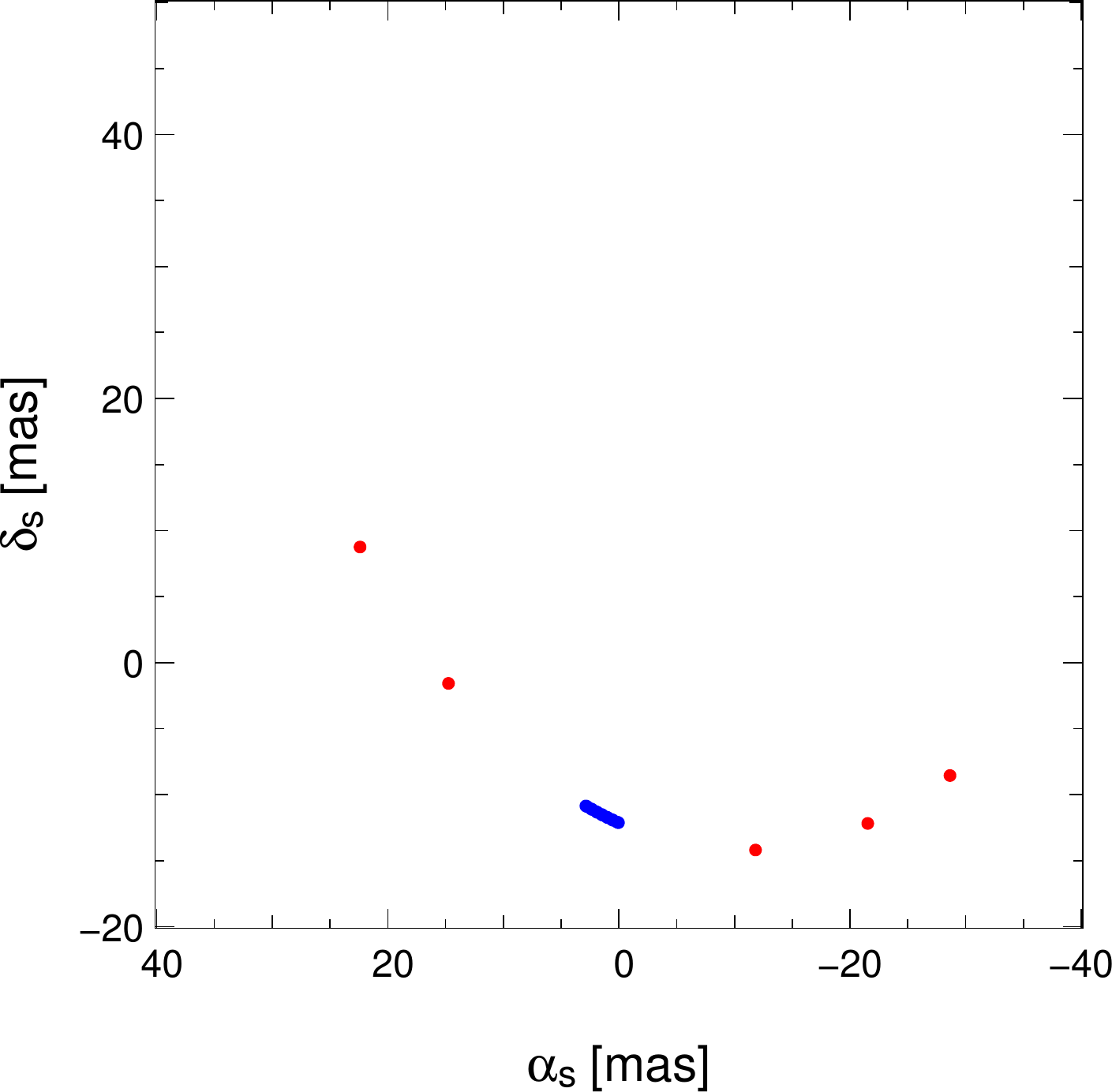}
      \quad
      \includegraphics[scale=0.45]{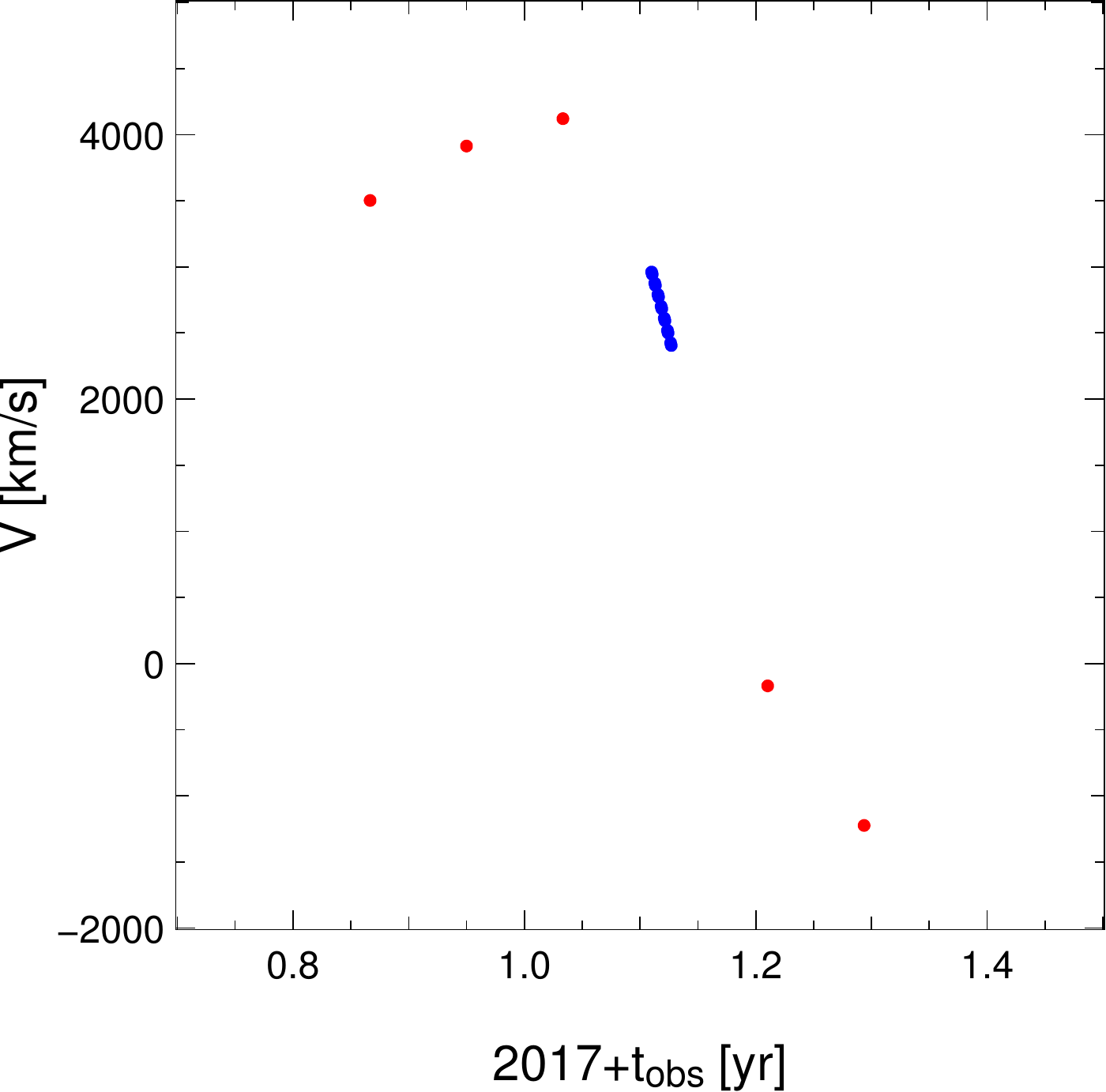}
   \caption{Astrometric and spectroscopic observations of the monitorings of one period (upper plots) and 6 months (lower plots) of the S2 star, simulated with the full-GR model. For one period of monitoring we consider one week of observation at pericenter passage (blue points), 6 months of observations around the pericenter passage (red points) and 14 years of observations for the rest of the orbit (black points). The 6 months of monitoring corresponds to the first nineteen points of the one-period run.}
        \label{fig:runs}
\end{figure*}

To compute the orbit of S2 with \textsc{Gyoto} we need the initial position and velocity of the star in Kerr-Schild coordinates. These coordinates are obtained using the two steps listed below:
\begin{itemize} 
\item Step 1: Get the three-dimensional positions $(\alpha_s,\delta_s,z_{s,\mathrm{obs}})$ and velocities $(v_{\alpha_s},v_{\delta_s},v_{z_{s,\mathrm{obs}}})$ of the star in the observer frame at a given observation date $t_\mathrm{obs}$, by using the Thiele-Innes formulas \citep{1985cmcg.book.....T} and the Keplerian orbital parameters: period $T$, semi-major axis $a_{\mathrm{sma}}$, eccentricity $e$, time of the pericenter passage $t_\mathrm{p}$, inclination $i$, angle of the line of nodes $\Omega$ and angle from ascending node to pericenter $\omega$ (see Fig.~\ref{frames_app} for an illustration of these three angles). See Appendix~\ref{app_B} for a demonstration of how we compute the star coordinates in the observer frame.
\item Step 2: Get the initial coordinates of the star in the black-hole frame. To do so, we apply a rotation matrix to the coordinates of the star obtained in the observer frame. The matrix depends on two angles $\Omega'$ and $i'$ giving the direction of the angular momentum of the black hole with respect to the observer (see Fig.~\ref{frames}). The rotation matrix is given by
\begin{equation}
\left(\begin{array}{ccc} \sin{(i'')}\sin{(\Omega')} & \sin{(i'')}\cos{(\Omega')} & -\cos{(i'')} \\-\cos{(\Omega')} & \sin{(\Omega')} & 0 \\ \cos{(i'')}\sin{(\Omega')}  & \cos{(i'')}\cos{(\Omega')} & \sin{(i'')} \end{array}\right)
,\end{equation}
where $i'' = 3\pi / 2 + i'$. We assume that the obtained position and velocity correspond to the initial Kerr-Schild coordinates of the star.
\end{itemize} 
Knowing the initial coordinates of the star, we can integrate the time-like geodesic in \textsc{Gyoto,} that is, obtain its GR orbit. We note that the Newtonian orbit can be considered as an osculating orbit of the GR one.

The initial coordinates of S2 are obtained considering the orbital parameters equal to the best-fit values evaluated by \citet{2009ApJ...692.1075G}: $T = 15.8$ yrs, $a_{\mathrm{sma}}~=~0.123''$, $e = 0.88$, $t_p = 2002.32$ yrs, $\Omega = 225.39^{\circ}$, $\omega = 63.56^{\circ}$, $i = 135.25^{\circ}$. We also consider the distance between the observer and the center of our galaxy equal to the best-fit found by these authors: $R_0 = 8.33$ kpc. For the angular momentum parameters of the black hole, we choose $a=0.99$, $i'=45^{\circ}$ and $\Omega'=160^{\circ}$ where $a$ is the dimensionless spin of the black hole.

The S2 noisy data are obtained by adding a Gaussian random noise to the full-GR observations whose distribution is parametrized by a standard deviation $\sigma_\mathrm{A}$ for the astrometry and $\sigma_{\mathrm{V}}$ for the spectroscopy. In this study, we consider different values for $\sigma_\mathrm{A}$: 10~$\mu$as, 30~$\mu$as, 50~$\mu$as and 100~$\mu$as; and  $\sigma_{\mathrm{V}}$: 1~km/s, 10~km/s and 100~km/s. We note that several of those accuracies can be reached by current and future instruments (see Table~\ref{table:instru}, whose principal accuracies are taken from \citealt{2005ApJ...622..878W,2009ApJ...707L.114G,2010RvMP...82.3121G,2011Msngr.143...16E}).

{\renewcommand{\arraystretch}{1.3}
\begin{table}[!t]
\begin{center}
\caption{Astrometric and spectroscopic accuracies of various current and future instruments, capable of observing in the near infrared. The instruments NACO, GRAVITY, NIRSPEC and SINFONI are already in use. The other instruments are supposed to be operational in $2020-2025$.}
\label{table:instru}
\begin{tabular}{l} 
\hline
\hline
\multicolumn{1}{c}{Astrometry} \\ 
 \hline
NACO$^{a}$ (VLT): $\sim 300~\mu$as \\
GRAVITY (VLT): $\sim 10~\mu$as \\
MICADO$^{b}$ (E-ELT): $50-100~\mu$as \\ 
TMT$^{c}$: $\sim 100~\mu$as \\
GMT$^{d}$: $\sim 100~\mu$as \\
\hline
\hline
\multicolumn{1}{c}{Spectroscopy} \\
\hline
NIRSPEC$^{e}$ (Keck): $\sim 10$~km/s \\
GRAVITY (VLT): $\gtrsim100$ km/s \\
MICADO (E-ELT): $\sim 1$~km/s \\ 
SINFONI$^{f}$ (VLT): $\sim 10$~km/s \\
\hline
\end{tabular}
\end{center}
\small
$^{a}$ Nasmyth Adaptive Optics System (NAOS) Near-Infrared Imager and Spectrograph (CONICA) \\
$^{b}$ Multi-AO Imaging Camera for Deep Observations \\
$^{c}$ Thirty Meter Telescope \\
$^{d}$ Giant Magellan Telescope \\
$^{e}$ Near InfRared echelle SPECtrograph\\
$^{f}$ Spectrograph for INtegral Field Observations in the Near Infrared
\end{table}
}

As shown in \cite{2010ApJ...711..157A}, \cite{2010ApJ...720.1303A}, \cite{2011ApJ...734L..19A} and \cite{2006ApJ...639L..21Z}, spectroscopic measurements obtained during the pericenter passage are a powerful tool to detect relativistic effects. In particular, \cite{2006ApJ...639L..21Z} showed that in the case of S2, the transverse Doppler shift and gravitational redshift represent a significant contribution to radial velocity of about $\approx 200$ km/s near pericenter. These are the reasons why we choose to better sample the mock observations at S2 pericenter passage. We consider different runs of observation ranging from one month to three periods ($\approx$ 47 years). All shorter runs are subsets of the three-period run. The latter run is sampled, for each period, as follows:
\begin{itemize} 
\item  Two points per night during one week at pericenter passage (starting in 2018.11 for the first period),
\item  One point per month during six months at pericenter passage (between 2017.78 and 2018.29 for the first period),
\item  One point every four months for the rest of the orbit (between 2018.63 and 2033.20 for the first period).
\end{itemize} 
All runs start between 2017.78 and 2018.11. As an illustration, the monitoring of one period and six months are visible in Fig.~\ref{fig:runs}. We note that astrometric and spectroscopic observations are supposed to be at the same dates, which could be more difficult in practice since radial velocity measurements should not be done with GRAVITY (because of its poorer spectroscopic accuracy, see Table~\ref{table:instru}).


\section{Models}
\label{models}

\begin{figure*}[!t]
\centering
      \includegraphics[scale=0.6]{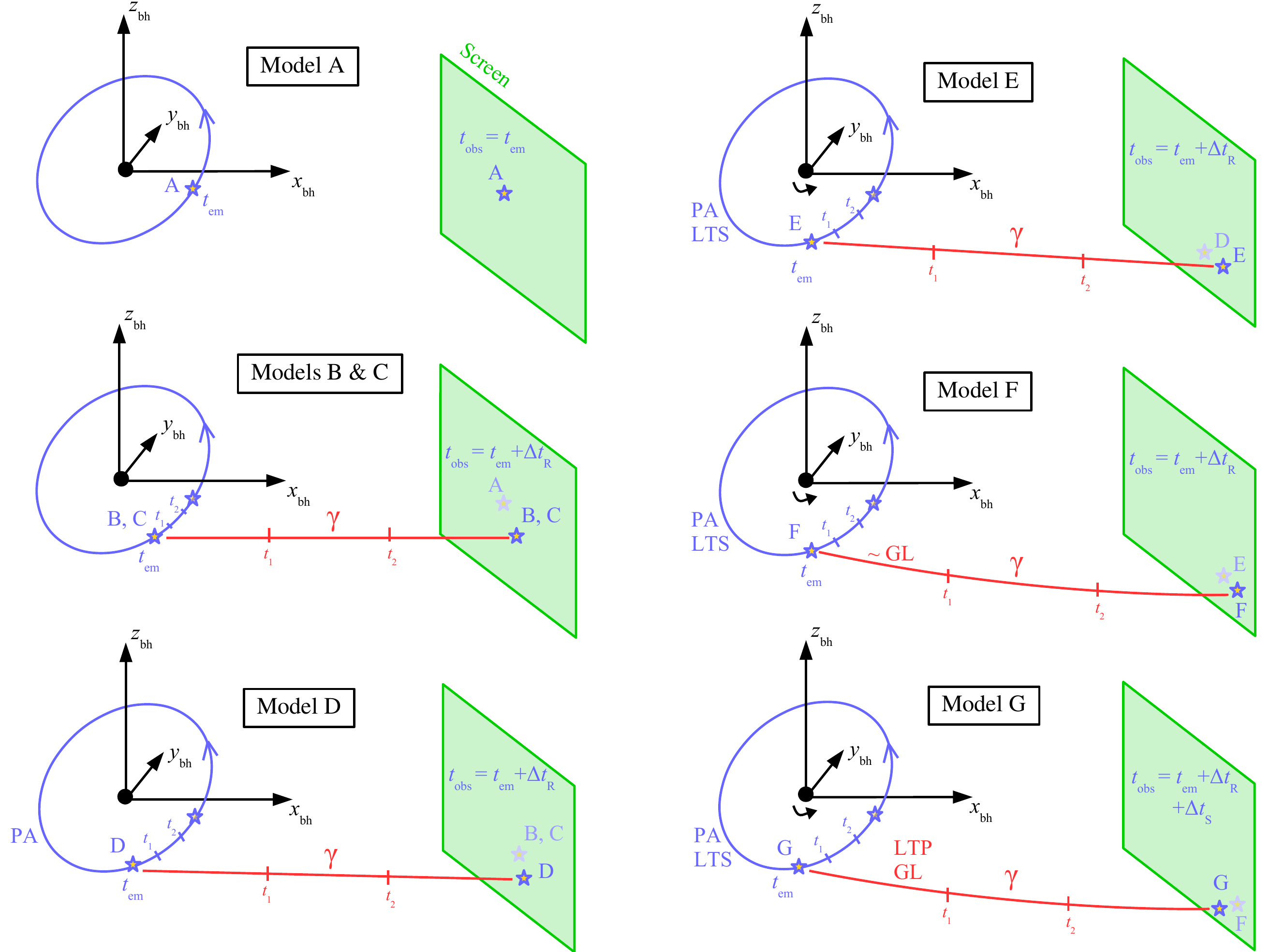}\\
      \caption{Illustration of each model described in Sect.~\ref{models_def}, only focusing on the different effects affecting the astrometric measurements of the S2 star. The positions of both the star (blue line) and the photon (red line) at two dates $t_1$ and $t_2$ are represented for models B to G. At some particular observation time $t_{\mathrm{obs}}$, the apparent position of the star on the observer screen, is highlighted by the symbols A to G. The corresponding emitting position of the star along its orbit is also represented. In Model A, the difference $\Delta t = t_{\mathrm{obs}} - t_{\mathrm{em}}$ is zero since the photon has an infinite speed. In models B to G, the Roemer time delay, noted $\Delta t_{\mathrm{R}}$, is included, thus $\Delta t$ depends on it. In Model D, the star trajectory is described by GR but only accounts for the pericenter advance (PA) since the Lense-Thirring effect is neglected. Contrary to Model D, Model E takes into account this latter effect on the time-like geodesic (LTS for Lense-Thirring on the Star). Model F is analogous to Model E but approximates the gravitational lensing effect (GL). Model G is the full-GR model; gravitational lensing is thus better estimated, the Lense-Thirring effect on the photon path (LTP) is naturally taken into account, and the quantity $\Delta t$ depends on the Shapiro time delay, $\Delta t_{\mathrm{S}}$, in addition to $\Delta t_{\mathrm{R}}$.}
        \label{Models}
\end{figure*}

\subsection{Definitions}
\label{models_def}

We want to estimate the minimal observation times needed to detect different effects affecting the S2 star observations, and considering different astrometric and spectroscopic accuracies. For doing so, we implement different models, in order of increasing complexity and computing time needed, including three \textit{Keplerian models} (whose star orbit is Keplerian) and four \textit{Relativistic models} (whose star orbit is relativistic) listed here:
\begin{itemize} 
\item Model A: Keplerian model without the Roemer time delay due to the finite speed of light, and without adding relativistic effects.
\item Model B: Keplerian model only considering the Roemer effect.
\item Model C: Keplerian model considering the Roemer effect, the transverse Doppler shift, and the gravitational redshift.
\item Model D: GR model without ray tracing. The orbit of the star is relativistic but not the photon trajectory. The Roemer effect, the transverse Doppler shift, the gravitational redshift and the pericenter advance are taken into account.
\item Model E: Similar to Model D but takes into account the Lense-Thirring effect on the star trajectory.
\item Model F: Similar to Model E but considers a supplement effect corresponding to an approximation of gravitational lensing.
\item Model G: Full-GR model described in Sect.~\ref{obs}. All previous effects are naturally taken into account in this model and additional effects are considered such as the Shapiro time delay and the Lense-Thirring effect on the photon trajectory.
\end{itemize}
An illustration of each model can be seen in Fig.~\ref{Models}. More detail on how the observations are generated in the different models are given below.\\

The astrometric and spectroscopic observations generated in Model A are simply obtained by using the orbital parameters and the Thiele-Innes formulas (see Appendix~\ref{app_B}).

For Model B, we use the same protocol as Model A but we include the Roemer time delay solving the following equation (see Appendix~\ref{app_C}):
\begin{equation}
\label{roemer}
t_{\mathrm{obs}} - t_{\mathrm{em}} + \frac{z_{s, \mathrm{obs}}(t_{\mathrm{em}})}{c} = 0
,\end{equation}
where $t_{\mathrm{em}}$ is the emission date, $z_{s, \mathrm{obs}}(t_{\mathrm{em}})$ is the position of the star along the line-of-sight at emission, and $c$ is the speed of light. Knowing the observation date $t_{\mathrm{obs}}$ and using this equation, we can determine the emission date $t_{\mathrm{em}}$ at which we need to compute the astrometric and spectroscopic data (obtained with formulas of Appendix~\ref{app_B}) to take into account the Newtonian time-traveling of the photon.

Model C is similar to Model B but considers two low-order relativistic effects only affecting the spectroscopy: the transverse Doppler shift and gravitational redshift. To implement these effects, we compute an approximated radial velocity allowing to simulate both the Newtonian Doppler shift (also called the longitudinal Doppler shift) and the two relativistic redshifts. This formula is given by (see Appendix~\ref{App_D} for a demonstration)
\begin{equation}
\label{eq:red}
\mathrm{V} \approx \left[ \frac{1}{\sqrt{1-\epsilon}} \times \frac{1+\mathcal{V}_{\mathrm{proj}}/c \times (1-\epsilon)^{-1/2}}{\sqrt{1- \left(\mathcal{V}/c\right)^2 \times(1-\epsilon)^{-1}}} - 1 \right] c
,\end{equation}
where $\epsilon = 2GM/(c^2r_{\mathrm{em}})$ with $r_{\mathrm{em}}$ being the radial coordinate of the star at emission in the black-hole frame given by equation~\eqref{eq:radius}, $\mathcal{V}$ being the velocity of the star in the black-hole frame and $\mathcal{V}_{\mathrm{proj}}$  the projection of $\mathcal{V}$ along the line-of-sight in the observer frame $(\alpha,\delta,z_{\mathrm{obs}})$. As Model C is a Keplerian model, the velocity $\mathcal{V}$ corresponds to the orbital velocity given by
\begin{equation}
\label{vorb}
\mathcal{V} = \sqrt{ \frac{2 G M}{r_{\mathrm{em}}} - \frac{G M}{a_{\mathrm{sma}}}}.
\end{equation}
We mention that the Roemer time delay affecting the spectroscopy is included since the three quantities $r_{\mathrm{em}}$, $\mathcal{V,}$ and $\mathcal{V}_{\mathrm{proj}}$ are computed considering the same protocol as Model B.

To generate the star orbit in Model D, we use a Schwarzschild metric and the procedure described in Sect.~\ref{obs}: compute a GR orbit with the ray-tracing code \textsc{Gyoto} considering an initial position of the star generated with the Keplerian orbital parameters and the Thiele-Innes formulas. In this model, we do not compute null geodesics, only the star trajectory is relativistic. The pericenter advance is thus naturally taken into account. As in Model C, we simulate the Roemer time delay, the transverse Doppler shift and the gravitational redshift by using the equations~\eqref{roemer} and~\eqref{eq:red}. More precisely, the emission date given by equation~\eqref{roemer} is used to recover the star coordinates (evaluated with \textsc{Gyoto}) allowing to simulate astrometric positions of this star (by projecting these coordinates in the plane of the sky) affected by the Newtonian travel time of the photon. In equation~\eqref{eq:red}, the orbital velocity $\mathcal{V}$  is computed through the GR expression $\sqrt{g_{ij}\mathcal{V}^{i}\mathcal{V}^{j}}$ where $g_{ij}$ are the spatial metric coefficients and $\mathcal{V}^{i}$ is the three-velocity of the star defined as $u^i/u^t$ with $u$ being the four-velocity of the star. The projection of this three-velocity, $\mathcal{V}_{\mathrm{proj}}$, is obtained considering a photon not affected by the spacetime curvature (i.e., we simply project the star velocity in Euclidian space along the line-of-sight). We note that the three effects mentioned above are better estimated in Model D than in Model C since the coordinates of the star (position and velocity) used in both equations~\eqref{roemer} and~\eqref{eq:red} are affected by the pericenter advance. 

Model E is equivalent to Model D but the star orbit is obtained by using a Kerr metric instead of a Schwarzschild metric. The Lense-Thirring effect on the star trajectory is thus naturally considered in such a model. 

Model F is similar to Model E but approximatively simulates the astrometric shift of the star induced by gravitational lensing. It is taken into account by using analytical approximations developed by \cite{2006PhRvD..74l3009S}. These formulas are obtained in weak-deflection limit, meaning that the minimal distance between a photon and the black hole is higher than the Schwarzschild radius $R_S = 2 G M /c^2$. In addition, they are developed in the weak-field regime, the observer and the emitter are thus considered in flat spacetime.

The different effects taken into account in each model are summarized in Table~\ref{table:models}.

{\renewcommand{\arraystretch}{1.1}
\begin{table}[!t]
\begin{center}
\caption{Effects considered in each model.}
\label{table:models}
\begin{tabular}{lccccccc}
        \hline
        \hline
         Effects & A & B & C & D & E & F & G\\  
        \hline
        Roemer & \textcolor{Red}{\xmark} & \textcolor{Green}{\checkmark} & \textcolor{Green}{\checkmark} & \textcolor{Green}{\checkmark} & \textcolor{Green}{\checkmark} & \textcolor{Green}{\checkmark} & \textcolor{Green}{\checkmark} \\ 
        TD$^a$ & \textcolor{Red}{\xmark} & \textcolor{Red}{\xmark} & \textcolor{Green}{\checkmark} & \textcolor{Green}{\checkmark} & \textcolor{Green}{\checkmark} & \textcolor{Green}{\checkmark} & \textcolor{Green}{\checkmark} \\
        Grav.$^b$ & \textcolor{Red}{\xmark} & \textcolor{Red}{\xmark} & \textcolor{Green}{\checkmark} & \textcolor{Green}{\checkmark} & \textcolor{Green}{\checkmark} & \textcolor{Green}{\checkmark} & \textcolor{Green}{\checkmark}\\
        PA$^c$ & \textcolor{Red}{\xmark} & \textcolor{Red}{\xmark} & \textcolor{Red}{\xmark} & \textcolor{Green}{\checkmark} & \textcolor{Green}{\checkmark} & \textcolor{Green}{\checkmark} & \textcolor{Green}{\checkmark}\\
        LTS$^d$ & \textcolor{Red}{\xmark} & \textcolor{Red}{\xmark} & \textcolor{Red}{\xmark} & \textcolor{Red}{\xmark} & \textcolor{Green}{\checkmark} & \textcolor{Green}{\checkmark} & \textcolor{Green}{\checkmark}\\
        GL$^e$ & \textcolor{Red}{\xmark} & \textcolor{Red}{\xmark} & \textcolor{Red}{\xmark} & \textcolor{Red}{\xmark} & \textcolor{Red}{\xmark} & \textcolor{Green}{\checkmark} & \textcolor{Green}{\checkmark}\\
        LTP$^f$ & \textcolor{Red}{\xmark} & \textcolor{Red}{\xmark} & \textcolor{Red}{\xmark} & \textcolor{Red}{\xmark} & \textcolor{Red}{\xmark} & \textcolor{Red}{\xmark} & \textcolor{Green}{\checkmark}\\
        Shapiro & \textcolor{Red}{\xmark} & \textcolor{Red}{\xmark} & \textcolor{Red}{\xmark} & \textcolor{Red}{\xmark} & \textcolor{Red}{\xmark} & \textcolor{Red}{\xmark} & \textcolor{Green}{\checkmark}\\
        \hline
\end{tabular}
\end{center}
\small
$^a$ Transverse Doppler shift\\
$^b$ Gravitational redshift\\
$^c$ Pericenter Advance\\
$^d$ Lense-Thirring on the Star\\
$^e$ Gravitational Lensing\\
$^f$ Lense-Thirring on the Photon
\end{table}
}

\subsection{Effects affecting the astrometry of the S2 star}
\label{models_astro}

In this Section we discuss the different effects that change the astrometric position of the S2 star. In particular, we focus on the pericenter advance, the Roemer and Shapiro time delay, the Lense-Thirring effect, and gravitational lensing. All plots in this Section and the following one, devoted to spectroscopy, are obtained with black hole and orbital parameters given in Sect.~\ref{obs}. See Appendix~\ref{app_E_2} for a brief explanation on how we evaluate the astrometric contribution (and spectroscopic one) of each effect.

\begin{figure}[!t]
    \centering
        \includegraphics[scale=0.5]{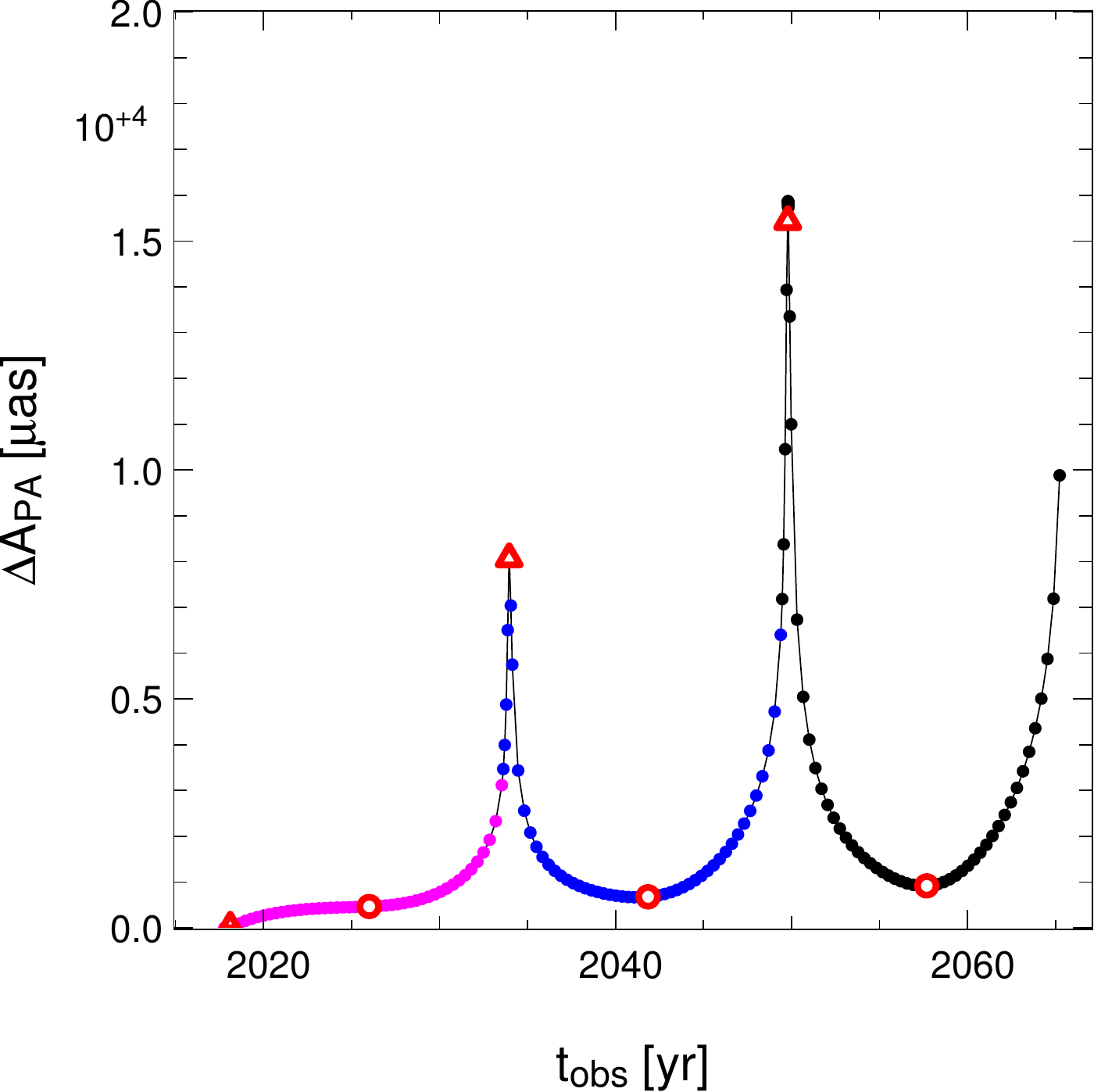}
    \caption{Astrometric impact of the pericenter advance on the S2 star observed during three orbital periods. Solid magenta, blue, and black circles correspond to the first, second, and third S2 period, respectively. Open red circles and triangles represent the apocenter and pericenter passages, respectively.}
    \label{fig:AAP}
\end{figure}

The pericenter advance is an effect which increases with the number of turns made by the star around the black hole. To evaluate this effect we can compare a Keplerian orbit with a GR orbit. That is what is presented in Fig.~\ref{fig:AAP}. We can see that the maximal astrometric difference is located near pericenter passages. At first pericenter passage, the difference is weak since it corresponds to the first data points: the apparent positions start to differ when the orbit has evolved. During the first S2 period, the maximal magnitude of the pericenter advance is reached near the second pericenter passage (last magenta dot in Fig.~\ref{fig:AAP}) and is equal to $\approx~3$~mas. At second and third periods, the maximal impact is of about 8~mas and 16~mas, respectively.

\begin{figure*}[htb]
\centering
      \includegraphics[scale=0.3]{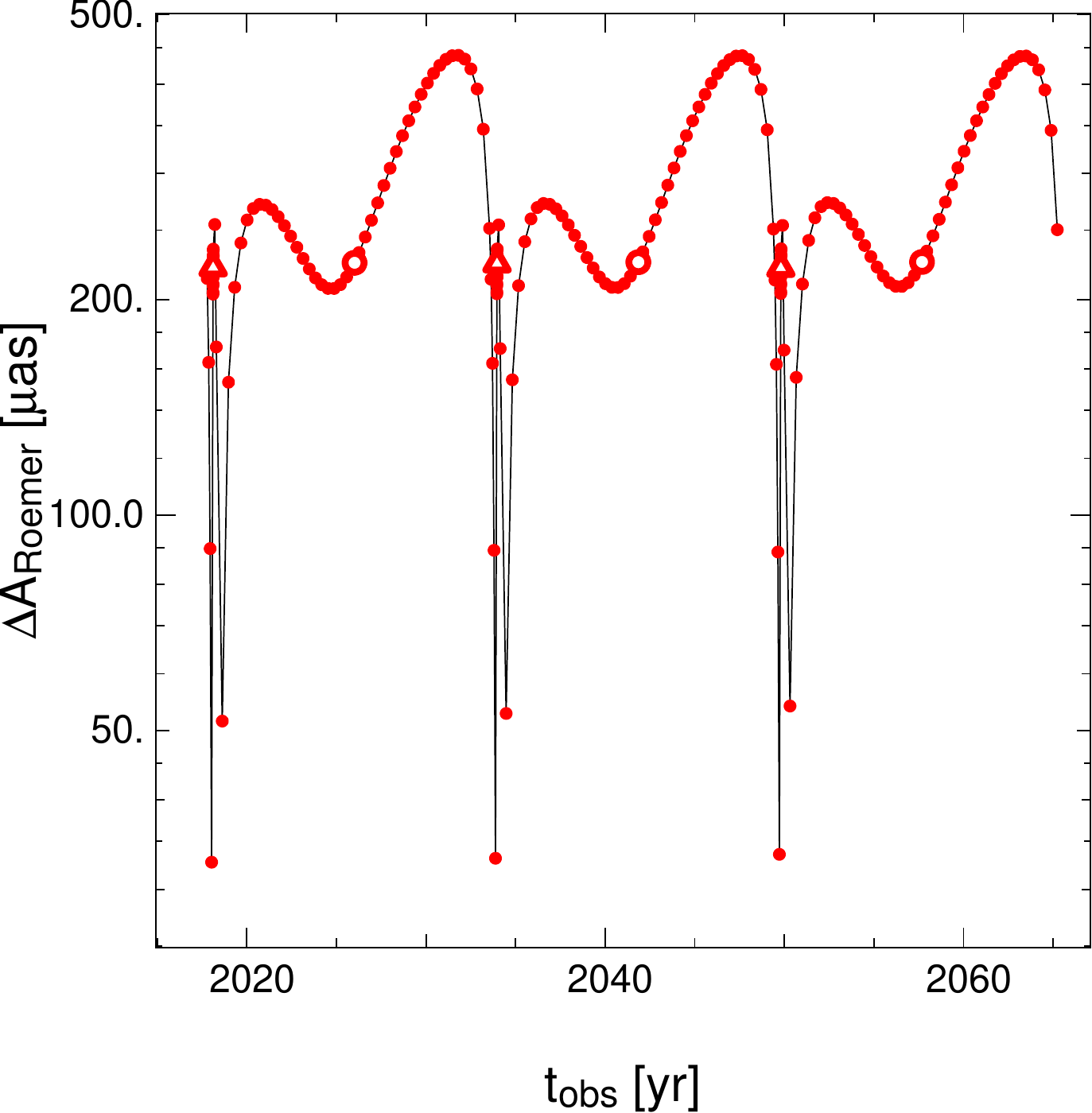}
      \quad
      \includegraphics[scale=0.3]{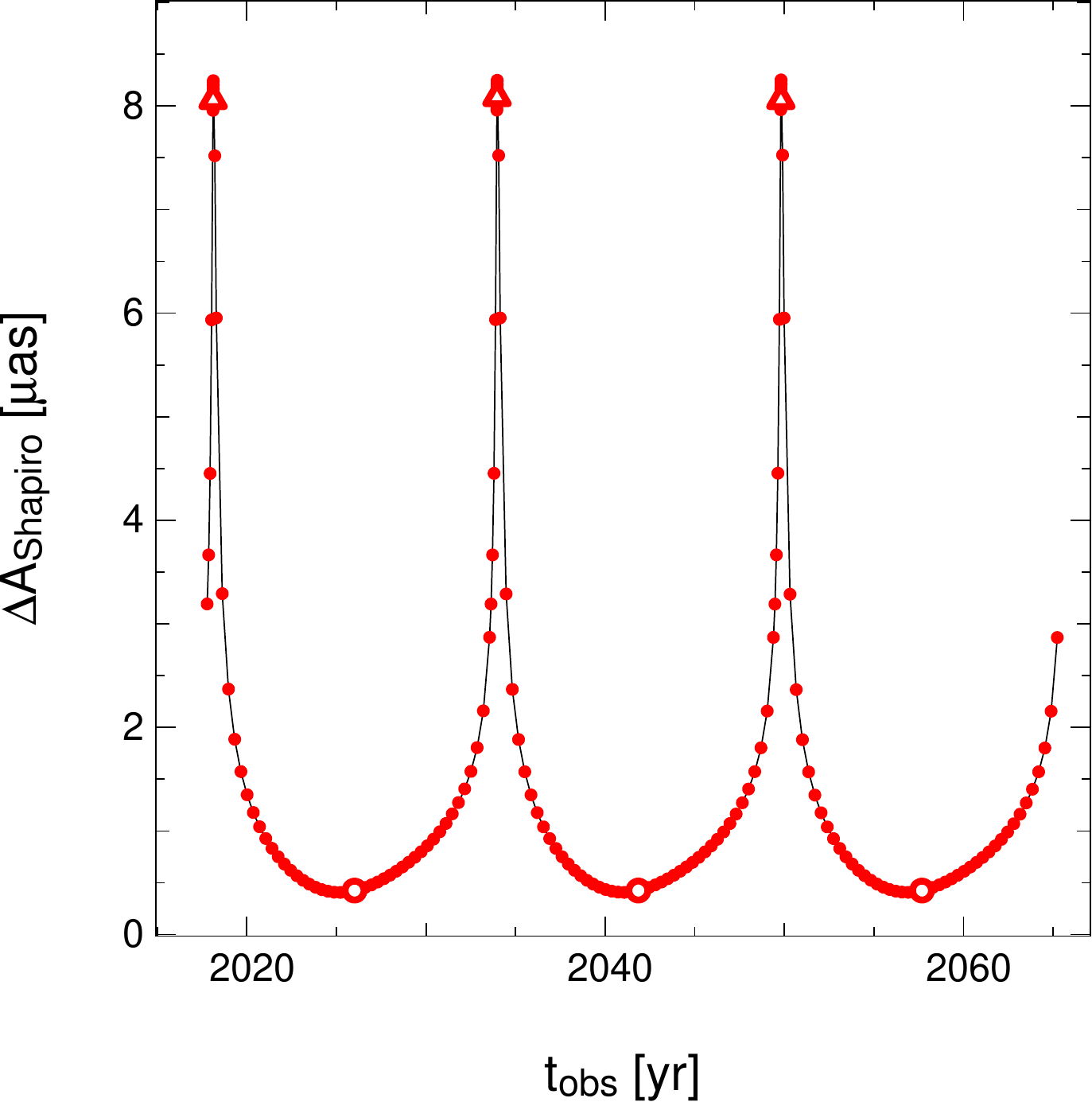}
            \quad
       \includegraphics[scale=0.3]{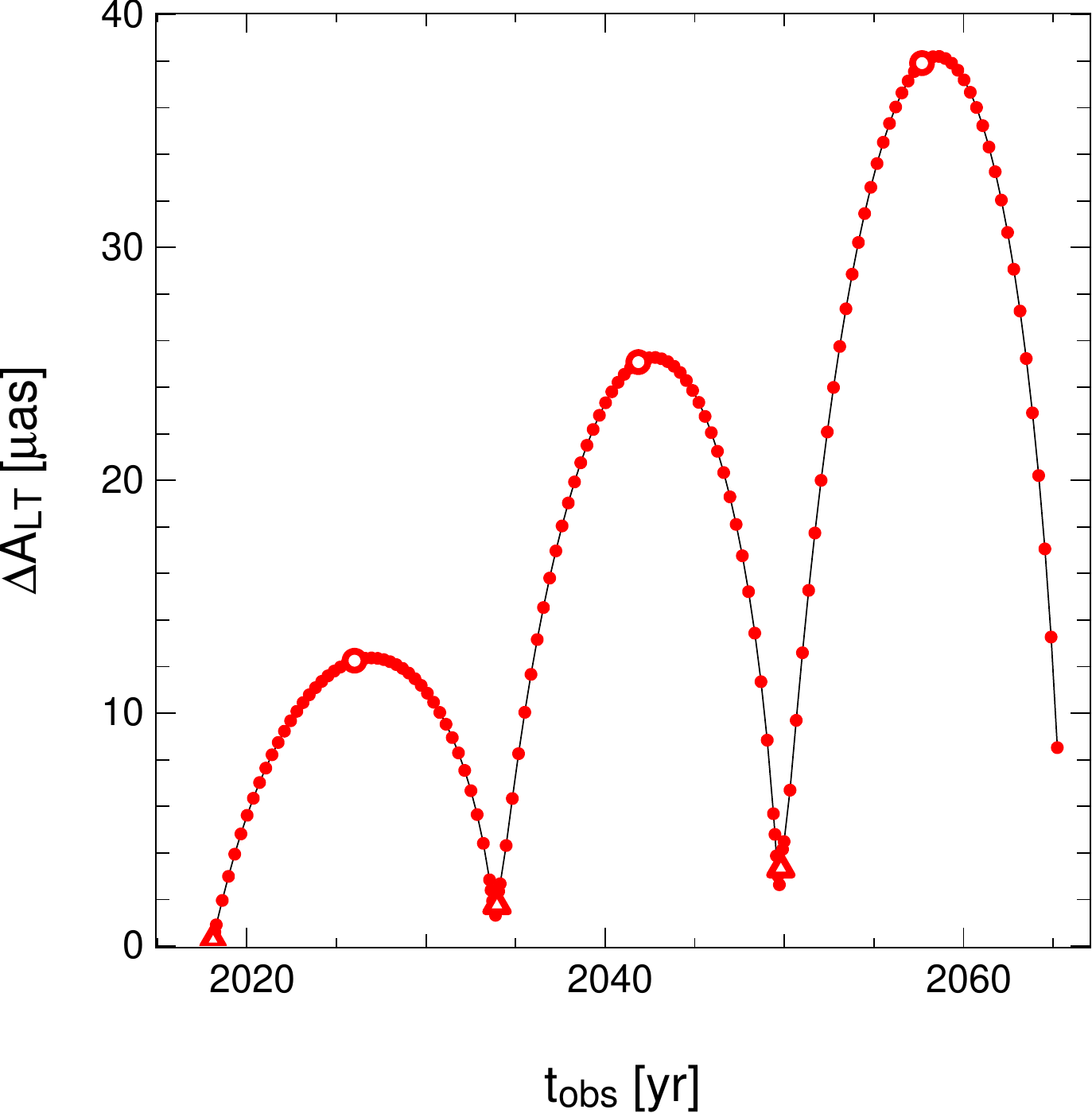}
       \quad
      \includegraphics[scale=0.3]{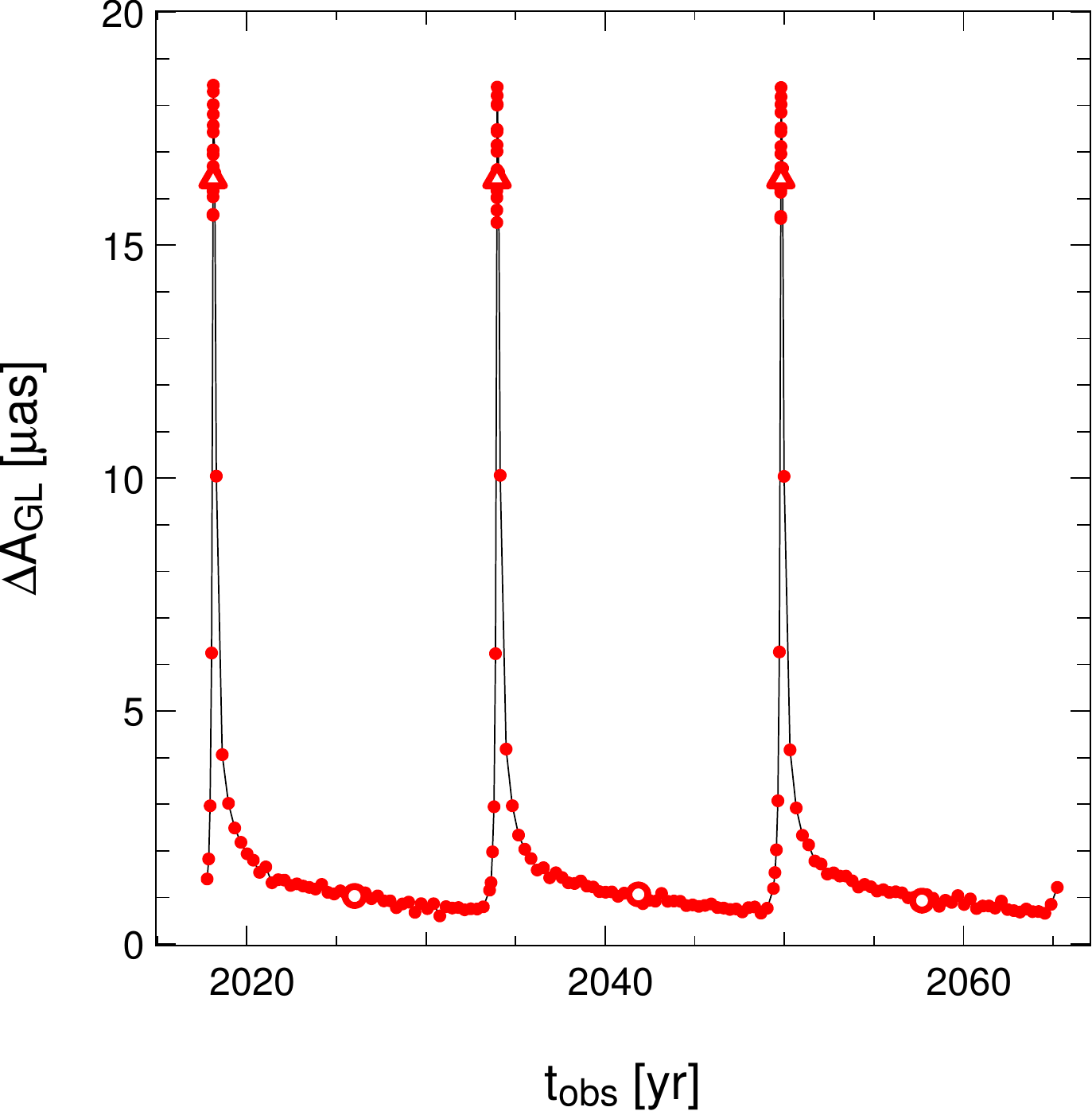}
   \caption{Astrometric impact of different effects on the S2 star observations obtained during three periods. \textit{First plot}: Roemer time delay. \textit{Second plot}: Shapiro time delay. \textit{Third plot:} Lense-Thirring effect considering the angular momentum parameters $a=0.99$, $i'=45^{\circ}$ and $\Omega'=160^{\circ}$. \textit{Fourth plot:} gravitational lensing. Open red circles and triangles denote the position of the apocenter and the pericenter, respectively. See Appendix~\ref{app_E_2} for the precise definition of the various quantities used in these plots.}
        \label{fig:astroS2}
\end{figure*}

The two first plots in Fig.~\ref{fig:astroS2} present the time delay effects on the S2 astrometric positions: the left plot shows the influence of the Roemer effect and the right plot gives the impact of the Shapiro effect. We can see that the influence of this first effect is always higher than the best astrometric accuracy of GRAVITY (10 $\mu$as) and can reach $\approx 450$ $\mu$as. It shows that this contribution cannot be neglected in stellar-orbit models used to interpret the GRAVITY data. We remind that this effect is not a GR effect. We note that the Roemer time delay has already been considered to treat the data obtained on the S stars by \cite{2008ApJ...689.1044G} and \cite{2009ApJ...692.1075G}. The shift induced by the Shapiro time delay is always lower than 10 $\mu$as. The maximal offsets are reached near pericenter at $\approx$ 8 $\mu$as.

On the third plot in Fig.~\ref{fig:astroS2} we can see the Lense-Thirring effect, where we considered $a=0.99$, $i'=45^{\circ}$ and $\Omega'=160^{\circ}$. It takes into account the black hole angular momentum effect on both the S2 stellar orbit and the photon path. However, the contribution of the Lense-Thirring effect on the null geodesic is negligible: the astrometric impact of this effect on both Shapiro time delay and gravitational lensing is $\ll1$ $\mu$as. Thus, the plot only shows the Lense-Thirring impact on the S2 time-like geodesic. This effect is similar to pericenter advance since it increases with the number of turns made by the star around the black hole. During the first, second and third period it reaches $\approx$~10~$\mu$as, $\approx$~25~$\mu$as and $\approx$~40~$\mu$as near apocenter passage, respectively. Near pericenter passages, the offset is always lower than 10 $\mu$as. It shows that in the black hole configuration considered here ($a=0.99$, $i'=45^{\circ}$ and $\Omega'=160^{\circ}$), the Lense-Thirring effect is negligible near pericenter meaning that it is important to observe near apocenter if we want to investigate a constraint on the black hole angular momentum parameters. A deeper analysis, considering different values for the parameters $a$, $i'$ and $\Omega'$, will be given in Sect.~\ref{cons_spin}. We note that the results discussed here are consistent with those given in \cite{2015ApJ...809..127Z} and \cite{2016ApJ...827..114Y}.

The last plot in Fig.~\ref{fig:astroS2} shows the gravitational lensing effect on apparent positions of the S2 star. The maximal shifts are reached near pericenter passages and are of about 20~$\mu$as. Most of the time, gravitational lensing is as low as 2 $\mu$as; close to apocenter passages, they reach $\approx$~1~$\mu$as. 

To summarize, maximal astrometric offsets due to each relativistic effect are listed in Table~\ref{OffAstro}.

{\renewcommand{\arraystretch}{1.1}
\begin{table}[!h]
\begin{center}
\caption{Maximal astrometric offsets in $\mu$as reached at first, second and third periods of the S2 star, due to different relativistic effects and considering the black hole angular momentum parameters $a=0.99$, $i'=45^{\circ}$, $\Omega'=160^{\circ}$. (Pe) and (Ap) denote the pericenter and apocenter passages, respectively. This table shows whether the maximal offset appears near the pericenter or the apocenter passages. When not mentioned, the shift appears for each observation date considered. See the legend of Table~\ref{table:models} for the acronyms used.
}
\label{OffAstro}
\begin{tabular}{lccc} 
        \hline\hline
         Effects  & $1^{\mathrm{st}}$ period  & $2^{\mathrm{nd}}$ period & $3^{\mathrm{rd}}$ period \\   
        \hline
        
        PA & 3000 (Pe) &  8000 (Pe) & $16 000$ (Pe) \\  
        Shapiro & 8 (Pe) & 8 (Pe) & 8 (Pe) \\
        LTS & 10 (Ap) & 25 (Ap) & 40 (Ap) \\
        LTP & $\ll 1$ & $\ll 1$ & $\ll1$ \\
        GL & 20 (Pe) & 20 (Pe) & 20 (Pe)\\
        \hline  
\end{tabular}
\end{center}
\end{table}
}

\subsection{Effects affecting the spectroscopy of the S2 star}
\label{models_radvel}

Here, we want to show the influence of different effects on the S2 radial velocity measurements. We focus on the pericenter advance, the Roemer time delay, the  transverse Doppler shift, the gravitational redshift and the Lense-Thirring effect. We also look for the impact of using an approximated radial velocity (see Eq.~\eqref{eq:red}). More precisely, it means that we are interested in the cumulative spectroscopic impact of high-order effects affecting the photon trajectory such as the Shapiro time delay; we denote these cumulative contributions High-Order Photon Curvature (HOPC).

\begin{figure}[!t]
    \centering
        \includegraphics[scale=0.5]{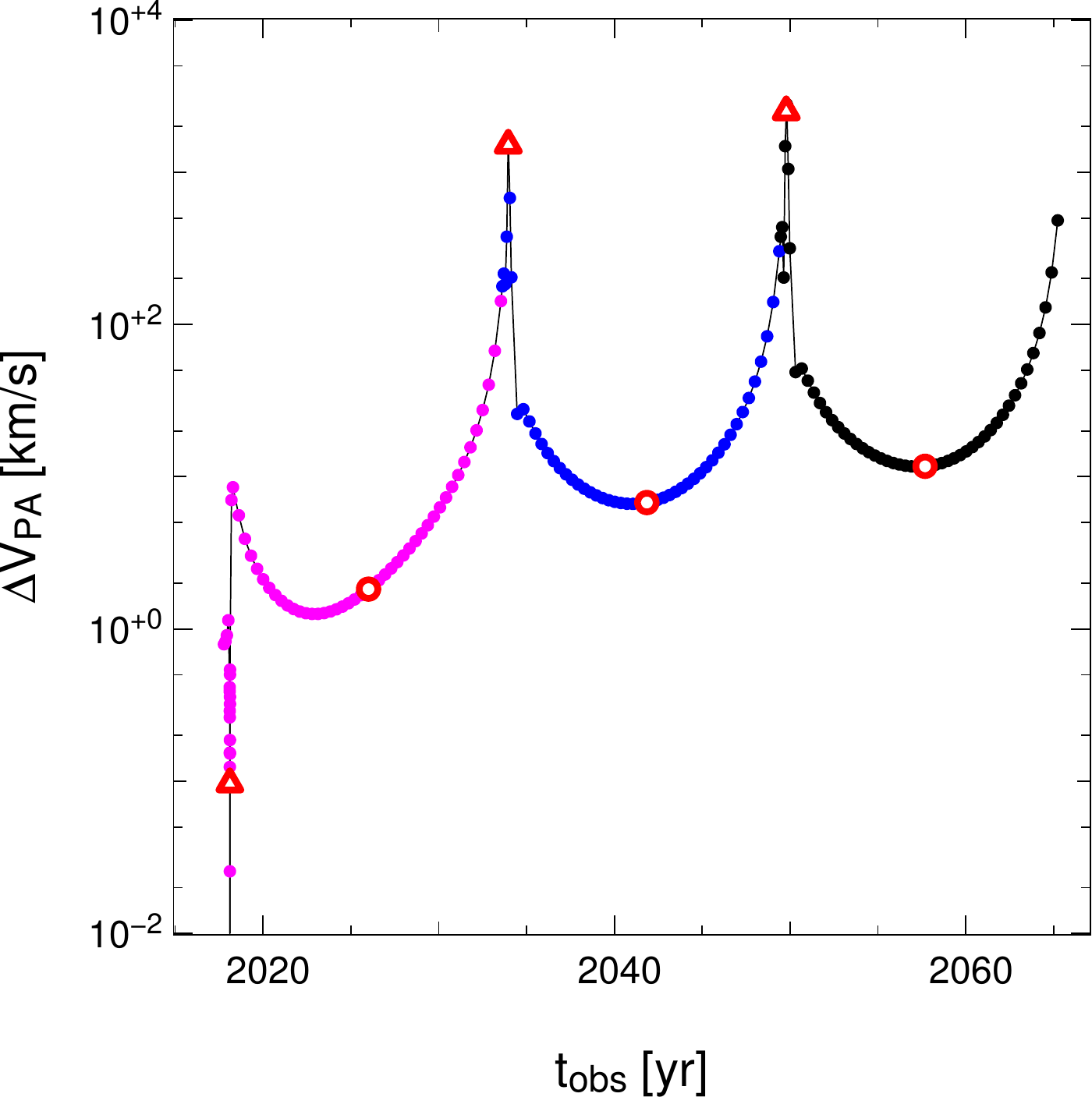}
    \caption{Spectroscopic impact of the pericenter advance on the S2 star observed during three orbital periods. Solid magenta, blue, and black circles correspond to the first, second, and third S2 period, respectively. Open red circles and triangles represent the apocenter and pericenter passages, respectively.}
    \label{fig:VAP}
\end{figure}

As for the astrometry, the impact of the pericenter advance on measured radial velocity increases with the number of orbits made by S2. Fig.~\ref{fig:VAP} shows the spectroscopic shift obtained when comparing radial velocities evaluated with a Keplerian orbit with those estimated with a GR orbit. The first maximal shift appears near the second pericenter passage (last magenta dot in Fig.~\ref{fig:VAP}) and is of about 140~km/s. At second and third periods, the maximal offset reaches $\approx$~1520~km/s and $\approx$~2800~km/s, respectively. 

\begin{figure*}[htb]
\centering
      \includegraphics[scale=0.3]{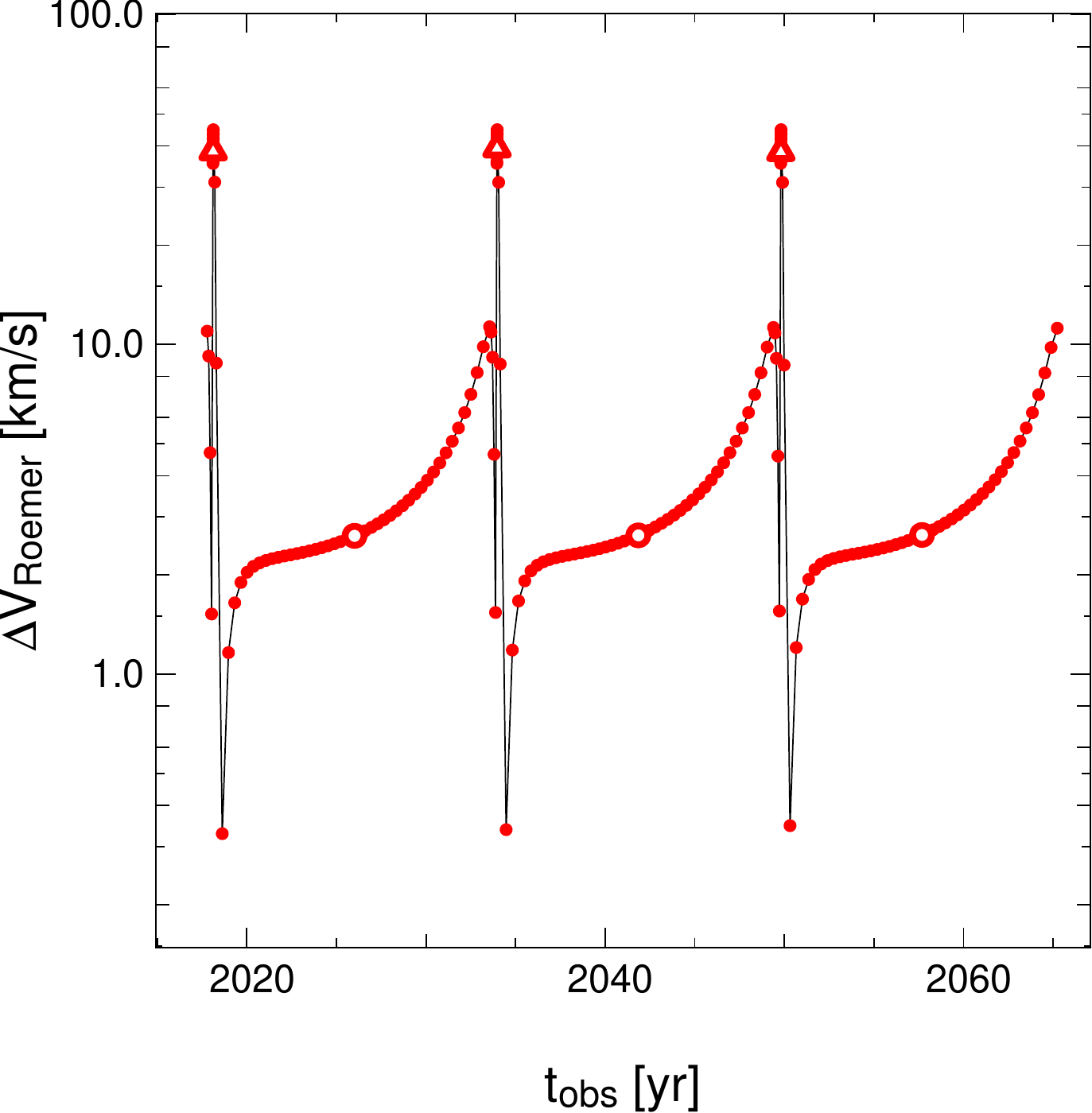}
       \quad
       \includegraphics[scale=0.3]{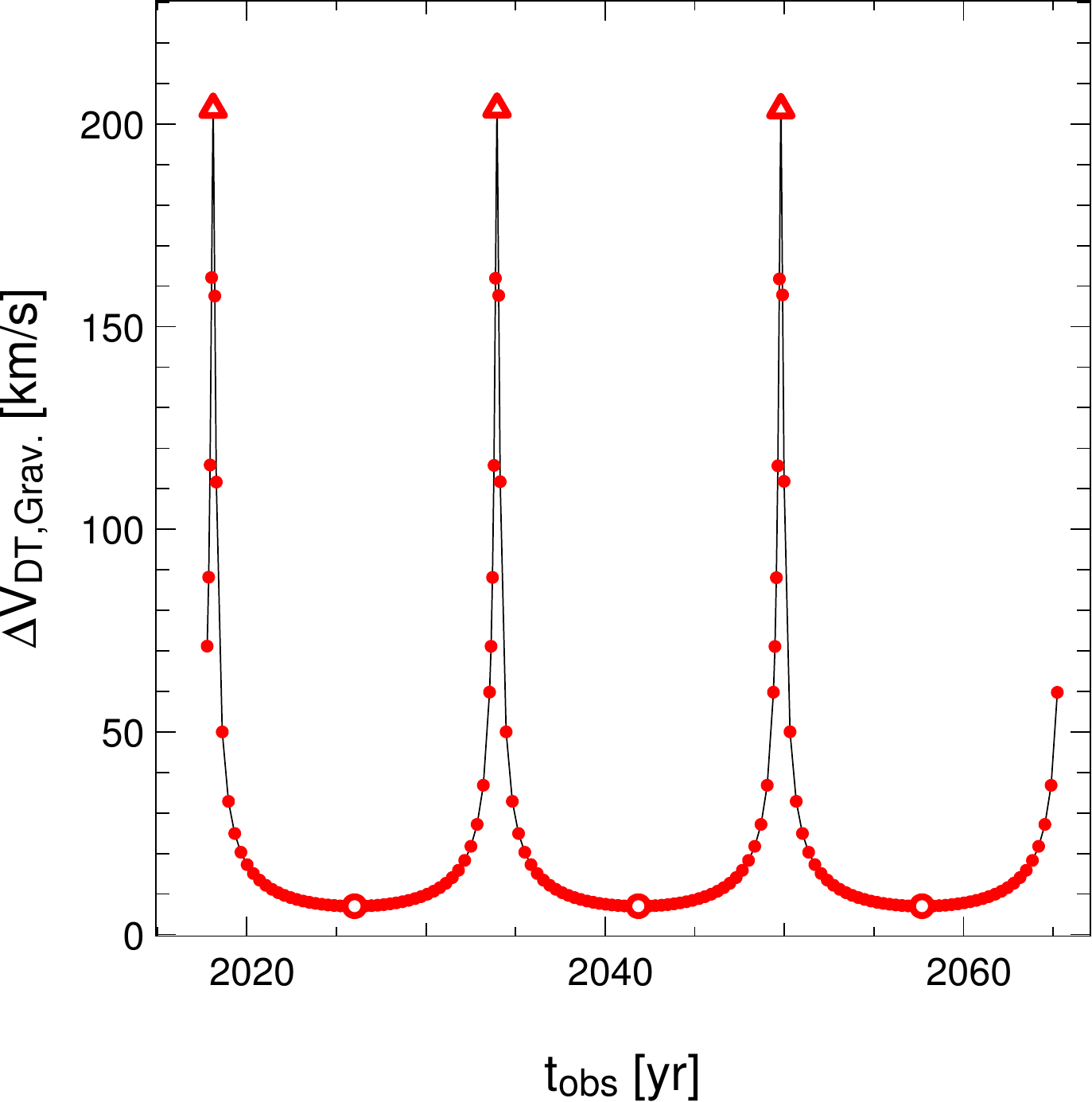}
       \quad
      \includegraphics[scale=0.3]{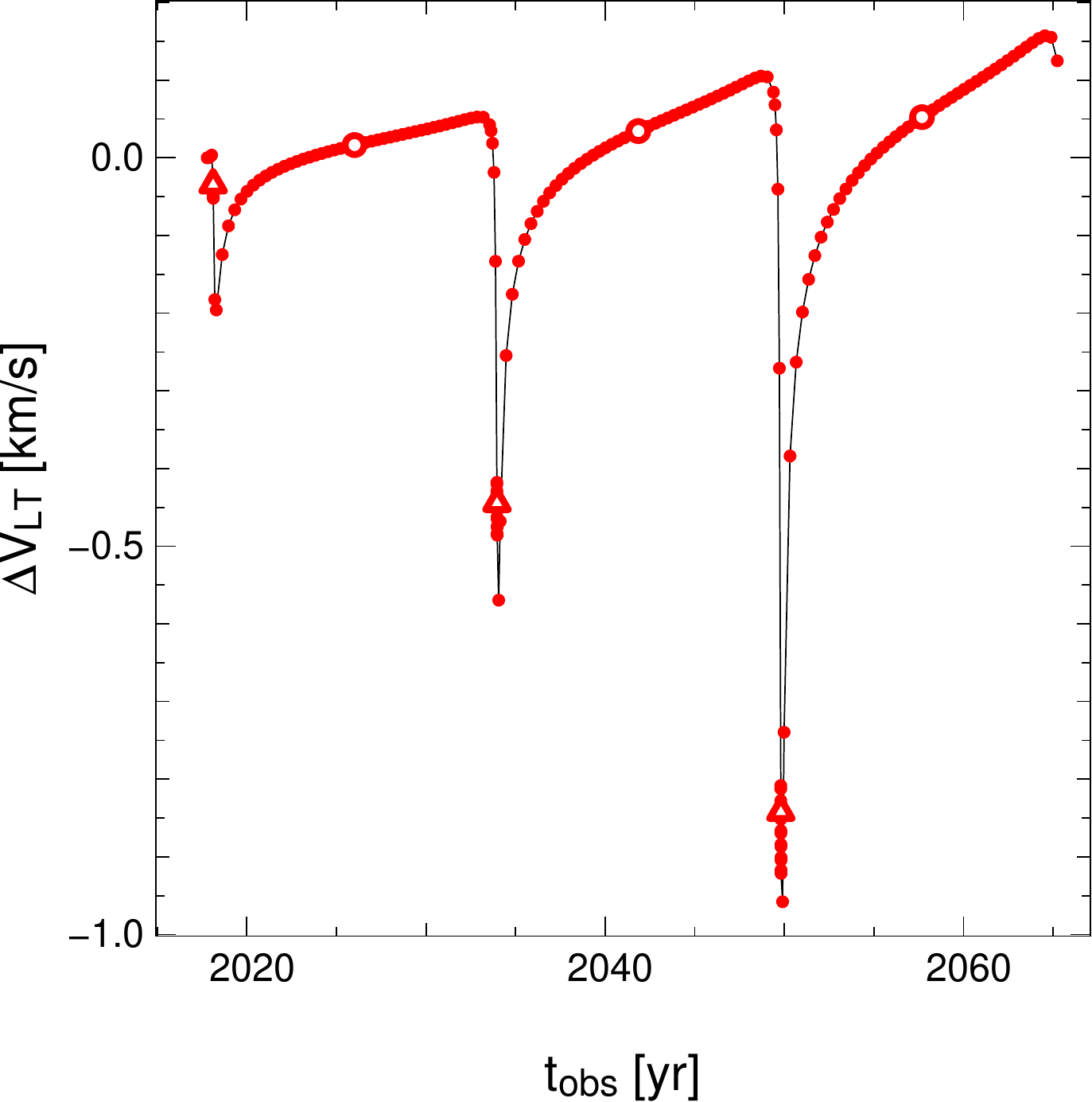}
      \quad
      \includegraphics[scale=0.3]{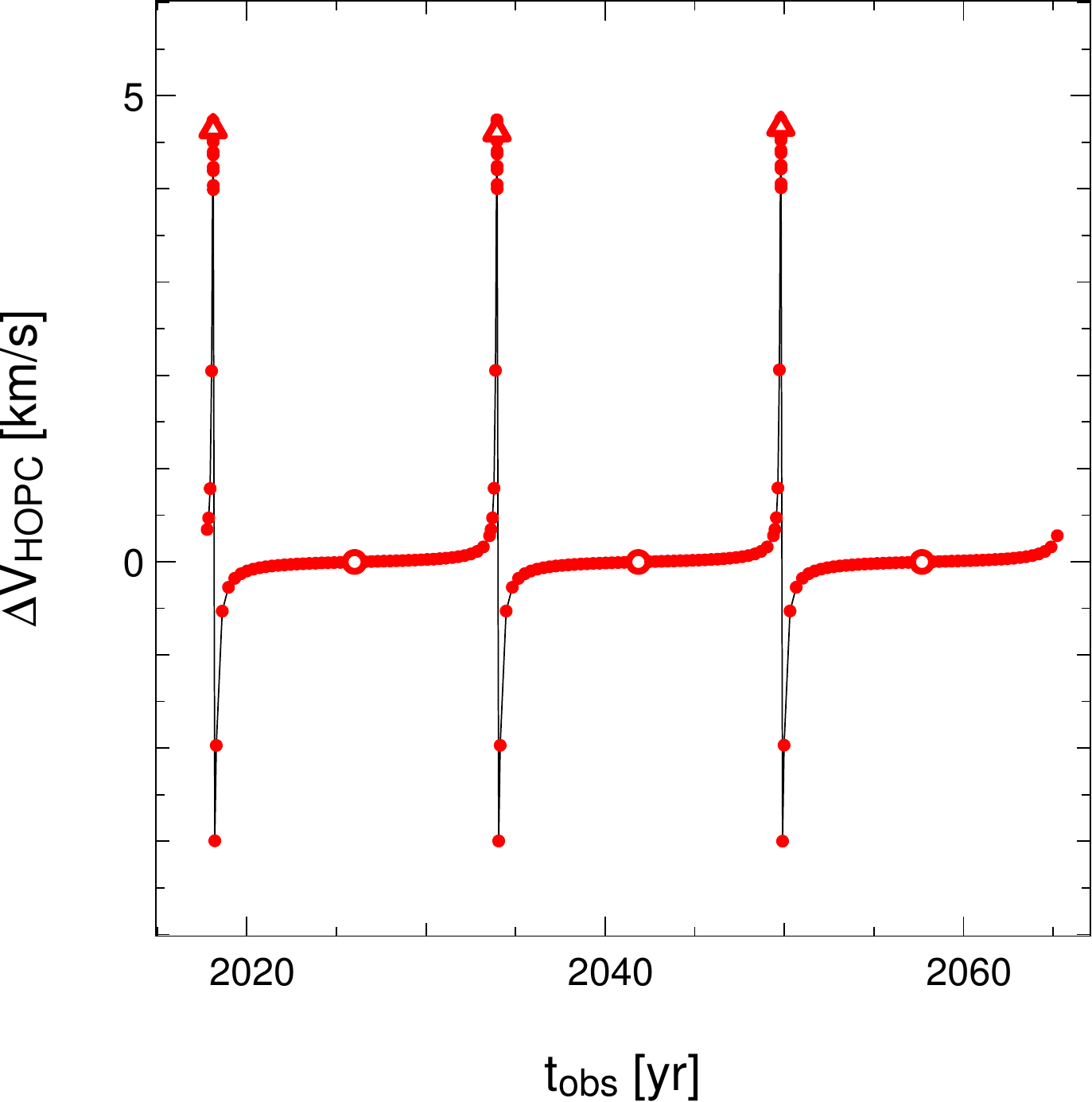}
   \caption{Spectroscopic impact of different effects on the S2 star observations obtained during three periods. \textit{First plot}: Roemer time delay. \textit{Second plot}: transverse Doppler shift and gravitational redshift. \textit{Third plot:} Lense-Thirring effect considering the angular momentum parameters $a=0.99$, $i'=45^{\circ}$ and $\Omega'=160^{\circ}$. \textit{Fourth plot:} HOPC contributions. Open red circles and triangles denote the position of the apocenter and the pericenter, respectively. See Appendix~\ref{app_E_2} for the precise definition of the various quantities used in these plots.}
        \label{fig:RVS2}
\end{figure*}

The first plot in Fig.~\ref{fig:RVS2} presents the influence of the Roemer time delay on radial velocity of the S2 star. The maximal shifts reach $\approx 50$ km/s near pericenter passages. 

The second plot in Fig.~\ref{fig:RVS2} shows the cumulative influence of the transverse Doppler shift and the gravitational redshift. Here again, the maximal offsets are reached near pericenter. The highest values are $\approx$~200~km/s which is consistent with \cite{2006ApJ...639L..21Z}. Near apocenter passages the shifts are lower than 10 km/s.

The Lense-Thirring effect on radial velocities is visible on the third plot in Fig.~\ref{fig:RVS2}. It essentially shows the shift due to its impact on the star trajectory. We see that at each period, the absolute maximal shifts are reached near pericenter passages, but the influence of this effect is very low and always below 1~km/s. These results are similar to those found by \cite{2010ApJ...711..157A}, \cite{2015ApJ...809..127Z} and \cite{2016ApJ...827..114Y}.

The last plot in Fig.~\ref{fig:RVS2} corresponds to the impact of the HOPC contributions. The maximal offsets are reached near pericenter and are $\approx$~5~km/s. As for astrometry, the Lense-Thirring effect on the photon trajectory essentially  does not modify the radial velocities of the S2 star ($\approx~10^{-2}$~km/s). The observed shift here is thus due to other approximations such as the Shapiro time delay. 

Maximal spectroscopic offsets obtained with each relativistic effect are listed in Table~\ref{OffRadVel}.\\

We can conclude that all relativistic effects on radial velocity are maximal near S2 pericenter passages (but not exactly at pericenter). However, considering the astrometry, the Lense-Thirring effect is maximal near apocenter passages. Moreover, the Lense-Thirring effect on the photon path is negligible for both astrometry and spectroscopy. Concurrently, the Schwarzschild spacetime curvature and the Lense-Thirring effect on the star trajectory are not negligible. In particular, this latter effect should be marginally detected at first apocenter passage by the GRAVITY instrument ($\approx$~10~$\mu$as $-$ in 2026). 
 
{\renewcommand{\arraystretch}{1.1}
\begin{table}[!h]
\begin{center}
\caption{As for Table~\ref{OffAstro} but for spectroscopy; the shifts are given in km/s. }
\label{OffRadVel}
\begin{tabular}{lccc} 
        \hline\hline
         Effects  & $1^{\mathrm{st}}$ period  & $2^{\mathrm{nd}}$ period & $3^{\mathrm{rd}}$ period \\   
        \hline
        PA & 140 (Pe) & 1520 (Pe) & 2800 (Pe) \\
        TD, Grav. & 200 (Pe) & 200 (Pe) & 200 (Pe) \\
        LTS & 0.2 (Pe) & 0.5 (Pe) & 0.9 (Pe) \\
        LTP & $10^{-2}$ & $10^{-2}$ & $10^{-2}$ \\
        HOPC & 5 (Pe) & 5 (Pe) & 5 (Pe) \\
        \hline
\end{tabular}
\end{center}
\end{table}
}

\subsection{Differences between models A to F and the full-GR model}

We report in Table~\ref{table:diffmodels} the maximal astrometric and spectroscopic differences between models A to F and model G. 

The maximal astrometric differences between Keplerian models (A, B and C) and the full-GR model are due to the fact that the pericenter advance is not taken into account in those models. Spectroscopic differences are also mainly dominated by the absence of the pericenter advance. However, we note variations between Keplerian models, which are explained by the fact that models B and C take into account different effects: Model B includes the Roemer time delay, and Model C considers the Roemer time delay, the transverse Doppler shift and the gravitational redshift.

At the first S2 period, the astrometric difference between models D and G is dominated by the absence of gravitational lensing. At second and third periods, it is dominated by the absence of the Lense-Thirring effect on the star trajectory. For Model E, the astrometric difference is only due to the fact that it neglects gravitational lensing. Finally, for Model F, the astrometric shift is induced by the fact that gravitational lensing is not completely reproduced by the approximations of \cite{2006PhRvD..74l3009S}. Radial velocities of models D to F are shifted from those computed with Model G by $\approx$~5~km/s. This offset is due to the HOPC contributions.

{\renewcommand{\arraystretch}{1.1}
\begin{table}[!h]
\begin{center}
\caption{Maximal astrometric and spectroscopic differences of each model with respect to the full-GR model (Model G) considering three orbital periods of the S2 star.}
\label{table:diffmodels}
\begin{tabular}{cccc} 
\hline\hline
         Models  & $1^{\mathrm{st}}$ period  & $2^{\mathrm{nd}}$ period & $3^{\mathrm{rd}}$ period \\     
        \hline 
        
        A & 3 mas & 8 mas & 16 mas \\
           & 250 km/s & 1760 km/s & 3010 km/s \\
          \hline 
        B & 3 mas & 8 mas & 16 mas \\
           & 210 km/s & 1700 km/s & 2980 km/s \\
        \hline 
        C & 3 mas & 8 mas & 16 mas \\
           & 140 km/s & 1520 km/s & 2800 km/s \\
         \hline
        D & 20 $\mu$as & 25 $\mu$as & 40 $\mu$as \\
           & 5 km/s & 5 km/s & 5 km/s \\
           \hline
        E & 20 $\mu$as & 20 $\mu$as & 20 $\mu$as \\
           & 5 km/s & 5 km/s & 5 km/s \\
           \hline
        F & 7 $\mu$as & 7 $\mu$as & 7 $\mu$as \\
           & 5 km/s & 5 km/s & 5 km/s \\
           \hline          
\end{tabular}
\end{center}
\end{table}
}


\section{Fitting}
\label{fitting}

In the following sections, we give an estimation of the threshold times above which we can detect the different effects discussed above with observations of the S2 star, and considering various astrometric and spectroscopic accuracies. We also focus on the constraint on the black hole angular momentum parameters with this star. In this section, we explain the methods used to estimate these different threshold times, and to constrain the angular momentum.

\subsection{Procedures}
\label{procedures}

Each model is described by eight parameters corresponding to the seven orbital parameters and the distance $R_0$ between the observer and the black hole. The mass of the black hole is not an individual parameter to fit but varies through the third Kepler's law: $4 \pi^2 a_{\mathrm{sma}}^3/T^2 = G M$. Models E, F and G are described by three other parameters since they take into account the Lense-Thirring effect. Those parameters are the norm and direction of the angular momentum of the black hole: $a$, $i'$ and $\Omega'$. \\

To estimate the different minimal observation times needed to detect the various effects, we determine the threshold times above which the models A to F fail to fit the full-GR observations generated with Model G. We mention that the fitting method used for Keplerian models and relativistic models will be different. The reason for this is explained below.

The fitting procedure used to fit models A, B and C is the Levenberg-Marquardt method \citep{citeulike:2946351} based on the least-squares method. The quantity to minimize is
\begin{equation}
\chi^2 = \sum_{i = 1}^{N}{\left[\frac{(\alpha_{\mathrm{obs},i}-\alpha_{\mathrm{m},i})^2 + (\delta_{\mathrm{obs},i}-\delta_{\mathrm{m},i})^2}{\sigma_{\mathrm{A},i}^2} + \frac{(\mathrm{V}_{\mathrm{obs},i}-\mathrm{V}_{\mathrm{m},i})^2}{\sigma_{\mathrm{V,i}}^2}\right]}
,\end{equation}
where $N$ is the number of observation dates. The quantities labeled \textit{obs} correspond to the mock observations generated with the full-GR model. Those labeled \textit{m} are obtained with models A, B or C. The different accuracies $(\sigma_\mathrm{A}, \sigma_{\mathrm{V}})$  considered to estimate the different threshold times are listed in Sect.~\ref{obs}. The initial parameters (initial guess) considered for the fitting of models A, B and C are given by $\mathbf{P}_\mathrm{init}~=~\mathbf{P}_{\mathrm{Gillessen}} + 1 \boldsymbol{\sigma}_{\mathrm{Gillessen}}$ where $\mathbf{P}_{\mathrm{Gillessen}}$ is a vector containing the best-fit parameters of \cite{2009ApJ...692.1075G} given in Sect.~\ref{obs}, and $1\boldsymbol{\sigma}_{\mathrm{Gillessen}}$ is the vector containing the $1\sigma$ error of each parameter also estimated by \cite{2009ApJ...692.1075G}: $\sigma_T = 0.11$ year, $\sigma_{a_{\mathrm{sma}}} = 0.001''$, $\sigma_e = 0.003$, $\sigma_{t_p} = 0.01$ year, $\sigma_{\Omega} = 0.84^\circ$, $\sigma_{\omega} = 0.84^\circ$, $\sigma_i = 0.47^\circ$, $\sigma_{R_0} = 0.48$~kpc.

\begin{figure}[!t]
\centering
      \includegraphics[scale=0.35]{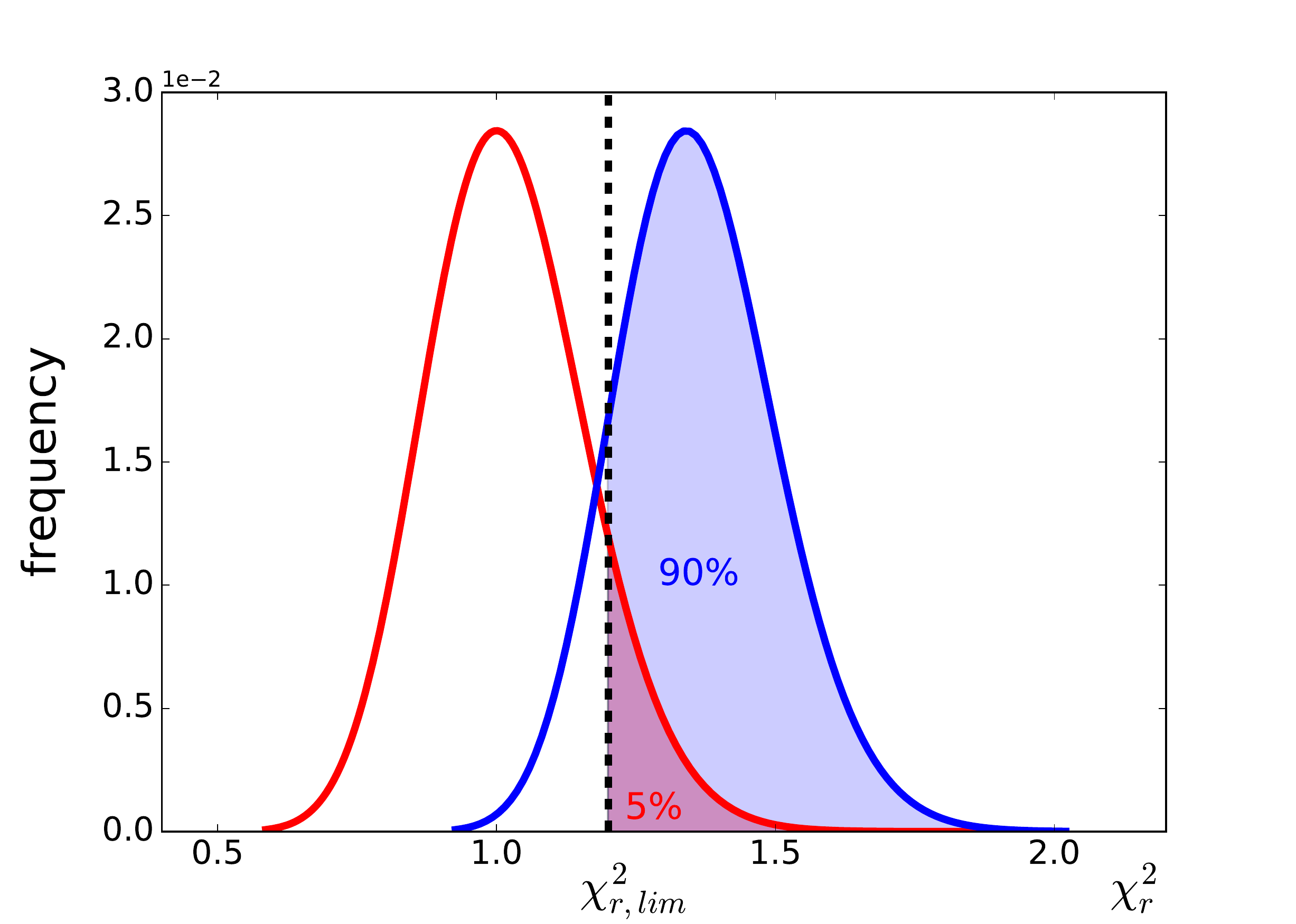} 
   \caption{\textit{Red curve:} distribution of $\chi_r^2$ obtained with a model describing the observations well. The distribution follows a reduced $\chi^2$ law. The $\chi^2_{r,lim}$ value (see the text for a definition of this quantity) is marked by the dashed line and is obtained considering $p=5\%$ in equation~\eqref{eq:proba}. \textit{Blue curve:} example of a $\chi_r^2$ distribution when the model does not describe the observations well.  $90\%$ of the $\chi_r^2$  are higher than the $\chi^2_{r,lim}$. For both curves, we consider $k=100$.}
        \label{chi2}
\end{figure}

If the model fits the data well, the distribution of $\chi^2$ obtained by evaluating this quantity  several times must follow a $\chi^2$ law with $k$ degrees of freedom\footnote{see \cite{2010arXiv1012.3754A} for more details on how to estimate the degrees of freedom of a model.} where $k = 3N - n$, with $n$ being the number of fitted parameters and where the factor three corresponds to the fact that we consider both astrometry ($\alpha$, $\delta$) and spectroscopy; see the red curve in Fig.~\ref{chi2} for an illustration of such a distribution, when considering a reduced $\chi^2$ defined by
\begin{equation}
\chi^2_r=\frac{\chi^2}{k}.
\end{equation}
 
To distinguish a bad fit from a good fit we can use the $\chi^2$ test. This test allows us to determine whether one evaluation of the quantity $\chi^2$ is consistent with the assumption that it is drawn from a $\chi^2$ law. In other words, it tests the null hypothesis given by
\begin{equation}
H_0 = \textrm{the model fits the observations well.}
\end{equation}
To test the null hypothesis $H_0$, we determine a limit $\chi^2$ noted $\chi^2_{lim}$ by solving the equation
\begin{equation}
\label{eq:proba}
P(\chi^2 > \chi_{lim}^2) = 1 - F_{\chi^2}(\chi_{lim}^2) = p
,\end{equation}
where $F_{\chi^2}(\chi_{lim}^2)$ is the cumulative distribution function (CDF) of the $\chi^2$ law with $k$ degrees of freedom, and $p$ is the probability of incorrectly rejecting $H_0$ , which we fix at $5\%$. Thus, if the condition $\chi_r^2 > \chi_{r,lim}^2$ is verified, knowing $H_0$, we incorrectly reject the model in $5\%$ of cases. It means that if we fit for instance a model 1000 times to data generated with this same model, there will be 5\% of the 1000 estimations of $\chi_r^2$ which will be superior to $\chi_{r,lim}^2$ (see the red curve in Fig.~\ref{chi2}). We specify that the 1000 $\chi_r^2$ are obtained by fitting the model to 1000 runs of observation which have the same duration in time, but whose noise differs by using a random draw of a normal law. On Table~\ref{chi2lim} several values of $\chi_{r,lim}^2$ are evaluated for the models A to F and different runs of observation.

The $\chi^2$ test requires only one estimation of $\chi^2$. However, we decide to use a method allowing to meaningfully reject the null hypothesis. More precisely, we chose to fit a Keplerian model 100 times to observations generated with Model G, and to reject the model if 90\% of the $\chi_r^2$ are superior to $\chi_{r,lim}^2$ (see the blue curve in Fig.~\ref{chi2}). The minimal observation time above which we consider that the model fails to describe the data is thus equal to the time duration of the run where 90\% of the $\chi_r^2$ satisfy $\chi_r^2 > \chi_{r,lim}^2$. Strictly speaking, if we consider an infinite number of $\chi_r^2$ realizations and that more than 5\% of the $\chi_r^2$ satisfy $\chi_r^2 > \chi_{r,lim}^2$, then the model can be rejected. However, in this paper we consider a finite number of fits, and we thus choose to be conservative and claim that a model fails when it reaches 90\%. In addition, we chose to report the minimal observation times obtained when 60\% of the $\chi_r^2$ verify $\chi_r^2 > \chi_{r,lim}^2$, since the model can already be rejected with sufficiently high confidence at such a percentage. 

In order to validate the procedure defined here, we fit Model C 100 times to observations obtained also with this model. The aim is to verify that we find 5\% of the $\chi_r^2$ satisfying $\chi_r^2~>~\chi_{r,lim}^2$. For doing so, we chose two runs of observation: 32~years ($\approx$ 2 periods of S2) with $(\sigma_\mathrm{A}, \sigma_{\mathrm{V}})~=~(10$~$\mu$as, 1~km/s) and 4~months with $(\sigma_\mathrm{A}, \sigma_{\mathrm{V}})~=~(100$~$\mu$as, 100~km/s). In both cases, we find $\approx~$ 6\% of the $\chi_r^2$ satisfying $\chi_r^2 > \chi_{r,lim}^2$ which is very close to the expected value. Such results show that 100 fits are sufficient to get a first estimation of the minimal observation times.
  
We also apply this test to Model E: fit 100 times Model E considering the Levenberg-Marquardt method, to observations generated with this model. In this case, we find that the fitting method is not appropriate since  we obtain $\approx$~35\% of the $\chi_r^2$ which satisfy $\chi_r^2 > \chi_{r,lim}^2$ when considering for instance the run of 32 years and $(\sigma_\mathrm{A}, \sigma_\mathrm{V})~=~(10$~$\mu$as, 1~km/s). This percentage is explained by the fact that the parameter space is more difficult to probe with relativistic models. Indeed, if we repeat the test but consider the initial guess to be equal to the parameters used to generate the observations instead of $\mathbf{P}_\mathrm{init} = \mathbf{P}_\mathrm{Gillessen} + 1 \boldsymbol{\sigma}_{\mathrm{Gillessen}}$, we find $\approx$~6\%. The Levenberg-Marquardt method is thus not appropriate for highly non-linear models such as relativistic models, since the initial guess needs to be chosen close to the solution. This is the reason why we chose another fitting method for models D, E and F. We point out that we do not use the same method for all models because the computing time needed by the fitting method used for relativistic models is more important.\\

To estimate the minimal observation times above which the relativistic models fail to describe the observations obtained with Model G, we use the same protocol as for Keplerian models but we consider a Monte Carlo Markov Chain (MCMC) method to fit the different models \citep{2002mcss.book.....B}. More precisely, we use the \texttt{emcee}\footnote{\url{http://dan.iel.fm/emcee/current/}} software allowing MCMC simulations using the Affine Invariant Ensemble Sample method proposed by \cite{2013PASP..125..306F}. We specify here that if we note the observations,  $\vec{O,}$  and the vector containing the
parameters of a model chosen during the MCMC, $\vec{P}$ , the posterior probability density $\pi(\vec{P}|\vec{O})$ is given  by (using the Bayesian theorem)
\begin{equation}
\ln{\pi(\vec{P}|\vec{O})} \propto \ln{f(\vec{O}|\vec{P})} + \ln{\pi(\vec{P})}
,\end{equation}
where $f(\vec{O}|\vec{P})$ is the likelihood function expressed as
\begin{align}
\ln{f(\vec{O}|\vec{P})} =& -\frac{1}{2} \sum_{i=1}^N \left[\frac{(\alpha_{\mathrm{obs},i}-\alpha_{\mathrm{m},i})^2 + (\delta_{\mathrm{obs},i}-\delta_{\mathrm{m},i})^2}{\sigma_{\mathrm{A},i}^2}\right], \notag \\
& -\frac{1}{2} \sum_{i=1}^N \left[\frac{(\mathrm{V}_{\mathrm{obs},i}-\mathrm{V}_{\mathrm{m},i})^2}{\sigma_{\mathrm{V,i}}^2} \right], \notag \\
& -\frac{1}{2} \sum_{i=1}^N \left[2\ln \left ( 2\pi\,\sigma_{\mathrm{A},i}^2 \right ) + \ln \left ( 2\pi\,\sigma_{\mathrm{V},i}^2 \right )\right],
\end{align}
and $\pi(\vec{P})$ is the prior probability density of the parameters $\vec{P}$ which we chose distributed according to a uniform law: the orbital parameters and $R_0$ are uniformly chosen between $\vec{P}_{\mathrm{Gillessen}} - 6\boldsymbol\sigma_{\mathrm{Gillessen}}$ and $\vec{P}_{\mathrm{Gillessen}} + 6\boldsymbol\sigma_{\mathrm{Gillessen}}$, and the angular momentum parameters are uniformly chosen in there own domain of variation: $a \in [0,1]$\footnote{the norm $a$ does not vary between -1 and 1 because the direction of the angular momentum of the black hole is already defined by using the angles $i'$ and $\Omega'$.}, $i' \in [0^\circ,180^\circ]$ and $\Omega' \in [0^\circ, 360^\circ]$.

In order to check the fitting method used for models D to F, we compare the percentages of $\chi_r^2$ verifying the condition $\chi_r^2 > \chi_{r,lim}^2$ and obtained with both the Levenberg-Marquardt and MCMC methods. For doing so, we fit Model C 100 times to observations generated with Model G and obtained during one period with $(\sigma_\mathrm{A}, \sigma_\mathrm{V})~=~(30$~$\mu$as, 10~km/s). For both methods we find that 45\% of $\chi_r^2$  are superior to the $\chi_{r,lim}^2$. These results show that the fitting obtained with \texttt{emcee} is in accordance with that obtained with Levenberg-Marquardt. 

Besides, we apply the same test as done in the Keplerian part: fit Model E 100 times to observations generated with this model and obtained during $\approx$~2 periods of S2 with $(\sigma_\mathrm{A}, \sigma_\mathrm{V})~=~(10$~$\mu$as, 1~km/s). In this case, we find $\approx$~6\% instead of the 35\% obtained with the Levenberg-Marquardt method. We can thus say that the MCMC method is more appropriate for relativistic models than that of Levenberg-Marquardt.\\

In the section devoted to the black hole angular momentum constraint, we also use the \texttt{emcee} software. Since the majority of the parameters are already constrained by \cite{2009ApJ...692.1075G}, we chose to vary the different parameters in a domain whose bounds are defined by $\mathbf{P}_{\mathrm{Gillessen}}~\pm~1\boldsymbol\sigma_{\mathrm{Gillessen}}$. For the black hole angular momentum  parameters, they vary in their domain of variation: $a \in [0,1]$, $i' \in [0^\circ,180^\circ]$ and $\Omega' \in [0^\circ, 360^\circ]$. The different fittings will be done considering accuracies that can be used with current instruments:$10-30$~$\mu$as and 10~km/s. The model used to constrain the eleven parameters is Model F. The advantage of this model is the computing time since it does not use ray tracing. We remind that the maximum differences between models F and G is of about 7~$\mu$as and 5~km/s. In spite of those differences, the percentage of $\chi_r^2$ verifying $\chi_r^2 > \chi_{r,lim}^2$ when considering the runs of observation of three periods with $(\sigma_\mathrm{A}, \sigma_\mathrm{V}) = (10$~$\mu$as, 10~km/s) is only $\approx$~10\%. This shows that Model F seems sufficient to describe observations obtained with Model G and thus appropriate to investigate the constraint on the norm and direction of the angular momentum of the black hole.

{\renewcommand{\arraystretch}{1.1}
\begin{table}[!t]
\begin{center}
\caption{$\chi_{r,lim}^2$ estimated for different models and runs of observation. The values are estimated by considering $p=5\%$ in equation~\eqref{eq:proba}. We recall that the parameters $n$, $N$ et $k$ correspond respectively to the number of parameters describing the model, the number of data points and the degrees of freedom.
}
\label{chi2lim}
\begin{tabular}{| l | c | cc | cc |} 
\cline{3-6}
        \multicolumn{1}{c}{} & &\multicolumn{2}{c|}{A, B, C, D ($n=8$)} & \multicolumn{2}{c|}{E, F ($n=11$)} \\
        \cline{1-6}
        \multicolumn{1}{|c}{Runs} & \multicolumn{1}{|c}{$N$} & \multicolumn{1}{|c|}{$k$}& \multicolumn{1}{c|}{$\chi_{r,lim}^2$} & \multicolumn{1}{c|}{$k$} & \multicolumn{1}{c|}{$\chi_{r,lim}^2$} \\      
         \hline
         1 month & 15 & \multicolumn{1}{c|}{37} & 1.4106 & \multicolumn{1}{c|}{34} & 1.4295 \\
         2 months & 16 & \multicolumn{1}{c|}{40} & 1.3940 & \multicolumn{1}{c|}{37} & 1.4106 \\
         4 months & 18 & \multicolumn{1}{c|}{46} & 1.3659 & \multicolumn{1}{c|}{43} & 1.3791 \\
         6 months & 20 & \multicolumn{1}{c|}{52} & 1.3429 & \multicolumn{1}{c|}{49} & 1.3538 \\
         10 months & 21 & \multicolumn{1}{c|}{55} & 1.3329 & \multicolumn{1}{c|}{52} & 1.3429 \\
         1 year & 22 & \multicolumn{1}{c|}{58} & 1.3238 & \multicolumn{1}{c|}{55} & 1.3329 \\
         4 years & 30 & \multicolumn{1}{c|}{82} & 1.2700 & \multicolumn{1}{c|}{79} & 1.2753 \\
         6 years & 36 & \multicolumn{1}{c|}{100} & 1.2434 & \multicolumn{1}{c|}{97} & 1.2473 \\
         12 years & 53 & \multicolumn{1}{c|}{151} & 1.1965 & \multicolumn{1}{c|}{148} & 1.1986 \\
         16 years & 67 & \multicolumn{1}{c|}{193} & 1.1731 & \multicolumn{1}{c|}{190} & 1.1745 \\
         18 years & 89 & \multicolumn{1}{c|}{259} & 1.1488 & \multicolumn{1}{c|}{256} & 1.1497 \\
         20 years & 95 & \multicolumn{1}{c|}{277} & 1.1437 & \multicolumn{1}{c|}{274} & 1.1446 \\
         \hline
\end{tabular}
\end{center}
\end{table}
}

\subsection{Results}

\subsubsection{Constraint on various effects}
\label{ModelFits}

{\renewcommand{\arraystretch}{1.1}
\begin{table*}[hbt]
\caption{Estimations of the threshold times needed to detect different effects with the S2 star, considering various astrometric and spectroscopic accuracies. The main effects that cause each model to fail to explain the full-GR observations are listed in the left column (see Table~\ref{table:models} for the different acronyms). Threshold times given in brackets correspond to those obtained considering a percentage of failure of 60\%, instead of 90\%. When there are no brackets it means that the thresholds are similar for both percentages. The times given in square brackets for Model C are obtained considering mock observations of the S2 star generated with Model D instead of Model G.}
\label{Time}
\begin{center}
\begin{tabular}{| c | l | c | c | c | c |} 
\cline{1-1}
        Detected effects \\
        \cline{1-6}
        & Model A & 10 $\mu$as & 30 $\mu$as & 50 $\mu$as & 100 $\mu$as \\
        \cline{2-6}  
         Roemer & 1 km/s & 1 month  & 1 month  & 2 months  & 4(2) months  \\
         TD & 10 km/s & 2 months  & 4 months  & 6(4) months  & 1 an (10 months) \\
         Grav. & 100 km/s & 4 months  & 10(6) months  & 4(1) years  & 16(4) years  \\
        \cline{1-6}
        & Model B & 10 $\mu$as & 30 $\mu$as & 50 $\mu$as & 100 $\mu$as \\
        \cline{2-6}  
         TD & 1 km/s & 1 month  & 1 month  & 2 months  & 4 months  \\
         Grav. & 10 km/s & 2 months  & 4 months  & 6(4) months  & 10 months  \\
         & 100 km/s & 4 years (10 months) & 18(10) years  & 18 years  & 18 years  \\
        \cline{1-6}
        & Model C & 10 $\mu$as & 30 $\mu$as & 50 $\mu$as & 100 $\mu$as \\
        \cline{2-6}  
        PA & 1 km/s & 10(6) months [8(2) years]  & 14(12) years  & 18(14) years  & 18 years  \\
        GL & 10 km/s & 6(4) years [8(6) years] & 18 years  & 18 years  & 20 years   \\
        HOPC & 100 km/s & 6 years  & 18 years  & 18 years  & 20 years  \\
        \cline{1-6} 
        & Model D & 10 $\mu$as & 30 $\mu$as & 50 $\mu$as & 100 $\mu$as \\
        \cline{2-6}  
        GL & 1 km/s & 6 months  & > 30 years  & / & / \\
        HOPC & 10 km/s & 18(4) years  & > 30 years  & / & / \\
        & 100 km/s & / & / & / & / \\
        \cline{1-6}
        & Model E & 10 $\mu$as & 30 $\mu$as & 50 $\mu$as & 100 $\mu$as \\
        \cline{2-6}  
        GL & 1 km/s & 6 months  & > 30 years  & / & / \\
        HOPC & 10 km/s & 18(6) years  & > 30 years  & / & / \\
        & 100 km/s & / & / & / & / \\
        \cline{1-6}
        & Model F & 10 $\mu$as & 30 $\mu$as & 50 $\mu$as & 100 $\mu$as \\
        \cline{2-6}  
        GL & 1 km/s & 18 years  (10 months) & > 30 years  & / & / \\
        HOPC & 10 km/s & > 30 years  & > 30 years  & / & / \\
        & 100 km/s & / & / & / & / \\
        \hline
        \end{tabular}
\end{center}
\end{table*}
}

The aim of this section is to estimate the minimal observation times required to detect different effects acting on the S2 star astrometric and spectroscopic observations, at 12 given pairs of accuracies $(\sigma_\mathrm{A}, \sigma_{\mathrm{V}})$. We will thus be capable of determining the threshold times needed for GRAVITY to detect relativistic effects. We remind that the different results are obtained by fitting models A to F to full-GR S2 observations generated with Model G. We mention that we do not compute threshold times of relativistic models for all pairs of accuracies $(\sigma_\mathrm{A}, \sigma_{\mathrm{V}})$ because the accuracies that we  consider are sufficient to make conclusions on the possibility of constraining different effects with these models.

Table~\ref{Time} gives the different threshold times obtained for models A to F and various pairs of accuracies. In what follows, we consider that an effect is detectable when a model X, neglecting this effect, fails after a significantly shorter period than a model X+1 more sophisticated, taking into account this effect. The detection of an effect is thus obtained by comparing two different models. The left column of Table~\ref{Time} gives some of the effects missing in a model and that cause its failure to reproduce the S2 observations. We mention that a conclusion of all results discussed in the following paragraphs is given at the end of this section.

First, Table~\ref{Time} clearly shows that the time threshold for telling an effect grows with poor spectral and astrometric accuracies. This is obvious: more observation time is needed to demonstrate that data with low quality are at odds with a given model.

Second, provided the spectroscopic accuracy is $\lesssim~10$~km/s, the threshold times increase mostly drastically for models C to F compared to models A and B. This means that models C to F become much better at describing the observations than models A and B. Models A and B are the only ones that do not contain any relativistic effects.
This shows that even for rather poor astrometric accuracies (of order 100~$\mu$as), lowest-order relativistic effects will be at hand after only a few months of monitoring, provided spectroscopic accuracy is $\lesssim~10$~km/s.

Finally, let us now compare the time thresholds for successive models in order to determine the minimum observation times needed to tell the various effects.

Roemer effect: This effect is tested by comparing the results of models A and B. We note that for spectroscopic accuracies $\lesssim~10$~km/s, models A and B have similar behaviors. This means that at such high spectroscopic accuracies, some relativistic effects dominate the Roemer time delay. Only at the lowest spectroscopic accuracy (100~km/s) is the Roemer effect strong enough to be detectable; within a few months to a few years depending on the astrometric accuracy. Indeed, the threshold times differ most significantly between models A and B at such accuracies. For instance, at $(\sigma_\mathrm{A}, \sigma_\mathrm{V}) = (10~\mu$as, 100~km/s), when Model A reaches 90\% rejection after 4 months, Model B has only reached 40\%
rejection. This shows that the detection of the effect is strong.

Relativistic redshifts: We now compare models B and C. For spectroscopic accuracies $\lesssim~10$~km/s, relativistic redshifts appear after only a few months of observations whatever the astrometric accuracy. Again, the very different time thresholds between the two models allows us to get a strong detection of the Doppler transverse shift and gravitational redshift. We add that the rejection percentages estimated with Model C at runs of observation where Model B fails to reproduce the observations at $(\sigma_\mathrm{A}, \sigma_\mathrm{V}) = (10 - 100$~$\mu$as, $1 - 10$~km/s) are all inferior to 20\%, which also supports the strong detection of the relativistic redshifts. This discussion also holds for Model A since it behaves very similarly to Model B at these accuracies.

Pericenter advance: this effect is tested by comparing models C and D. At very high astrometric accuracy (10~$\mu$as), models C and D behave either very similarly (for high spectral resolution) or rather similarly (for medium spectral resolution). For this latter case, the threshold times at 90\% rejection are different by a factor of three, but the thresholds at 60\% rejection are the same. This shows that the effect is detected only weakly. Moreover, it highlights that higher-order effects such as the gravitational lensing and HOPC contributions dominate the pericenter advance since threshold times are mainly equal for both models. Only at a lower astrometric accuracy of 30~$\mu$as can the pericenter advance be detected clearly, within $\sim$~15 to 20 years (i.e., around one orbital period of S2) depending on the spectroscopic accuracy. If we want to conclude on the detection of the pericenter advance at accuracies $(\sigma_\mathrm{A}, \sigma_\mathrm{V}) = (10~\mu$as, $\leqslant~10$~km/s), we need to compare models C and D taking into account all high-order effects in both models by using the ray-tracing code \textsc{Gyoto}. In such a case, Model D becomes a full-GR model and the only missing effect in Model C is the pericenter advance. In order to get an estimation of the threshold times that could be obtained with such models at those accuracies, we fit the current Model C to observations of the S2 star generated with the current Model D. Indeed, the only missing effect needed to reproduce the observations of S2 with Model C will be the pericenter advance. The different results are given in square brackets in Table~\ref{Time}. They show that the pericenter advance should be detectable within a few years when considering accuracies $(\sigma_\mathrm{A}, \sigma_\mathrm{V}) = (10~\mu$as, $\leqslant~10$~km/s).

Lense-Thirring: Let us now compare models D and E. The threshold times are very similar for all accuracies. This demonstrates that the Lense-Thirring effect is not detectable for runs of observation $\leqslant$~30~years, and for the considered astrometric and spectroscopic accuracies. The obtained threshold times also show that both models D and E fail to reproduce the data due to high-order effects corresponding to the gravitational lensing and the HOPC contributions.

Gravitational lensing: this effect can be detected by comparing models E and F and by providing a very high astrometric accuracy (10~$\mu$as). A strong detection within a few years ($p~=~60\%$) is possible for a spectroscopic accuracy of 10~km/s. We add that at $(\sigma_\mathrm{A}, \sigma_\mathrm{V}) = (10$~$\mu$as, 10~km/s), when Model E reaches 60\% rejection after 6~years, Model F has only reached 20\% rejection which confirms the strong detection of gravitational lensing. Its detection is weaker at 1 km/s, showing that higher-order effects, corresponding to the HOPC contributions, are involved.

\begin{figure*}[htb]
\centering
      \includegraphics[scale=0.3]{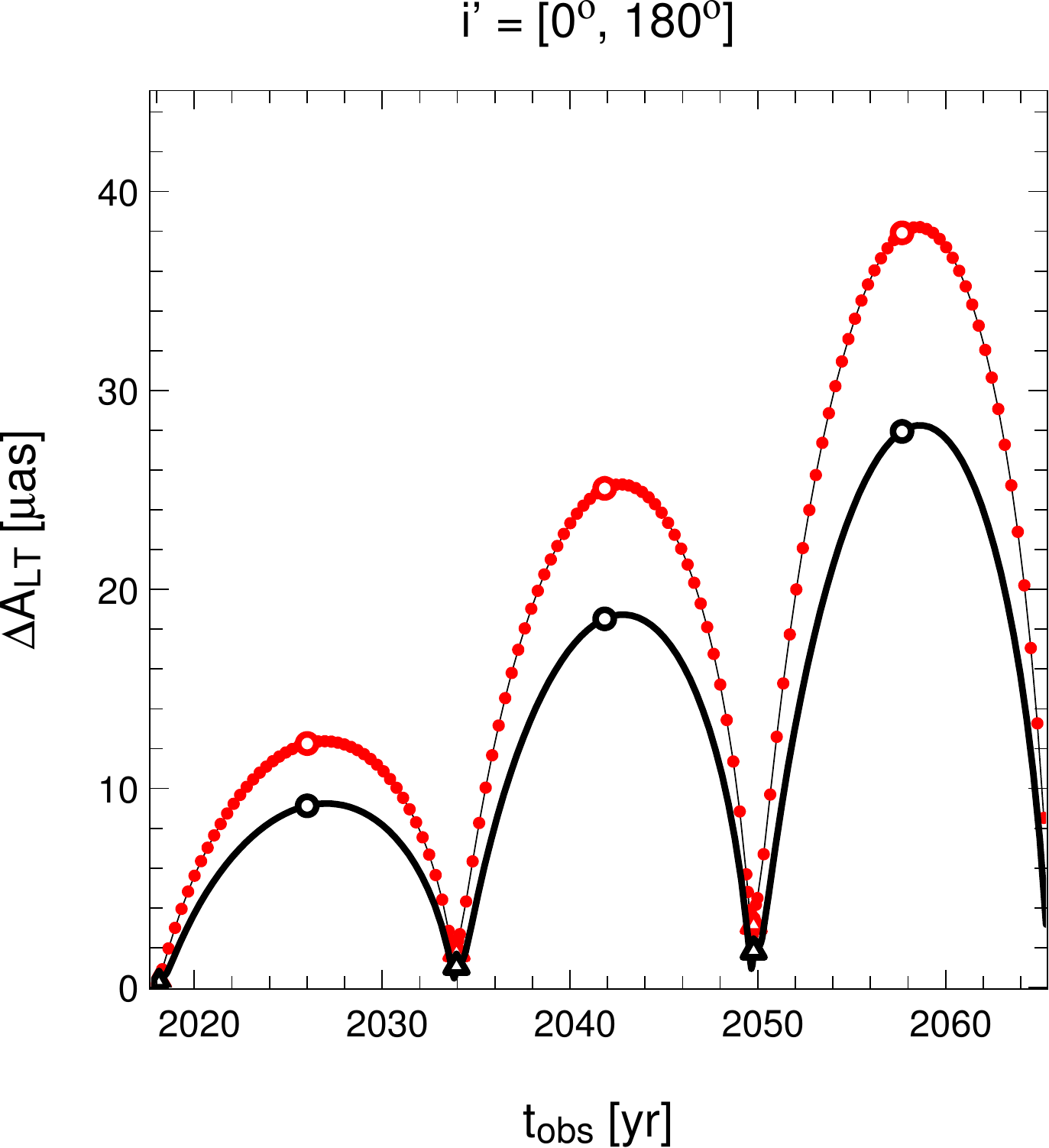}
      \quad
      \includegraphics[scale=0.3]{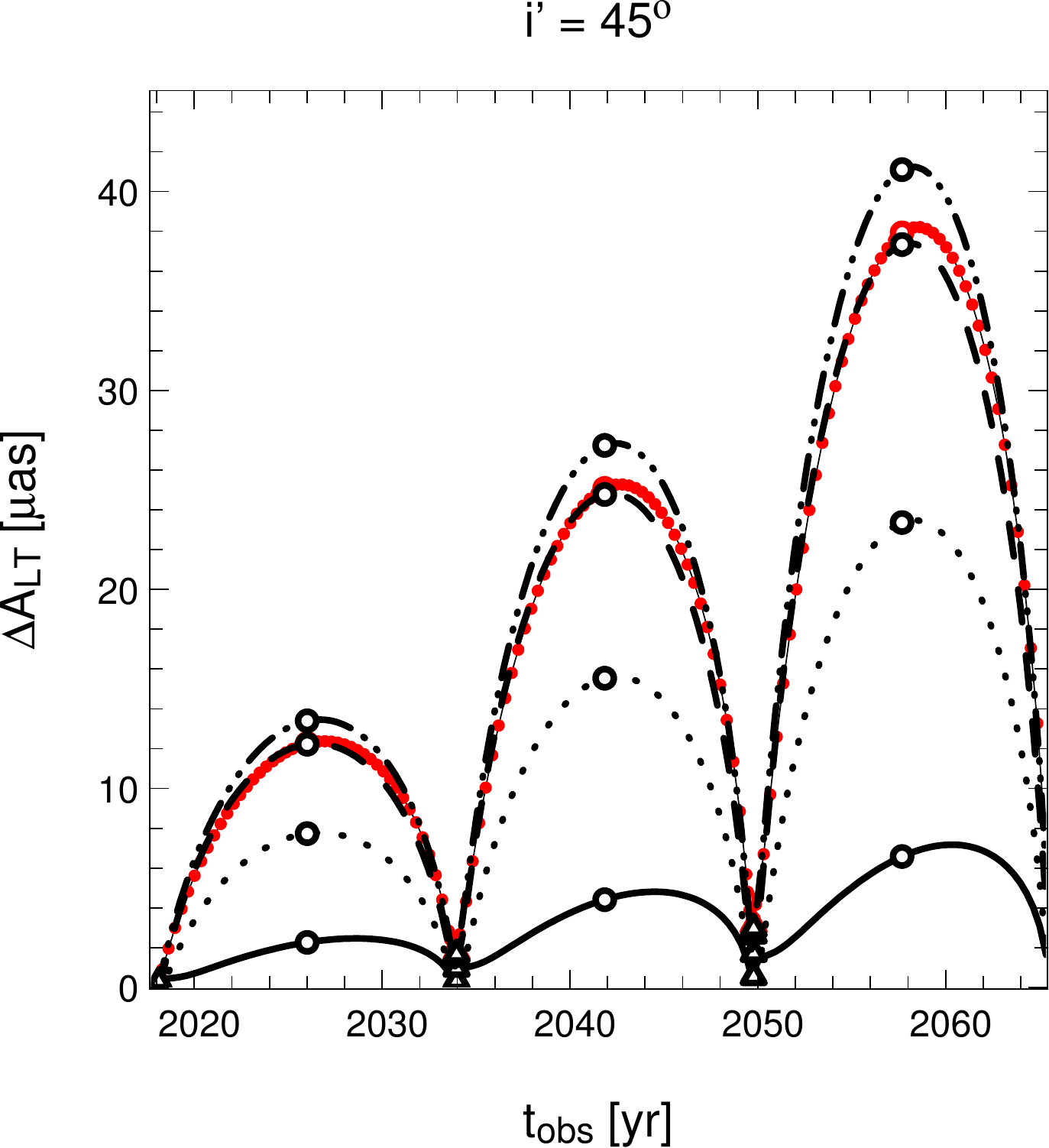} 
      \quad
      \includegraphics[scale=0.3]{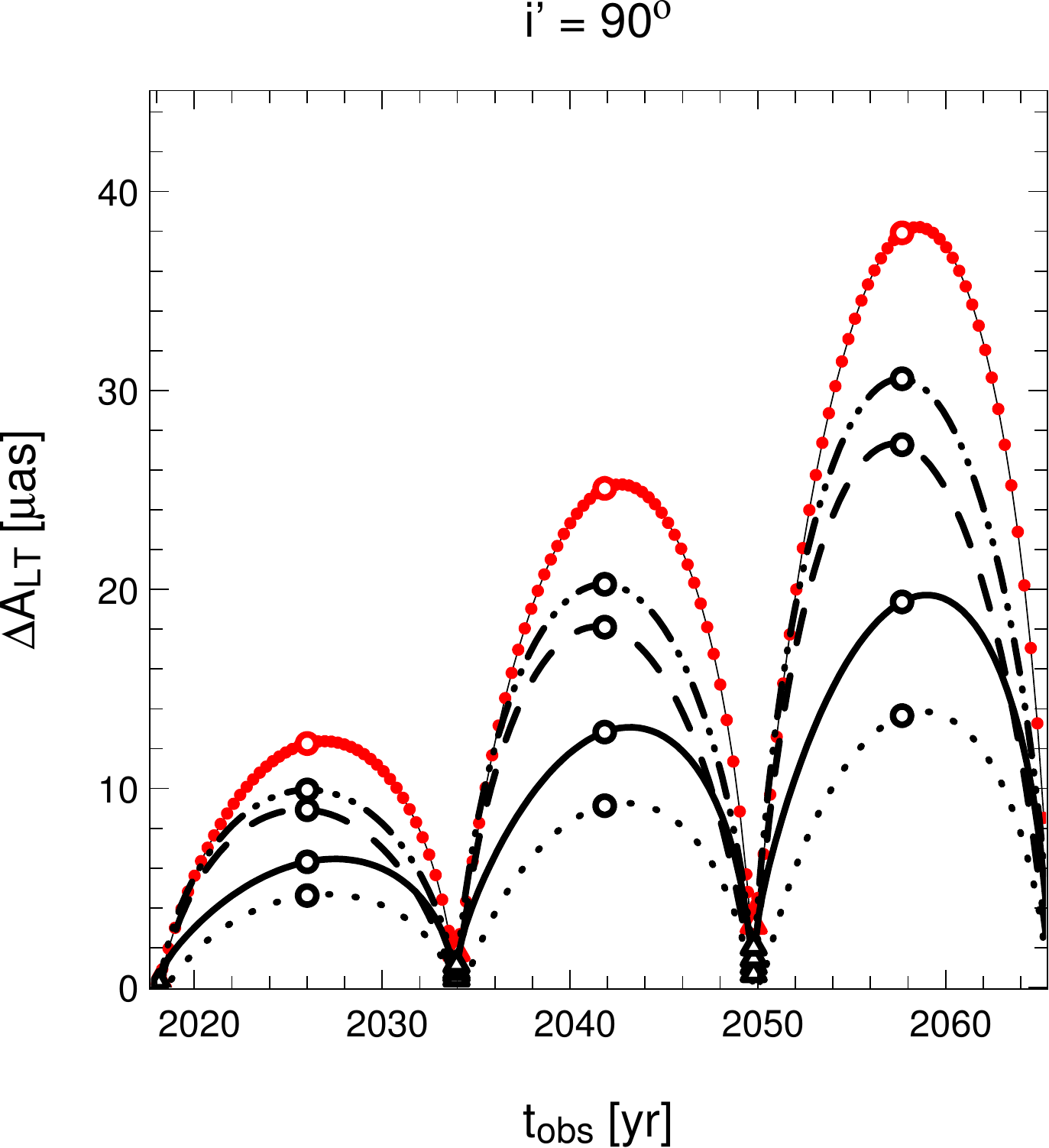}
      \quad
      \includegraphics[scale=0.3]{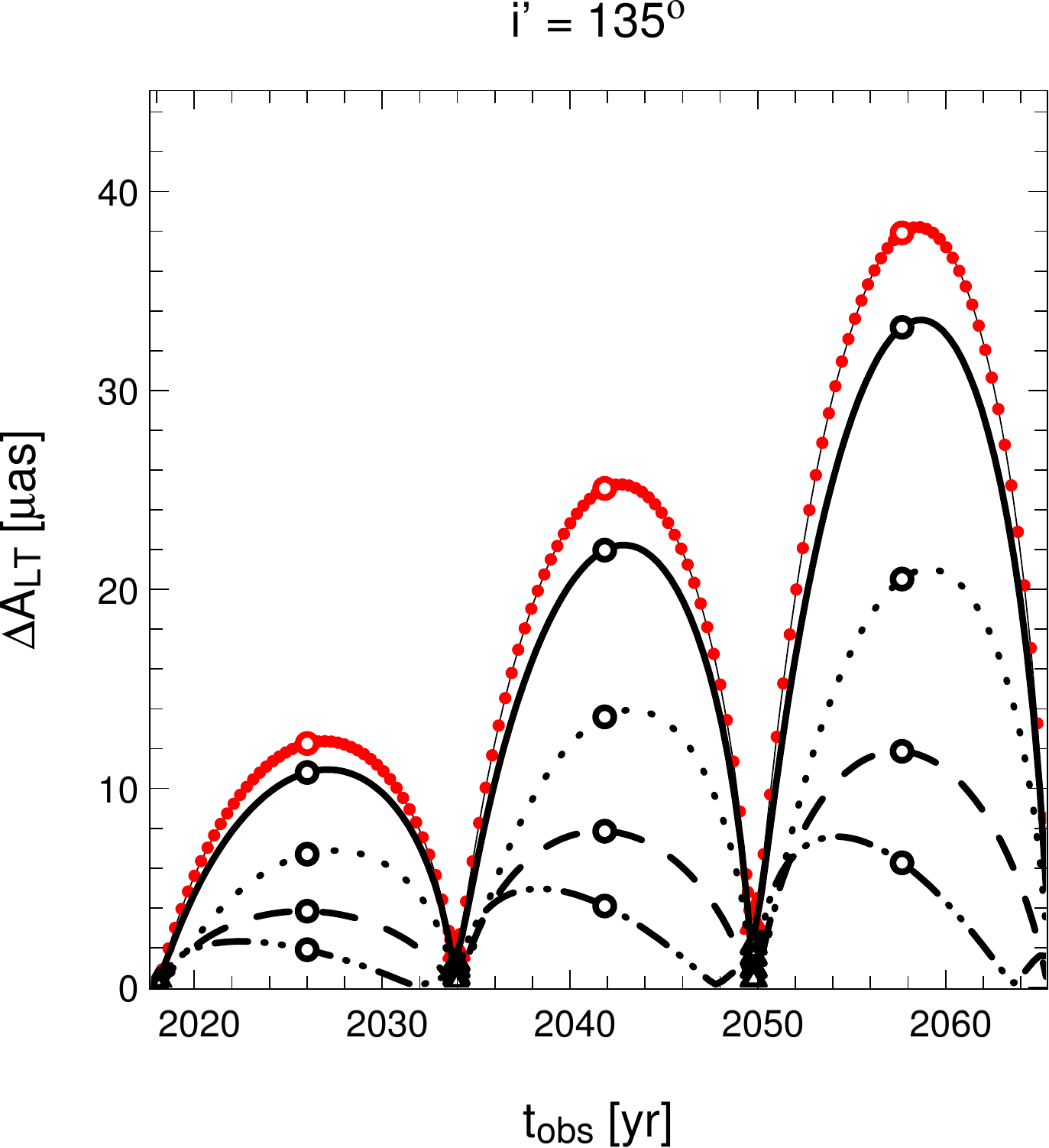}\\
      \quad
      
      \includegraphics[scale=0.3]{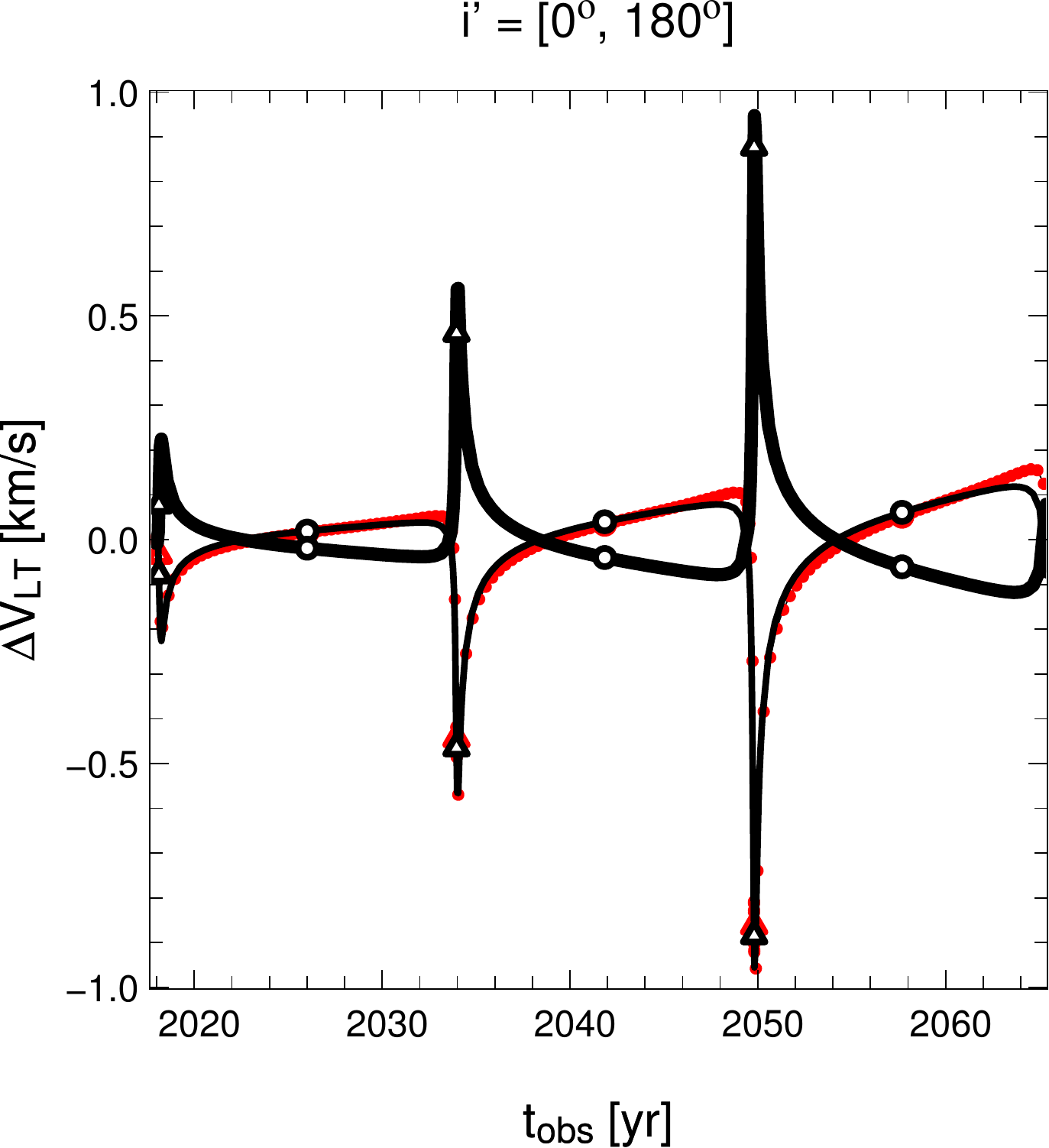}
      \quad
      \includegraphics[scale=0.3]{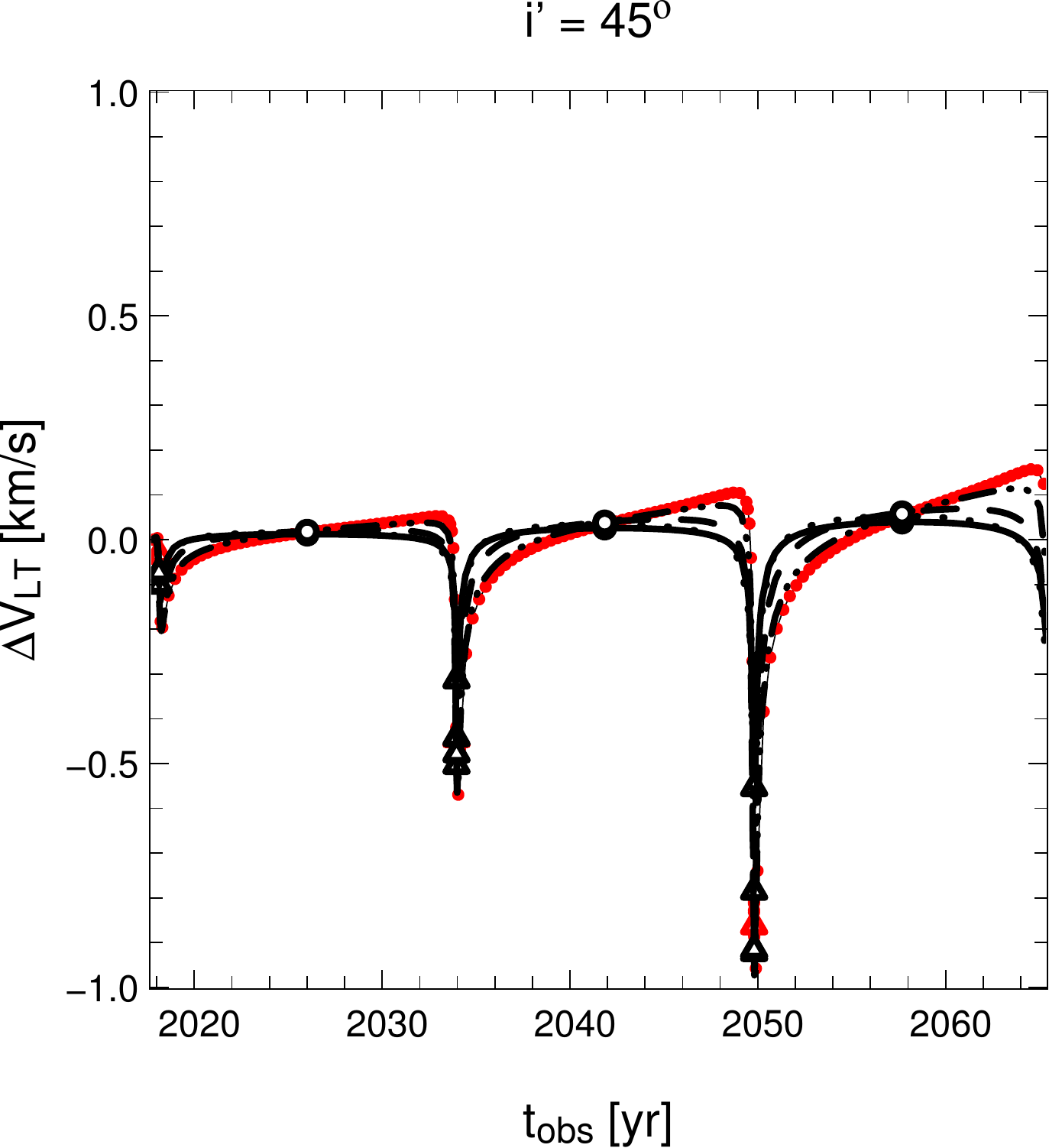} 
      \quad
      \includegraphics[scale=0.3]{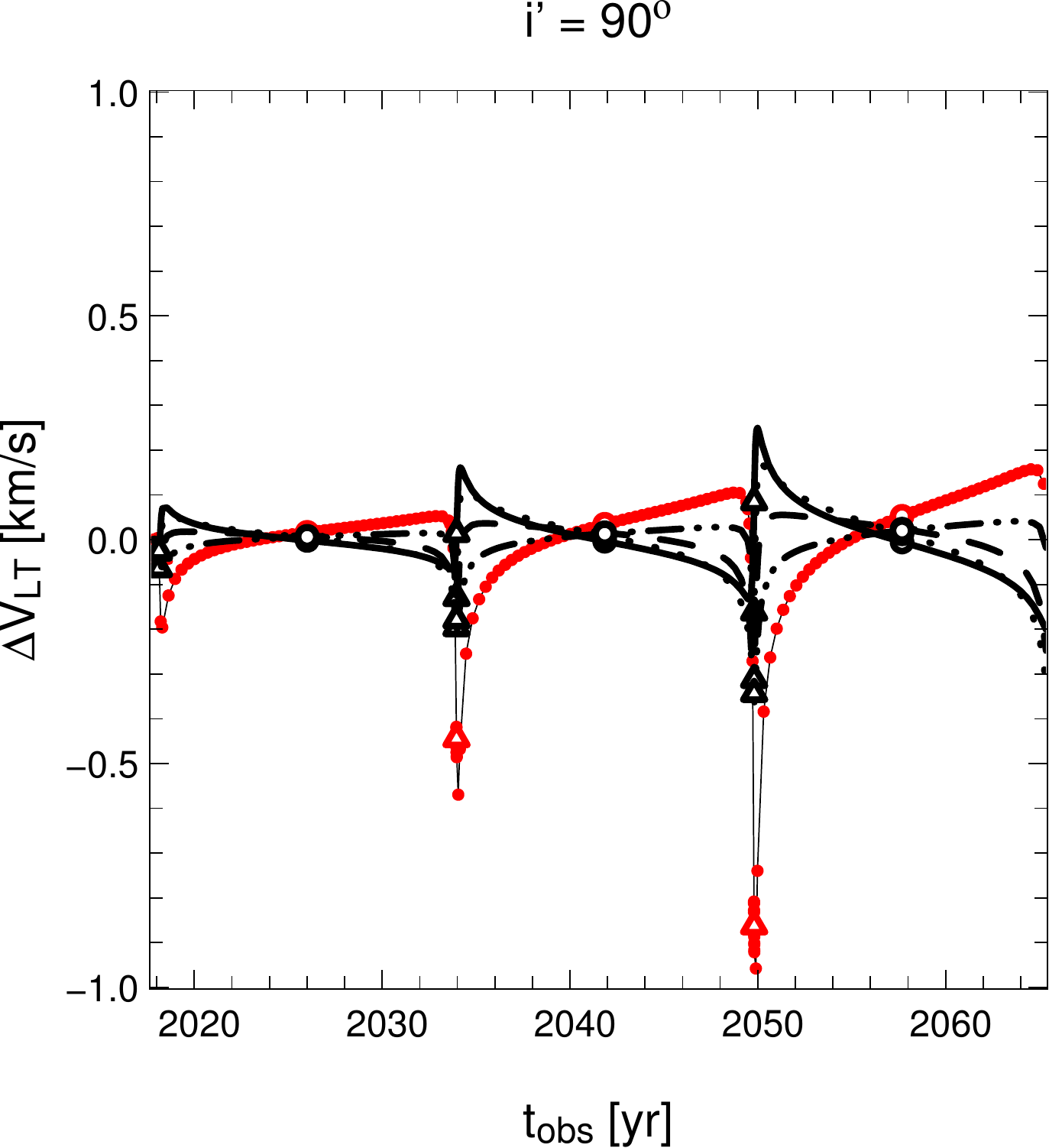}
      \quad
      \includegraphics[scale=0.3]{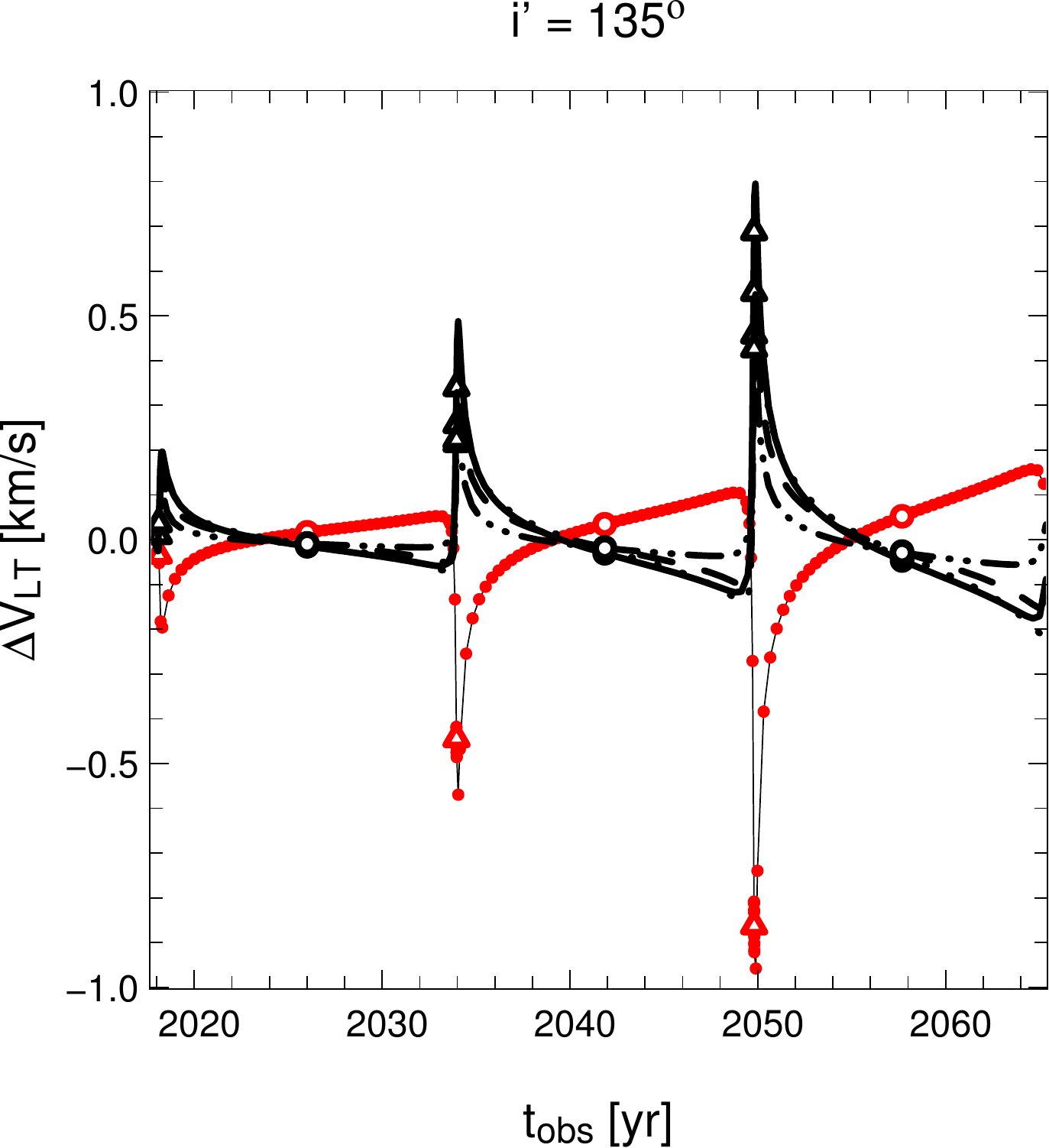}\\
   \caption{Lense-Thirring effect on astrometric (upper plots) and spectroscopic (lower plots) observations of the S2 star, considering $a=0.99$ and various $\Omega'$ and $i'$. The angular momentum direction $i' = 45^{\circ}$ and $\Omega' = 160^{\circ}$ considered for the mock observations are visible on each plot and are denoted with the solid red circles. The different types of curves on each plot correspond to different values of the angle $\Omega'$. For $i' = [0^{\circ}, 180^{\circ}]$, we obtain the same curves whatever $\Omega'$ for the astrometry but they are different when considering the spectroscopy: thin and thick lines correspond to $i' = 180^{\circ}$ and $i' = 0^{\circ}$, respectively. On the other plots, solid, dotted, dashed and dash-dot-dotted curves correspond to $\Omega'$ equal to $0^{\circ}$, $45^{\circ}$, $90^{\circ}$ and $135^{\circ}$, respectively. Open circles on all plots denote the position of the apocenter and open triangles denote the pericenter passages.}
\label{fig:LTMCMC}
\end{figure*}

In order to determine whether the threshold times obtained with models A, B and C depend on the initial guess used in the Levenberg-Marquardt method, we chose these parameters far from the solution: $\mathbf{P}_\mathrm{init} = \mathbf{P}_{\mathrm{Gillessen}} + 5\boldsymbol\sigma_{\mathrm{Gillessen}}$ (instead of $\mathbf{P}_\mathrm{init} = \mathbf{P}_{\mathrm{Gillessen}} + 1\boldsymbol\sigma_{\mathrm{Gillessen}}$, see Sect.~\ref{procedures}). For this test, we consider Model C and the pairs of accuracies $(\sigma_\mathrm{A}, \sigma_\mathrm{V}) = (10$~$\mu$as, $1 - 100$~km/s). We find similar results to those estimated considering the previous initial guess: 10~months for $\sigma_\mathrm{V}~=~1$~km/s and 6~years for $\sigma_\mathrm{V}~=~10 - 100$~km/s. This test tends to show that fittings of models A, B and C have converged to the global minimum since the threshold times are similar for both initial guesses considered. We also reiterate that we showed in Sect.~\ref{procedures} that the percentage found with Model C and the Levenberg-Marquardt method, at $(\sigma_\mathrm{A}, \sigma_\mathrm{V})~=~(30$~$\mu$as, 10~km/s), and the run of observation of one orbital period, was similar to the percentage found with the MCMC method (45\%), which supports the fact that the fitting method used for Keplerian models converges to the global minimum. 

Moreover, we also investigate the influence of the pericenter passages sampling on the threshold time. For doing so, we again use Model C and accuracies $(\sigma_\mathrm{A}, \sigma_\mathrm{V})~=~(10$~$\mu$as, $1 - 100$ km/s). The new sampling is similar to the previous one described in Sect.~\ref{obs} but we consider only one point per night during the three weeks where the pericenter passages are observed, instead of two points: there are a total of 7 data points instead of 14 at S2 pericenter passages. The minimal observation times obtained are similar to those estimated with the previous sampling: 6~months for $\sigma_\mathrm{V}~=~1$~km/s and 8~years for $\sigma_\mathrm{V}~=~10 - 100$~km/s with the new sampling, and 10~months for $\sigma_\mathrm{V}~=~1$~km/s and 6~years for $\sigma_\mathrm{V}~=~10 - 100$~km/s with the previous sampling. This shows that sampling at pericenter weakly impacts the results. However, it remains essential to correctly sample during pericenter since the majority of relativistic effects are maximal near S2 pericenter passages (see Tables~\ref{OffAstro} and \ref{OffRadVel}).

To summarize, we can say that if we consider S2 astrometric and spectroscopic observations starting in 2018, we can detect the Roemer effect within 4~months by using models A and B at $(\sigma_\mathrm{A}, \sigma_\mathrm{V})~=~(10$~$\mu$as, $100$~km/s). Relativistic redshifts can be detected within 2~months by using models B (or A) and C with $(\sigma_\mathrm{A}, \sigma_\mathrm{V})~=~(10$~$\mu$as, $10$~km/s). Gravitational lensing can be detected by using models C (or D, or E) and F within $\approx~4$~years at $(\sigma_\mathrm{A}, \sigma_\mathrm{V})~=~(10$~$\mu$as, $10$~km/s). Pericenter advance should be detected within 8~years with $(\sigma_\mathrm{A}, \sigma_\mathrm{V})~=~(10$~$\mu$as, $1-10$~km/s) and using modified models C and D taking into account the computation of null geodesics. HOPC contributions can be detected within $6-10$~months with $(\sigma_\mathrm{A}, \sigma_\mathrm{V})~=~(10$~$\mu$as, $1$~km/s) and with Model C (or D, or E, or F). Regarding the Lense-Thirring effect, it is not detectable if we consider observations obtained during 2 periods of S2. However, this result does not exclude the possibility of getting a first constraint on the angular momentum parameters of the black hole, which is discussed in the following section.

\subsubsection{Constraint on the black hole angular momentum}
\label{cons_spin}

{\renewcommand{\arraystretch}{1.5}
\begin{table*}[!t]
\begin{center}
\caption{Constraints obtained by fitting Model F to astrometric and spectroscopic observations of the S2 star, generated with Model G. The values are given (except for the norm of the angular momentum of the black hole) by computing the difference between parameters found by the fitting method and those used to get the mock observations. Errors associated to each parameter are estimated at $1\sigma$ by the MCMC method. Four runs of observation are considered where the astrometric accuracy, the norm $a$ and the duration of the run (mentioned by the parameter $N_P$ corresponding to the number of orbital periods of S2) vary. All runs are generated considering a direction for the angular momentum of the black hole: $(i', \Omega')=(45^\circ, 160^\circ)$.}
\label{table:res}
\begin{tabular}{ccccc} 
\hline\hline
$(a, N_P, \sigma_\mathrm{A} [\mu$as$], \sigma_\mathrm{V} [$km/s$])$ & $(0.99, 3, 10, 10)$ & $(0.7, 3, 10, 10)$ & $(0.99, 3, 30, 10)$ & $(0.99, 2, 10, 10)$ \\
\hline

\multicolumn{1}{l}{$\delta T$ [hr]} & $-0.205^{+0.316}_{-0.318}$ & $-0.375^{+0.313}_{-0.361}$ & $-0.535^{+0.355}_{-0.389}$ & $-1.007^{+0.357}_{-0.384}$\\ 
\hline
\multicolumn{1}{l}{$\delta a_\mathrm{sma}$ [$\mu$as]} & $-2.082^{+3.006}_{-3.164}$ & $0.617^{+3.327}_{-3.394}$ & $9.968^{+9.038}_{-9.309}$ & $4.660^{+3.950}_{-4.284}$ \\ 
\hline
\multicolumn{1}{l}{$\delta e \times 10^{-6}$} & $-1.069^{+6.021}_{-6.265}$ & $7.716^{+7.323}_{-6.599}$ & $25.093^{+18.953}_{-18.545}$ & $18.203^{+8.296}_{-9.440}$ \\ 
\hline
\multicolumn{1}{l}{$\delta t_p$ [hr]} & $0.339^{+0.344}_{-0.353}$ & $0.622^{+0.387}_{-0.334}$ & $1.034^{+0.487}_{-0.503}$ & $1.376^{+0.405}_{-0.407}$ \\ 
\hline
\multicolumn{1}{l}{$\delta \Omega \times 10^{-2}$ [$^{\circ}$]} & $0.156^{+0.286}_{-0.290}$ & $0.336^{+0.303}_{-0.310}$ & $0.645^{+0.867}_{-0.844}$ & $0.393^{+0.401}_{-0.414}$ \\ 
\hline
\multicolumn{1}{l}{$\delta \omega \times 10^{-2}$ [$^{\circ}$]} & $0.098^{+0.264}_{-0.261}$ & $0.186^{+0.307}_{-0.308}$ & $1.025^{+0.761}_{-0.842}$ & $0.570^{+0.364}_{-0.400}$ \\ 
\hline
\multicolumn{1}{l}{$\delta i  \times 10^{-2}$ [$^{\circ}$]} & $0.145^{+0.196}_{-0.196}$ & $-0.023^{+0.206}_{-0.207}$ & $-0.755^{+0.522}_{-0.513}$ & $-0.355^{+0.240}_{-0.237}$ \\ 
\hline
\multicolumn{1}{l}{$\delta R_0$ [pc]} & $3.094^{+3.416}_{-3.280}$ & $4.100^{+3.765}_{-3.509}$ & $6.134^{+3.873}_{-3.712}$ & $9.301^{+3.957}_{-3.805}$ \\ 
\hline
\multicolumn{1}{l}{$\delta M_{\mathrm{TN}} \times 10^{4}$ [$M_\odot$]} &$0.458^{+0.529}_{-0.508}$ & $0.565^{+0.585}_{-0.548}$ & $1.068^{+0.572}_{-0.566}$ & $1.494^{+0.618}_{-0.576}$ \\ 
\hline
\multicolumn{1}{l}{$a$} & $0.931_{-0.113}$ & $0.770^{+0.111}_{-0.139}$ & $0.986_{-0.255}$ & $0.980_{-0.212}$ \\ 
\hline
\multicolumn{1}{l}{$\delta \Omega'$ [$^{\circ}$]} & $35.182^{+26.249}_{-21.530}$ & $51.049^{+27.432}_{-28.105}$ & $82.122^{+41.169}_{-40.054}$ & $96.461^{+45.497}_{-32.060}$ \\ 
\hline
\multicolumn{1}{l}{$\delta i'$ [$^{\circ}$]} & $-0.334^{+14.990}_{-13.335}$ & $7.623^{+15.189}_{-14.763}$ & $-1.981^{+21.828}_{-37.070}$ & $-3.166^{+27.761}_{-24.978}$ \\ 
\hline
\end{tabular}
\end{center}
\end{table*}
}

 \begin{figure*}[!t]
 \centering
        \includegraphics[scale=0.3]{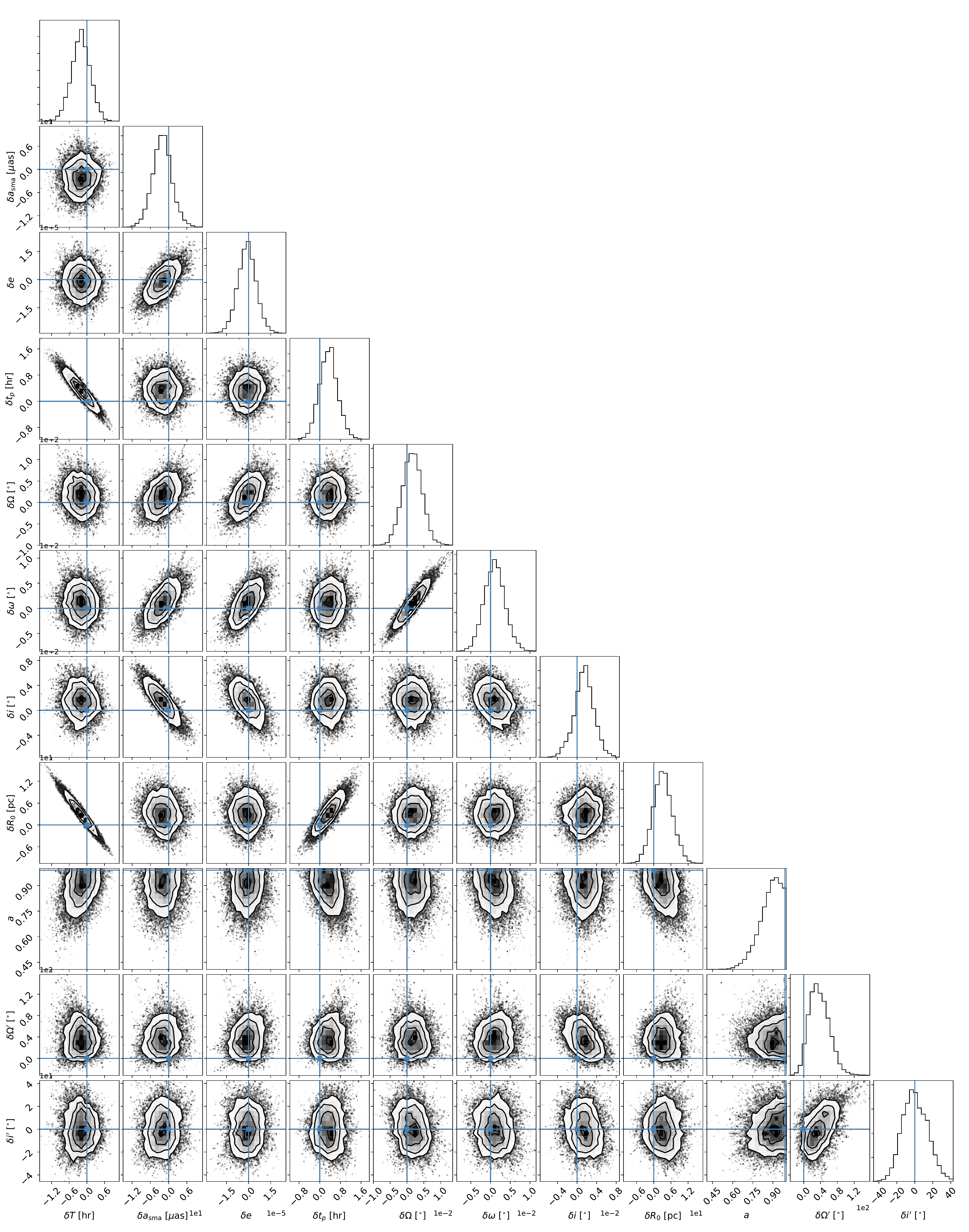}
    \caption{1D and 2D probability distributions obtained by fitting Model F to observations during three orbital periods of the S2 star generated with Model G and $(\sigma_\mathrm{A}, \sigma_\mathrm{V}) = (10$~$\mu$as, 10~km/s). The angular momentum parameters of the black hole are $(a, i', \Omega')=(0.99, 45^\circ, 160^\circ)$. Blue dots and lines on each plot correspond to the parameters used to generate the mock observations.The darker the delineated area on 2D distributions, the denser the area.}
    \label{fig:fit099}
\end{figure*}

 \begin{figure*}[!t]
 \centering
        \includegraphics[scale=0.3]{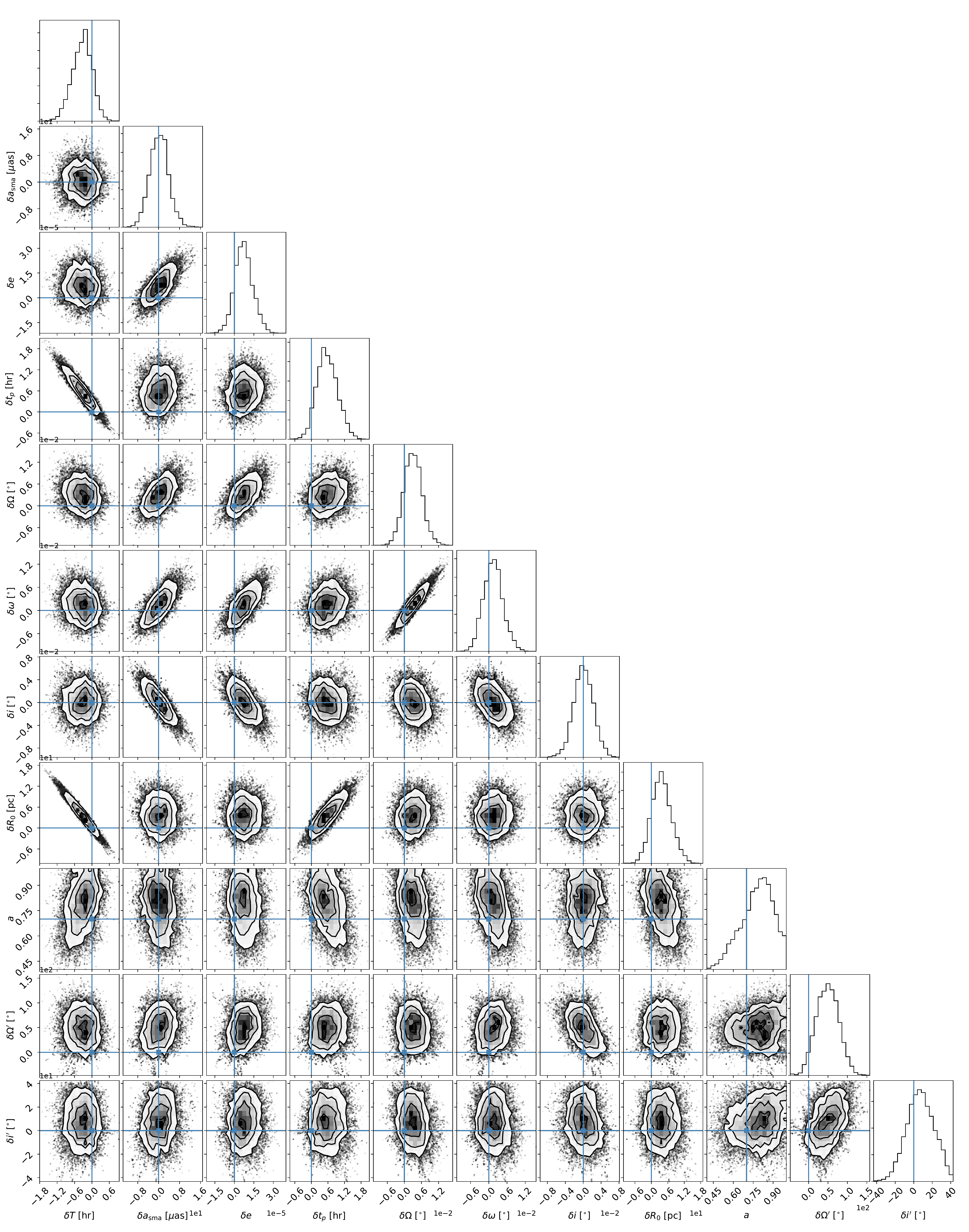}
    \caption{As in Fig.~\ref{fig:fit099} but for $a=0.7$.}
        \label{fig:fit07}
\end{figure*}

As noticed in Tables~\ref{OffAstro} and \ref{OffRadVel}, the Lense-Thirring effect impacts weakly the photon trajectory. However, these results are obtained only considering the angular momentum direction $(i,\Omega'')~=~(45^\circ,160^{\circ})$. In order to validate the fact that null geodesics are weakly affected by this effect we compute its impact on both astrometry and spectroscopy considering various pairs $(i', \Omega')$. We found a negligible shift on both observables: $< 1$~$\mu$as et $<1$~km/s, which shows that ray tracing is not primordial to constrain the angular momentum parameters of the black hole with S2. These results are in accordance with \cite{2015ApJ...809..127Z} and \cite{2016ApJ...827..114Y}.

Now, we are interested in the Lense-Thirring effect on the star trajectory, and its effect on the astrometric observations of the S2 star. We remind that the astrometric shift is maximal near the three apocenter passages (see the third plot in Fig.~\ref{fig:astroS2}). However, we again considered one direction for the angular momentum of the black hole. The upper plots in Fig.~\ref{fig:LTMCMC} give astrometric shifts induced by the Lense-Thirring effect and obtained for several directions $(i', \Omega')$. We note in each case that the shift is maximal near apocenter passages. It is thus important to observe near to S2 apocenter to constrain the parameters $a$, $i'$ and $\Omega'$ with this star. However, as mentioned previously, the majority of the relativistic effects are maximal near pericenter. Astrometric observations near pericenter are thus also necessary to investigate a constraint on angular momentum parameters. More precisely, we need to get a strong constraint on orbital parameters and both distance and mass of the black hole by using relativistic effects observed near pericenter if we want to constrain the Lense-Thirring effect near apocenter. On the upper plots in Fig.~\ref{fig:LTMCMC}, we note that the Lense-Thirring effect reaches between 10~$\mu$as and 40~$\mu$as during the three orbital periods for some values of $(i', \Omega')$. However, the astrometric shift can be less than 10~$\mu$as throughout the three S2 periods. This shows that the detection of this effect will be possible only for particular parameters $(a, i', \Omega')$. See for instance \cite{2016ApJ...827..114Y} to get pairs of angles $(i', \Omega')$ that could be favorable for detecting the Lense-Thirring effect when considering $a=0.99$. If we refer to the results found by \cite{2011ApJ...735..110B}, the angular momentum parameters of the black hole candidate located at the center of our galaxy should be: $a~=~0,0^{+0,64+0,86}$, $i'~=~{68^\circ }^{+5^\circ +9^\circ }_{-20^\circ -28^\circ }$ and $\Omega'~=~{-52^\circ }^{+17^\circ +33^\circ }_{-15^\circ -24^\circ }$ where the errors are those obtained at $1\sigma$ and $2\sigma$. We therefore determined the maximal astrometric shift induced by the Lense-Thirring effect and obtained considering the values of $(a, i', \Omega')$ allowed by \cite{2011ApJ...735..110B}. First, if we consider the set of values of $(a, i', \Omega')$ contained in the interval $\pm1\sigma$;  the impact of this effect is maximal for $(a,i',\Omega')~=~(0.64, {68^\circ+5^\circ }, {-52^\circ -15^\circ})$. The shift reaches 5~$\mu$as, 10~$\mu$as and 15~$\mu$as at first, second and third apocenter passages, respectively. If we consider now the set of values of $(a, i', \Omega')$ contained in the interval $\pm2\sigma$, the impact of this effect is maximal for $(a,i',\Omega')~=~(0.86, {68^\circ+9^\circ }, {-52^\circ -24^\circ})$: the shift reaches 5~$\mu$as, 12~$\mu$as and 18~$\mu$as at first, second and third apocenter passages, respectively. These results show that it seems difficult to strongly constrain the black hole angular momentum parameters with the GRAVITY instrument, if we consider astrometric observations obtained on three orbital periods of S2 and values of $(a, i', \Omega')$ predicted by \cite{2011ApJ...735..110B}.

Degeneracies of the astrometric shift associated to the angles $i'$ and $\Omega'$ are also observed, but not all of these degeneracies are visible on the upper plots in Fig.~\ref{fig:LTMCMC}. First, if we observe the rotation axis of the black hole from the top ($i'=0^\circ$) or from the bottom ($i'=180^\circ$), the shift is the same whatever the angle $\Omega'$ (see solid curve in Fig.~\ref{fig:LTMCMC} for $i' = [0^{\circ}, 180^{\circ}]$). If we observe the rotation axis edge on ($i'=90^\circ$), there is a central symmetry with respect to the center of the plane of the sky. Indeed, the shifts at $(i'=90^\circ, \Omega')$ and $(i'~=~90^{\circ}, 180^{\circ}~+~\Omega')$ are similar. Finally, the astrometric shifts are similar when considering $(i', \Omega')~=~(45^\circ, 135^\circ)$ and $(i', \Omega')~=~(135^\circ, 315^\circ)$: there is a degeneracy between $(i', \Omega')~=~(45^\circ, \Omega')$ and $(i', \Omega')~=~(135^\circ, 180^\circ + \Omega')$. To summarize, there are three degeneracy groups present whatever the norm of the angular momentum of the black hole (expect for $a=0$):
\begin{itemize}
\item $(i'=0^{\circ}, \forall \Omega')$ and $(i'=180^{\circ}, \forall \Omega')$,
\item $(i'=90^{\circ}, \Omega')$ and  $(i'=90^{\circ}, 180^{\circ} + \Omega')$,
\item $(i', \Omega')$ and $(180^{\circ} - i', 180^{\circ}+\Omega')$ if $ 0^\circ < i' < 90^{\circ}$ or $90^\circ~<~i'~<~180^{\circ}$.
\end{itemize}
These degeneracies have also been noticed independently by \cite{2016ApJ...827..114Y}. In addition to these degeneracies, shifts obtained with different triplets $(a, i', \Omega')$ are also similar. For instance, both shifts obtained with $(a, i', \Omega')~=~(0.99, 135^\circ, 45^\circ)$ and $(a, i', \Omega')~=~(0.7, 90^\circ, 135^\circ)$ reach $\approx 6~\mu$as, 15~$\mu$as and 20~$\mu$as at first, second and third periods, respectively.

On the bottom plots in Fig.~\ref{fig:LTMCMC} the spectroscopic shifts are visible, due to the Lense-Thirring effect for various directions $(i', \Omega')$. The maximal shift is in this case located near pericenter of the S2 star. Moreover, whatever the angles $i'$ and $\Omega'$, the shift is always inferior to 1~km/s. It is thus necessary to get an accuracy better than (or close to) 1~km/s if we want to investigate the constraint on the angular momentum parameters with spectroscopy. Such accuracy is not available with current instruments. We mention nonetheless that the spectroscopic accuracy intended for the E-ELT is of about 1~km/s. Degeneracies listed above are almost all broken when considering spectroscopic measurements. For instance, the shifts are different (up to $\approx$~2~km/s) for $(a, i', \Omega')~=~(0.99, 45^{\circ}, 135^{\circ})$ and $(a, i', \Omega')~=~(0.99, 180^{\circ}-45^{\circ}, 135^{\circ})$, which is not the case for astrometry. The only degeneracy is observed for $i'~=~0^{\circ}$ or $i'~=~180^{\circ}$. Indeed, we obtain the same shifts whatever the angle $\Omega'$ at those $i'$ (see the thick and thin lines on the first bottom plot in Fig.~\ref{fig:LTMCMC}).

To summarize, to constrain the black hole parameters $(a, i', \Omega')$ with current instruments it is better to consider both S2 astrometric and spectroscopic observations obtained at both pericenter and apocenter. These two observables obtained at pericenter should allow for optimization of the constraint on the orbital parameters, and the mass and distance of the black hole. The astrometric data obtained at apocenter should allow us to constrain the angular momentum parameters of the black hole. However, the detection of the Lense-Thirring effect will only be possible for privileged triplets $(a, i', \Omega')$ such as $(0.99, 45^{\circ}, 160^{\circ})$. By using results of \cite{2016ApJ...827..114Y}, we can estimate that if we consider a high norm for the angular momentum, $\approx$ 40\% of pairs $(i', \Omega')$  seem favorable to constrain the Lense-Thirring effect.

The aim now is to determine whether it is possible to strongly constrain the angular momentum parameters of the black hole by using astrometric and spectroscopic observations of the S2 star, and considering accuracies available with current instruments: $10-30$~$\mu$as and 10~km/s. Contrary to \cite{2015ApJ...809..127Z} and \cite{2016ApJ...827..114Y}, we want to estimate these parameters without using ray tracing, which means using a stellar-orbit model that does not require an important amount of time for the computation. More precisely we chose to use Model F allowing to reproduce at best the Model G observations (see the end of Sect.~\ref{procedures}). 

Table~\ref{table:res} gives the results of the different fittings, where we considered various norm $a$, observation time, and astrometric accuracies for the run. On Figs.~\ref{fig:fit099} and \ref{fig:fit07} we can see the 1D and 2D probability distributions of the different parameters for $(\sigma_\mathrm{A}, \sigma_\mathrm{V}) = (10$~$\mu$as, 10~km/s), and $a=0.99$ and $a=0.7$, respectively. We can see that all parameters on both Figures are well fitted. In particular, the $1\sigma$ error of the norm $a$ for both fittings is of about 0.1. For the angles $i'$ and $\Omega'$ the $1\sigma$ error is around $15^\circ$ and $30^\circ$, respectively. Other black hole parameters such as the mass and the distance are also correctly constrained by Model F: $\sigma_{R_0}/R_0 \approx 4 \times 10^{-4}$ and $\sigma_{M}/M \approx 10^{-3}$ with $\sigma_{R_0}$ and $\sigma_{M}$ the $1\sigma$ errors; the constraints are improved by a factor $\approx~100$ with respect to \cite{2009ApJ...692.1075G}. These results are similar to those estimated by \cite{2016ApJ...827..114Y} whose constraints are obtained by fitting a stellar-orbit model including ray tracing, on mock observations also generated with this model. We mention that the parameters used by these authors to simulate the S2 observations are nearly similar to those used for our study. The difference between these authors and this study is the number of data points: 120 in \cite{2016ApJ...827..114Y} and 192 in this paper, and the fact that we neglect the photon path curvature but approximate the gravitational lensing effect by using analytical formulas. 

The $1\sigma$ errors of the angular momentum parameters obtained when considering $(a, N_p, \sigma_\mathrm{A}, \sigma_\mathrm{V})~=~(0.99, 3, 30~\mu$as, 10~km/s) are $\sigma_a~\approx~0.26$, $\sigma_{i'}~\approx~30^\circ$ and $\sigma_{\Omega'}~\approx~40^\circ$. Those estimated for $(a, N_p, \sigma_\mathrm{A}, \sigma_\mathrm{V})~=~(0.99, 2, 10~\mu$as, 10~km/s) are $\sigma_a~\approx~0.2$, $\sigma_{i'}~\approx~25^\circ$ and $\sigma_{\Omega'}~\approx~40^\circ$. For both runs of observation we get $\sigma_{R_0}/R_0 \approx 5 \times 10^{-4}$ and $\sigma_{M}/M \approx 10^{-3}$. Such results show that improving the astrometric accuracy or increasing the duration of the run of observation mainly allows us to obtain a better constraint on the black hole angular momentum. The results obtained for the mass and the distance of the black hole are only weakly (or not) modified. In addition, as demonstrated by \cite{2016ApJ...827..114Y} , a better spectroscopic accuracy (e.g., $\sigma_\mathrm{V}~=~1~$km/s instead of 10~km/s) allows us to improve the constraints on both the mass and the distance of the black hole, but does not allow us to obtain better estimations of the angular momentum parameters. This is due to the fact that spectroscopic impact of the Lense-Thirring effect is weak (see the bottom plots in Fig.~\ref{fig:LTMCMC}). In addition, we mention that the $1\sigma$ error obtained on the norm $a$ reaches $\approx~0.4$ when considering a run of observation of one orbital period and parameters $(a, \sigma_\mathrm{A}, \sigma_\mathrm{V})~=~(0.99, 10~\mu$as, 10~km/s). Such constraint thus needs to be considered since it allows us to decrease the parameter space of $a$ for future fittings made with longer monitorings of S2 or with other S stars data. However, to strongly constrain the norm $a$ through the observation of stellar orbits located at the Galactic center in a suitable time, it is necessary to detect new stars with GRAVITY closer to Sgr~A* than S2. As mentioned by \cite{2016ApJ...827..114Y}, it should be possible to reach a $1\sigma$ uncertainty of $\sigma_a \lesssim 0.02$ if we observe stars during a period $\lesssim 10$~years with $(\sigma_\mathrm{A}, \sigma_\mathrm{V}) = (10$~$\mu$as, $1-10$~km/s), and whose semi-major axis and eccentricity satisfy $a_{\mathrm{sma}} \lesssim 40$~mas and $e \gtrsim 0.95$, respectively.

The study performed here shows that, even with a stellar-orbit model, without using a ray-tracing code but with considering approximated formulas to simulate gravitational lensing, it is possible to strongly constrain the norm of the angular momentum, the mass and the distance of the black hole, and to get non negligible constraints on the direction $(i', \Omega')$. In particular, we found similar results as \cite{2016ApJ...827..114Y} which uses a model with ray tracing. However, astrometric and spectroscopic differences between models F and G are not negligible when considering mock observations obtained for $(\sigma_\mathrm{A}, \sigma_\mathrm{V})~=~(10$~$\mu$as, 1~km/s). Indeed, as mentioned in Table~\ref{Time}, Model F fails to describe the observations within 18~years for $p~=~90\%$. With such spectroscopic accuracy it is thus necessary to use Model G.


\section{Conclusions and discussions}
\label{conclusion}

To conclude, the studies made in this paper show that various relativistic effects can be detected with current instruments by using astrometric and spectroscopic observations of the S2 star beginning near its next pericenter passage in 2018. In particular, transverse Doppler shift and gravitational redshift can be detected within a few months by using observations obtained with accuracies $(\sigma_\mathrm{A}, \sigma_\mathrm{V})=(10-100~\mu$as, $1-10$~km/s). If we consider the pair of accuracies $(\sigma_\mathrm{A}, \sigma_\mathrm{V})=(10~\mu$as, 10~km/s), these effects can be detected by combining observations obtained with GRAVITY and SINFONI at VLT. Spectroscopic measurements can also be obtained with NIRSPEC at Keck. Gravitational lensing can be detected within a few years for $(\sigma_\mathrm{A}, \sigma_\mathrm{V})=(10~\mu$as, 10~km/s). The GRAVITY and SINFONI instruments can also be used to detect this effect. It should be possible to detect the pericenter advance within a few years for $(\sigma_\mathrm{A}, \sigma_\mathrm{V})=(10~\mu$as, $10$~km/s), also by using these two instruments. The HOPC contributions can be observed within several months when considering accuracies $(\sigma_\mathrm{A}, \sigma_\mathrm{V})=(10~\mu$as, 1~km/s). These effects can be detected using GRAVITY and MICADO. The first light of this latter instrument is expected to be in 2024, detection of the HOPC contributions will thus be obtained later than the previous effects.

Contrary to the other relativistic effects, the astrometric impact of the Lense-Thirring effect is maximal near apocenter passages of the S2 star, whatever the direction of the angular momentum of the black hole. The influence of this effect on the photon path is negligible which is not the case for the time-like geodesic, when considering high spin rate and some specific directions of the angular momentum. Indeed, the trajectory of the star is affected by this effect that can lead to a maximal astrometric shift of about 10~$\mu$as, 25~$\mu$as and 40~$\mu$as at the first, second and third orbital period of S2, respectively. The maximal spectroscopic shift is of about 1~km/s near the three pericenter passages. Strong constraint on the angular momentum parameters of the black hole with S2 observations generated by current instruments is thus only possible using astrometric measurements (near apocenter passages) since spectrographs reaching accuracies of about 1~km/s or better are not yet available. However, spectroscopic (and astrometric) observations are also important to constrain the angular momentum since the other relativistic effects are maximal near pericenter passages. More precisely, such observations will allow us to bring strong constraint on the orbital parameters, and the mass and the distance of the black hole, and thus lead to constrain the Lense-Thirring effect. 

As null geodesics are weakly affected by the Lense-Thirring effect, we investigated the possibility of constraining the parameters of the angular momentum of the black hole without considering ray tracing, and by using analytical approximations of the gravitational lensing from \cite{2006PhRvD..74l3009S} since such an effect can reach an amount of $\approx 20~\mu$as near pericenter of the S2 star. We showed that with such a stellar-orbit model it is possible to constrain the parameter space with small uncertainties. In particular, if we consider observations obtained during three S2 orbital periods and $(a, \Omega',i')~=~(0.99,160^{\circ},45^\circ)$ with accuracies $(\sigma_\mathrm{A}, \sigma_\mathrm{V})~=~(10~\mu$as, 10~km/s), we find very good $1\sigma$ errors of $\sigma_a~=~0.1$, $\sigma_{i'}~=~15^\circ$ and $\sigma_{\Omega'}~=~30^\circ$. We found similar constraints even with a smaller norm of $a~=~0.7$. If we consider $\sigma_\mathrm{A}~=~30~\mu$as or an observing time of one or two orbital periods instead of three, the constraints on the norm $a$ are still very good: $\sigma_a~=~0.2 - 0.4$. The constraint on the Lense-Thirring effect is thus possible with S2 observations obtained with GRAVITY and SINFONI, and by using a model that does not use ray tracing of photons, and is thus much faster. However, long monitorings are necessary to highly constrain the angular momentum parameters with S2. Detection by GRAVITY of closer stars to the Galactic center than S2 would allow us to obtain similar constraints within a few years.

We showed that the GRAVITY instrument will be capable of detecting low- and high-order relativistic effects with the S2 star. This instrument will thus allow to test GR in the strong-field regime, and possibly constrain the properties of the compact source Sgr~A*. We note however that the detected effects could be explained by other theories of gravitation. It is thus necessary to go further ahead in developing methods used to test GR and to investigate how it is possible to highlight deviations of those alternative theories from GR. Moreover, even if the central object is described by GR, other exotic objects also described by this theory can explain the mass at the Galactic center, such as the boson stars or the gravastar \citep{1969PhRv..187.1767R,2001gr.qc.....9035M}. Comparisons between Kerr and exotic metrics have already been explored but further studies need to be performed \citep{2014PhRvD..90b4068G,2014PhRvD..90j4013S}. In particular to determine whether degeneracies between the two types of objects can appear. For instance, the possibility that stellar orbits obtained in strong-field regime with a Kerr black hole and a boson star are similar has not been ruled out, even when the compact object parameters and the orbital parameters used to generate such orbits are different.

It is important to state that the work done in this paper is a preliminary study on detection of relativistic effects with the S2 star. Indeed, it neglects several contributions such as the extended mass that may be present in the Galactic center and composed of stars, stellar remnants or dark matter. It is possible that this mass modifies the star trajectory and induces a Newtonian precession which is opposite to the one due to the pericenter advance. Thus, the astrometric and spectroscopic impacts of this latter effect on S2 observations will be decreased. Several works have been done on this topic in order to determine whether this mass could be constrained and to evaluate its impact on stellar orbits observed at the Galactic center. 

In particular, \cite{2001A&A...374...95R} showed that the Newtonian effect can either partially or entirely compensate for the relativistic precession. The authors determined that in the particular case of the S2 star, an extended mass equal to 0.1\% of the black hole mass is needed to dominate the pericenter advance effect. \cite{2005ApJ...622..878W} showed that it should be possible to constrain this extended mass if we observe the motion of 100 stars located in the 0.4 central parsec for a period of  10~years. In their study these authors considered an extended mass of about $10^3~M_\odot$, and uncertainties of 0.5~mas and 10~km/s. Moreover, they claimed that we will be able to detect relativistic effects such as the pericenter advance in spite of the Newtonian precession induced by this mass. This will require the use of the same conditions as used to constrain the mass. The detection of the Lense-Thirring effect will require consideration of astrometric accuracies less than 0.05~mas. Other studies have been performed on the detection of the Lense-Thirring effect in the presence of gravitational perturbations generated by stars and stellar remnants \citep{2010PhRvD..81f2002M, 2011CQGra..28v5029S}. In particular, \cite{2010PhRvD..81f2002M} demonstrated that the detection of this effect should be possible when considering astrometric observations of GRAVITY, if the semi-major axis of the star is less than 0.5~mpc (the semi-major of S2 is around 5~mpc). In a more recent work made by \cite{2017ApJ...834..198Z}, the authors studied the influence of the S102 star on S2 astrometric and spectroscopic observations. They concluded that this star will very likely obscure the angular momentum-induced effect of the black hole. However, the authors found that this effect dominates the stellar perturbations if the observed stars have semi major-axes smaller than $0.5 - 2~$mpc and if the black hole is maximally spinning. 

All of these studies highlight the importance of taking into account the hypothetical extended mass in future stellar-orbit models used to interpret the S2 star observations. If GRAVITY does not discover stars closer to the Galactic center, the detection of the different relativistic effects with S2 will be possible only if we succeed in constraining the extended mass by using observations of several stars obtained at the Galactic center, and if this mass is sufficiently weak to not dominate all relativistic effects. Besides, as mentioned in various papers \citep[see e.g.,][]{2010PhRvD..81f2002M, 2010ApJ...720.1303A}, we expect that the Newtonian effect can be dissociated from the relativistic effects if we consider runs of observation that include several orbital periods of the star, because those types of effects have different temporal evolutions.


\section{Acknowledgements}
This work was supported by ASHRA (Action Sp\'ecifique Haute R\'esolution Angulaire). We thank Didier Pelat for helpful discussions on statistical tests used in model-fitting processes. We also thank Guillaume Schworer for both having presented us the MCMC method used in this paper, and for advice and explanations regarding this method.

\bibliographystyle{aa}
\bibliography{biblio}

\begin{thebibliography}{54}
\expandafter\ifx\csname natexlab\endcsname\relax\def\natexlab#1{#1}\fi

\bibitem[{{Andrae} {et~al.}(2010){Andrae}, {Schulze-Hartung}, \&
  {Melchior}}]{2010arXiv1012.3754A}
{Andrae}, R., {Schulze-Hartung}, T., \& {Melchior}, P. 2010, ArXiv e-prints

\bibitem[{{Ang{\'e}lil} \& {Saha}(2010)}]{2010ApJ...711..157A}
{Ang{\'e}lil}, R. \& {Saha}, P. 2010, \apj, 711, 157

\bibitem[{{Ang{\'e}lil} \& {Saha}(2011)}]{2011ApJ...734L..19A}
{Ang{\'e}lil}, R. \& {Saha}, P. 2011, \apjl, 734, L19

\bibitem[{{Ang{\'e}lil} {et~al.}(2010){Ang{\'e}lil}, {Saha}, \&
  {Merritt}}]{2010ApJ...720.1303A}
{Ang{\'e}lil}, R., {Saha}, P., \& {Merritt}, D. 2010, \apj, 720, 1303

\bibitem[{{Binder}(2002)}]{2002mcss.book.....B}
{Binder}, Heermann, D.~W.~K. 2002, {Monte Carlo simulation in statistical
  physics: an introduction}

\bibitem[{{Boehle} {et~al.}(2016){Boehle}, {Ghez}, {Sch{\"o}del}, {Meyer},
  {Yelda}, {Albers}, {Martinez}, {Becklin}, {Do}, {Lu}, {Matthews}, {Morris},
  {Sitarski}, \& {Witzel}}]{2016ApJ...830...17B}
{Boehle}, A., {Ghez}, A.~M., {Sch{\"o}del}, R., {et~al.} 2016, \apj, 830, 17

\bibitem[{{Bozza} \& {Mancini}(2004)}]{2004ApJ...611.1045B}
{Bozza}, V. \& {Mancini}, L. 2004, \apj, 611, 1045

\bibitem[{{Bozza} \& {Mancini}(2005)}]{2005ApJ...627..790B}
{Bozza}, V. \& {Mancini}, L. 2005, \apj, 627, 790

\bibitem[{{Bozza} \& {Mancini}(2012)}]{2012ApJ...753...56B}
{Bozza}, V. \& {Mancini}, L. 2012, \apj, 753, 56

\bibitem[{{Broderick} {et~al.}(2011){Broderick}, {Fish}, {Doeleman}, \&
  {Loeb}}]{2011ApJ...735..110B}
{Broderick}, A.~E., {Fish}, V.~L., {Doeleman}, S.~S., \& {Loeb}, A. 2011, \apj,
  735, 110

\bibitem[{{Broderick} {et~al.}(2014){Broderick}, {Johannsen}, {Loeb}, \&
  {Psaltis}}]{2014ApJ...784....7B}
{Broderick}, A.~E., {Johannsen}, T., {Loeb}, A., \& {Psaltis}, D. 2014, \apj,
  784, 7

\bibitem[{{Broderick} {et~al.}(2009){Broderick}, {Loeb}, \&
  {Narayan}}]{2009ApJ...701.1357B}
{Broderick}, A.~E., {Loeb}, A., \& {Narayan}, R. 2009, \apj, 701, 1357

\bibitem[{{Catanzarite}(2010)}]{2010arXiv1008.3416C}
{Catanzarite}, J.~H. 2010, ArXiv e-prints

\bibitem[{{Doeleman} {et~al.}(2009){Doeleman}, {Fish}, {Broderick}, {Loeb}, \&
  {Rogers}}]{2009ApJ...695...59D}
{Doeleman}, S.~S., {Fish}, V.~L., {Broderick}, A.~E., {Loeb}, A., \& {Rogers},
  A.~E.~E. 2009, \apj, 695, 59

\bibitem[{{Eckart} \& {Genzel}(1997)}]{1997MNRAS.284..576E}
{Eckart}, A. \& {Genzel}, R. 1997, \mnras, 284, 576

\bibitem[{{Eisenhauer} {et~al.}(2003){Eisenhauer}, {Abuter}, {Bickert},
  {Biancat-Marchet}, {Bonnet}, {Brynnel}, {Conzelmann}, {Delabre}, {Donaldson},
  {Farinato}, {Fedrigo}, {Genzel}, {Hubin}, {Iserlohe}, {Kasper},
  {Kissler-Patig}, {Monnet}, {Roehrle}, {Schreiber}, {Stroebele}, {Tecza},
  {Thatte}, \& {Weisz}}]{2003SPIE.4841.1548E}
{Eisenhauer}, F., {Abuter}, R., {Bickert}, K., {et~al.} 2003, in \procspie,
  Vol. 4841, Instrument Design and Performance for Optical/Infrared
  Ground-based Telescopes, ed. M.~{Iye} \& A.~F.~M. {Moorwood}, 1548--1561

\bibitem[{{Eisenhauer} {et~al.}(2011){Eisenhauer}, {Perrin}, {Brandner},
  {Straubmeier}, {Perraut}, {Amorim}, {Sch{\"o}ller}, {Gillessen}, {Kervella},
  {Benisty}, {Araujo-Hauck}, {Jocou}, {Lima}, {Jakob}, {Haug}, {Cl{\'e}net},
  {Henning}, {Eckart}, {Berger}, {Garcia}, {Abuter}, {Kellner}, {Paumard},
  {Hippler}, {Fischer}, {Moulin}, {Villate}, {Avila}, {Gr{\"a}ter}, {Lacour},
  {Huber}, {Wiest}, {Nolot}, {Carvas}, {Dorn}, {Pfuhl}, {Gendron}, {Kendrew},
  {Yazici}, {Anton}, {Jung}, {Thiel}, {Choquet}, {Klein}, {Teixeira}, {Gitton},
  {Moch}, {Vincent}, {Kudryavtseva}, {Str{\"o}bele}, {Sturm}, {F{\'e}dou},
  {Lenzen}, {Jolley}, {Kister}, {Lapeyr{\`e}re}, {Naranjo}, {Lucuix},
  {Hofmann}, {Chapron}, {Neumann}, {Mehrgan}, {Hans}, {Rousset}, {Ramos},
  {Suarez}, {Lederer}, {Reess}, {Rohloff}, {Haguenauer}, {Bartko}, {Sevin},
  {Wagner}, {Lizon}, {Rabien}, {Collin}, {Finger}, {Davies}, {Rouan},
  {Wittkowski}, {Dodds-Eden}, {Ziegler}, {Cassaing}, {Bonnet}, {Casali},
  {Genzel}, \& {Lena}}]{2011Msngr.143...16E}
{Eisenhauer}, F., {Perrin}, G., {Brandner}, W., {et~al.} 2011, The Messenger,
  143, 16

\bibitem[{{Foreman-Mackey} {et~al.}(2013){Foreman-Mackey}, {Hogg}, {Lang}, \&
  {Goodman}}]{2013PASP..125..306F}
{Foreman-Mackey}, D., {Hogg}, D.~W., {Lang}, D., \& {Goodman}, J. 2013, \pasp,
  125, 306

\bibitem[{{Fragile} \& {Mathews}(2000)}]{2000ApJ...542..328F}
{Fragile}, P.~C. \& {Mathews}, G.~J. 2000, \apj, 542, 328

\bibitem[{{Genzel} {et~al.}(2010){Genzel}, {Eisenhauer}, \&
  {Gillessen}}]{2010RvMP...82.3121G}
{Genzel}, R., {Eisenhauer}, F., \& {Gillessen}, S. 2010, Reviews of Modern
  Physics, 82, 3121

\bibitem[{{Genzel} {et~al.}(1996){Genzel}, {Thatte}, {Krabbe}, {Kroker}, \&
  {Tacconi-Garman}}]{1996ApJ...472..153G}
{Genzel}, R., {Thatte}, N., {Krabbe}, A., {Kroker}, H., \& {Tacconi-Garman},
  L.~E. 1996, \apj, 472, 153

\bibitem[{{Ghez} {et~al.}(1998){Ghez}, {Klein}, {Morris}, \&
  {Becklin}}]{1998ApJ...509..678G}
{Ghez}, A.~M., {Klein}, B.~L., {Morris}, M., \& {Becklin}, E.~E. 1998, \apj,
  509, 678

\bibitem[{{Ghez} {et~al.}(2008){Ghez}, {Salim}, {Weinberg}, {Lu}, {Do}, {Dunn},
  {Matthews}, {Morris}, {Yelda}, {Becklin}, {Kremenek}, {Milosavljevic}, \&
  {Naiman}}]{2008ApJ...689.1044G}
{Ghez}, A.~M., {Salim}, S., {Weinberg}, N.~N., {et~al.} 2008, \apj, 689, 1044

\bibitem[{{Gillessen} {et~al.}(2009{\natexlab{a}}){Gillessen}, {Eisenhauer},
  {Fritz}, {Bartko}, {Dodds-Eden}, {Pfuhl}, {Ott}, \&
  {Genzel}}]{2009ApJ...707L.114G}
{Gillessen}, S., {Eisenhauer}, F., {Fritz}, T.~K., {et~al.} 2009{\natexlab{a}},
  \apjl, 707, L114

\bibitem[{{Gillessen} {et~al.}(2009{\natexlab{b}}){Gillessen}, {Eisenhauer},
  {Trippe}, {Alexander}, {Genzel}, {Martins}, \& {Ott}}]{2009ApJ...692.1075G}
{Gillessen}, S., {Eisenhauer}, F., {Trippe}, S., {et~al.} 2009{\natexlab{b}},
  \apj, 692, 1075

\bibitem[{{Gillessen} {et~al.}(2017){Gillessen}, {Plewa}, {Eisenhauer}, {Sari},
  {Waisberg}, {Habibi}, {Pfuhl}, {George}, {Dexter}, {von Fellenberg}, {Ott},
  \& {Genzel}}]{2017ApJ...837...30G}
{Gillessen}, S., {Plewa}, P.~M., {Eisenhauer}, F., {et~al.} 2017, \apj, 837, 30

\bibitem[{{Grandcl{\'e}ment} {et~al.}(2014){Grandcl{\'e}ment}, {Som{\'e}}, \&
  {Gourgoulhon}}]{2014PhRvD..90b4068G}
{Grandcl{\'e}ment}, P., {Som{\'e}}, C., \& {Gourgoulhon}, E. 2014, \prd, 90,
  024068

\bibitem[{{Grould} {et~al.}(2016){Grould}, {Paumard}, \&
  {Perrin}}]{2016A&A...591A.116G}
{Grould}, M., {Paumard}, T., \& {Perrin}, G. 2016, \aap, 591, A116

\bibitem[{{Jaroszynski}(1998)}]{1998AcA....48..653J}
{Jaroszynski}, M. 1998, \actaa, 48, 653

\bibitem[{{Johannsen}(2016)}]{2016CQGra..33k3001J}
{Johannsen}, T. 2016, Classical and Quantum Gravity, 33, 113001

\bibitem[{{J{\o}rgensen} {et~al.}(2016){J{\o}rgensen}, {Bj{\ae}lde}, \&
  {Hannestad}}]{2016MNRAS.458.3614J}
{J{\o}rgensen}, J.~H., {Bj{\ae}lde}, O.~E., \& {Hannestad}, S. 2016, \mnras,
  458, 3614

\bibitem[{{Kannan} \& {Saha}(2009)}]{2009ApJ...690.1553K}
{Kannan}, R. \& {Saha}, P. 2009, \apj, 690, 1553

\bibitem[{{Kraniotis}(2007)}]{2007CQGra..24.1775K}
{Kraniotis}, G.~V. 2007, Classical and Quantum Gravity, 24, 1775

\bibitem[{Levenberg(1944)}]{citeulike:2946351}
Levenberg, K. 1944, Quart. Applied Math., 2, 164

\bibitem[{{Mazur} \& {Mottola}(2001)}]{2001gr.qc.....9035M}
{Mazur}, P.~O. \& {Mottola}, E. 2001, ArXiv General Relativity and Quantum
  Cosmology e-prints

\bibitem[{{Merritt} {et~al.}(2010){Merritt}, {Alexander}, {Mikkola}, \&
  {Will}}]{2010PhRvD..81f2002M}
{Merritt}, D., {Alexander}, T., {Mikkola}, S., \& {Will}, C.~M. 2010, \prd, 81,
  062002

\bibitem[{{Nishiyama} {et~al.}(2017){Nishiyama}, {Saida}, {Takamori},
  {Takahashi}, {Schoedel}, {Najarro}, {Hamano}, {Omiya}, {Tamura}, {Takahashi},
  {Gorin}, {Nagatomo}, \& {Nagata}}]{2017arXiv170901598N}
{Nishiyama}, S., {Saida}, H., {Takamori}, Y., {et~al.} 2017, ArXiv e-prints

\bibitem[{{Parsa} {et~al.}(2017){Parsa}, {Eckart}, {Shahzamanian}, {Karas},
  {Zaja{\v c}ek}, {Zensus}, \& {Straubmeier}}]{2017ApJ...845...22P}
{Parsa}, M., {Eckart}, A., {Shahzamanian}, B., {et~al.} 2017, \apj, 845, 22

\bibitem[{{Rubilar} \& {Eckart}(2001)}]{2001A&A...374...95R}
{Rubilar}, G.~F. \& {Eckart}, A. 2001, \aap, 374, 95

\bibitem[{{Ruffini} \& {Bonazzola}(1969)}]{1969PhRv..187.1767R}
{Ruffini}, R. \& {Bonazzola}, S. 1969, Physical Review, 187, 1767

\bibitem[{{Sadeghian} \& {Will}(2011)}]{2011CQGra..28v5029S}
{Sadeghian}, L. \& {Will}, C.~M. 2011, Classical and Quantum Gravity, 28,
  225029

\bibitem[{{Sakai} {et~al.}(2014){Sakai}, {Saida}, \&
  {Tamaki}}]{2014PhRvD..90j4013S}
{Sakai}, N., {Saida}, H., \& {Tamaki}, T. 2014, \prd, 90, 104013

\bibitem[{{Sereno} \& {de Luca}(2006)}]{2006PhRvD..74l3009S}
{Sereno}, M. \& {de Luca}, F. 2006, \prd, 74, 123009

\bibitem[{{Taff}(1985)}]{1985cmcg.book.....T}
{Taff}, L.~G. 1985, {Celestial mechanics: A computational guide for the
  practitioner}

\bibitem[{{Vincent} {et~al.}(2016){Vincent}, {Meliani}, {Grandcl{\'e}ment},
  {Gourgoulhon}, \& {Straub}}]{2016CQGra..33j5015V}
{Vincent}, F.~H., {Meliani}, Z., {Grandcl{\'e}ment}, P., {Gourgoulhon}, E., \&
  {Straub}, O. 2016, Classical and Quantum Gravity, 33, 105015

\bibitem[{{Vincent} {et~al.}(2011){Vincent}, {Paumard}, {Gourgoulhon}, \&
  {Perrin}}]{2011CQGra..28v5011V}
{Vincent}, F.~H., {Paumard}, T., {Gourgoulhon}, E., \& {Perrin}, G. 2011,
  Classical and Quantum Gravity, 28, 225011

\bibitem[{{Visser}(2007)}]{2007arXiv0706.0622V}
{Visser}, M. 2007, ArXiv e-prints

\bibitem[{{Weinberg} {et~al.}(2005){Weinberg}, {Milosavljevi{\'c}}, \&
  {Ghez}}]{2005ApJ...622..878W}
{Weinberg}, N.~N., {Milosavljevi{\'c}}, M., \& {Ghez}, A.~M. 2005, \apj, 622,
  878

\bibitem[{{Will}(2008)}]{2008ApJ...674L..25W}
{Will}, C.~M. 2008, \apjl, 674, L25

\bibitem[{{Wollman} {et~al.}(1977){Wollman}, {Geballe}, {Lacy}, {Townes}, \&
  {Rank}}]{1977ApJ...218L.103W}
{Wollman}, E.~R., {Geballe}, T.~R., {Lacy}, J.~H., {Townes}, C.~H., \& {Rank},
  D.~M. 1977, \apjl, 218, L103

\bibitem[{{Yu} {et~al.}(2016){Yu}, {Zhang}, \& {Lu}}]{2016ApJ...827..114Y}
{Yu}, Q., {Zhang}, F., \& {Lu}, Y. 2016, \apj, 827, 114

\bibitem[{{Zhang} \& {Iorio}(2017)}]{2017ApJ...834..198Z}
{Zhang}, F. \& {Iorio}, L. 2017, \apj, 834, 198

\bibitem[{{Zhang} {et~al.}(2015){Zhang}, {Lu}, \& {Yu}}]{2015ApJ...809..127Z}
{Zhang}, F., {Lu}, Y., \& {Yu}, Q. 2015, \apj, 809, 127

\bibitem[{{Zucker} {et~al.}(2006){Zucker}, {Alexander}, {Gillessen},
  {Eisenhauer}, \& {Genzel}}]{2006ApJ...639L..21Z}
{Zucker}, S., {Alexander}, T., {Gillessen}, S., {Eisenhauer}, F., \& {Genzel},
  R. 2006, \apjl, 639, L21

\end{thebibliography}


\newpage
\appendix


\section{Obtaining the astrometric positions and radial velocities of a moving star with the ray-tracing code \textsc{Gyoto}}
\label{app_A}

\subsection{Astrometry}
\label{app_A1}

\textsc{Gyoto} is a ray-tracing code integrating null geodesics backwards in time. It works on the basis that each photon is initially located at the observer screen: one pixel of the screen corresponds to the final direction of one photon. When the photon reaches the star during its integration, the pixel corresponding to this photon illuminates. It is thus possible to obtain the image of the star. However, in our study we only use one photon of the primary image and consider the final direction of this photon $(\alpha_{\gamma}, \delta_{\gamma})$ as being the astrometric position of the star $(\alpha_{s}, \delta_{s})$. The aim of this method is to decrease the computation time without any loss of precision on the simulated apparent position of the star. This position is obtained when the Euclidian distance between the radial coordinate of the photon and the surface of the star reaches $\approx 10^{-2} M$. The corresponding maximal astrometric and spectroscopic errors of the star are about $10^{-1}$$\mu$as and $10^{-4}$~km/s, respectively.

\begin{figure}[!t]
\centering
    \begin{minipage}[c]{.9\linewidth}
      \includegraphics[scale=0.5]{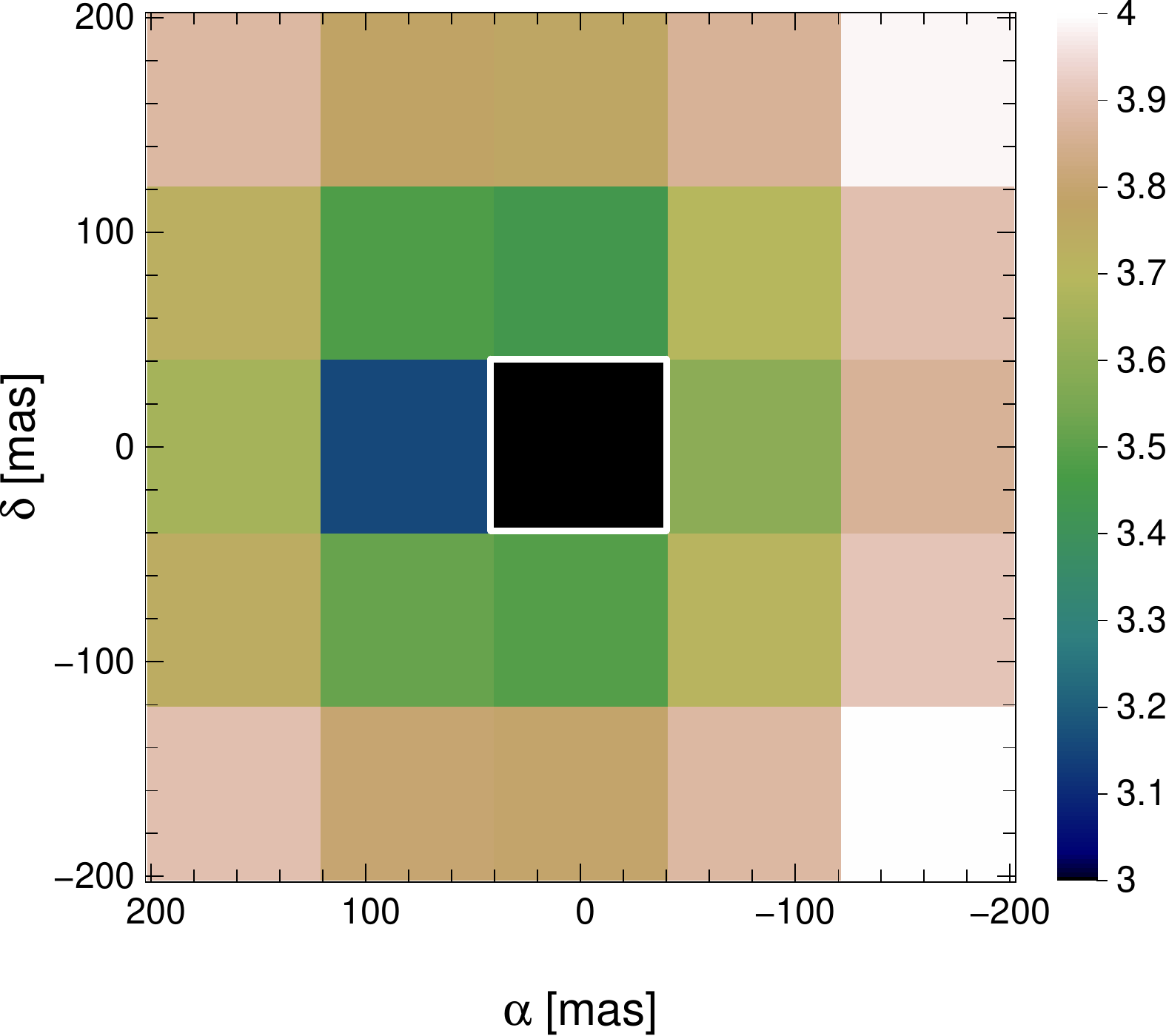}\\
   \end{minipage}
          \begin{minipage}[c]{.9\linewidth}
      \includegraphics[scale=0.5]{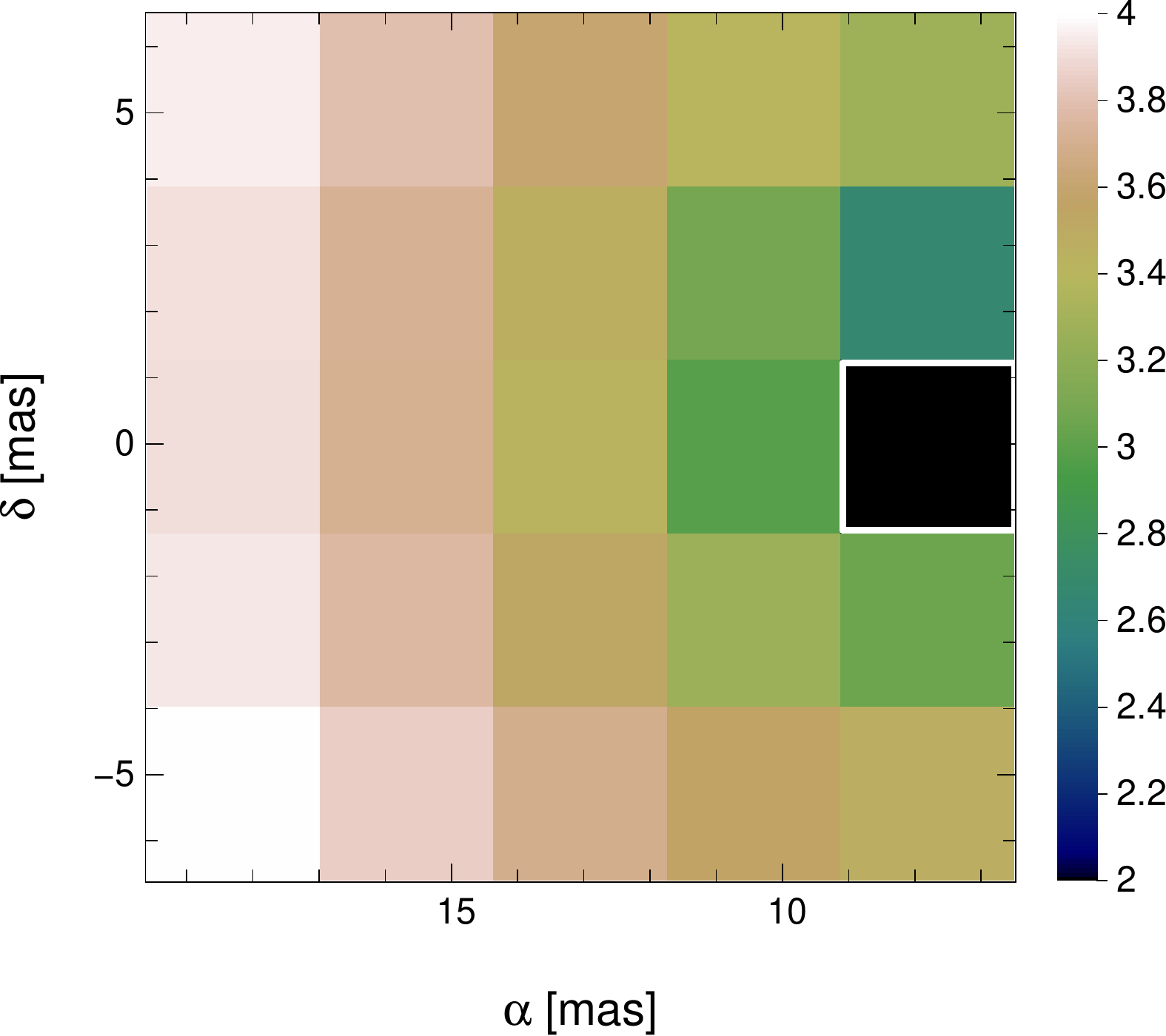}
   \end{minipage}
   \caption{Maps of the logarithmic \texttt{MinDistance} function. \textit{Top}: map obtained considering a field-of-view of 200 mas. The darker the pixels, the smaller the distance between the photon and the surface of the star. The pixel where the \texttt{MinDistance} function is minimal is marked by a white square. \textit{Bottom}: map obtained considering the previous white square as field-of-view. The color-bars are labeled in $M$ unit.}
        \label{MD}
\end{figure}

As we consider a full-GR model to generate the S2 observations, we need to consider a moving star in the ray-tracing code \textsc{Gyoto}. At a given observation date, we do not know the position of the star relative to the black hole. In other words, we cannot predict where the image of the star will be in the observer screen. We thus need to implement a procedure to get the star position $(\alpha_s,\delta_s)$. To do so, we use a  mathematical function named \texttt{MinDistance} in \textsc{Gyoto}. This function gives the squared minimum Euclidian distance between the photon and the surface of the star. Zeroes of this function correspond to photons that have reached the surface of the star. The procedure that we use to converge to the closest photon from the surface of the star ($\approx~10^{-2} M$) with the \texttt{MinDistance} function is described below:
\begin{itemize} 
\item step 1: we compute a \texttt{MinDistance} map considering an initial screen with a sufficiently big field-of-view to contain the full S2 orbit. A resolution of $5 \times 5$ pixels is taken for this first image (see the top image in Fig.~\ref{MD}). Then, we locate the pixel where the value of the \texttt{MinDistance} function is the smallest: this minimal pixel is associated with the photon which passes the closest to the star. This corresponds to the black pixel marked by a white square on the top image in Fig.~\ref{MD}. 
\item Step 2: we compute another \texttt{MinDistance} map also considering a resolution of $5 \times 5$ pixels but with a field-of-view both centered on the pixel previously found and equal to the size of this pixel. This new map is thus a zoom of the former map. In this image, we again locate the minimal pixel (see the bottom image in Fig.~\ref{MD}).
\item Step 3: we repeat the step 2 until the minimal distance between the photon and the surface of the star is of about $10^{-2} M$. When this distance is reached, we obtain the astrometric position $(\alpha_s,\delta_s)$ of the star which is equal to the final direction of the photon $(\alpha_{\gamma}, \delta_{\gamma})$.
\end{itemize} 

The computing time needed to reach the distance $\approx 10^{-2} M$ is of about 3 seconds.

\subsection{Spectroscopy}
\label{App_A2}

The energy $E$ of a photon measured by an observer is given by
\begin{equation}
E = - \vec{u} \cdot \vec{p}
,\end{equation}
where $\vec{u}$ and $\vec{p}$ are the four-velocity of the observer and the four-momentum of the photon along its geodesic, respectively. The emitted $E_{\mathrm{em}}$ and received $E_{\mathrm{obs}}$ energies of the photon are related as
\begin{equation}
E_{\mathrm{obs}} = g E_{\mathrm{em}},
\end{equation}
with 
\begin{eqnarray}
  g & = & \frac{\vec{u_{\mathrm{obs}}} \cdot \vec{p_{\mathrm{obs}}}}{\vec{u_{\mathrm{em}}} \cdot \vec{p_{\mathrm{em}}}}
,\end{eqnarray}
where $\vec{u_{\mathrm{obs(em)}}}$ and $\vec{p_{\mathrm{obs(em)}}}$ are the four-velocity of the observer and the four-momentum of the photon at reception(emission), respectively. To evaluate the quantities at emission we need to know the coordinates of the star and the photon at emission. To do so, we consider the photon used to estimate the astrometric position of the star. Indeed, by using this photon we have access to its emission date and we thus can obtain the position of both the photon and the star at this date. The radial velocity of the star can also be obtained since it is related to the $g$ factor by
\begin{equation}
\mathrm{V} = \left(\frac{1}{g} - 1\right)c = \mathcal{Z}c
,\end{equation}
where $\mathcal{Z}$ is the redshift of the star.


\begin{figure}[!t]
\centering
      \includegraphics[scale=0.5]{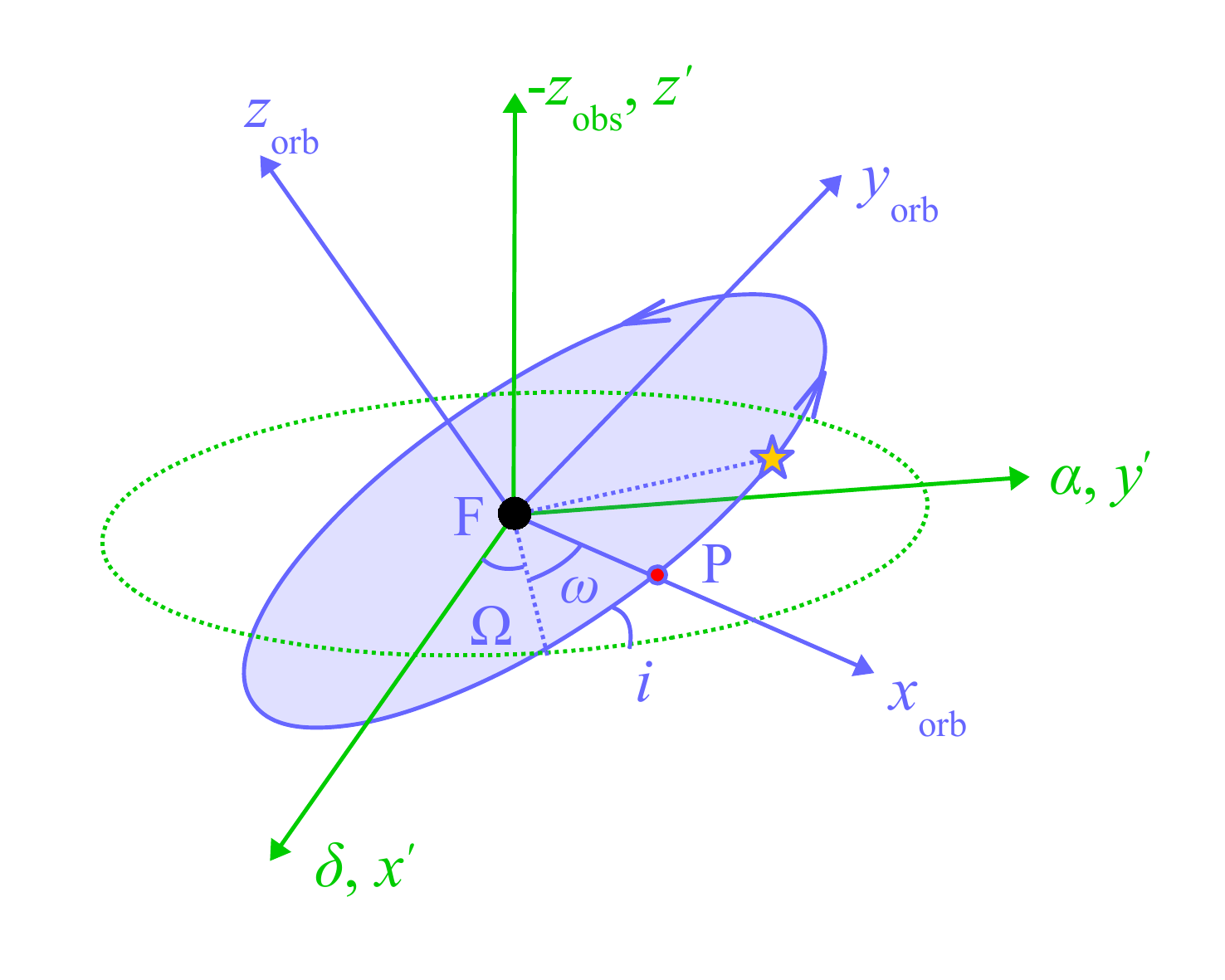}
   \caption{Illustration of the orbital parameters $i$, $\Omega$ and $\omega$ corresponding to the inclination of the orbit, the angle of the line of nodes and the argument of pericenter, respectively. The frame $(x',y',z')$ is a second observer frame and is related to the observer frame $(\alpha,\delta,z_{\mathrm{obs}})$ defined in Sect.~\ref{obs}.}
   \label{frames_app}
\end{figure}

\section{Recovering the star coordinates in the observer coordinates $(\alpha,\delta,z_{\mathrm{obs}})$ from its orbital parameters}
\label{app_B}

\begin{figure}[!t]
\centering
      \includegraphics[scale=0.5]{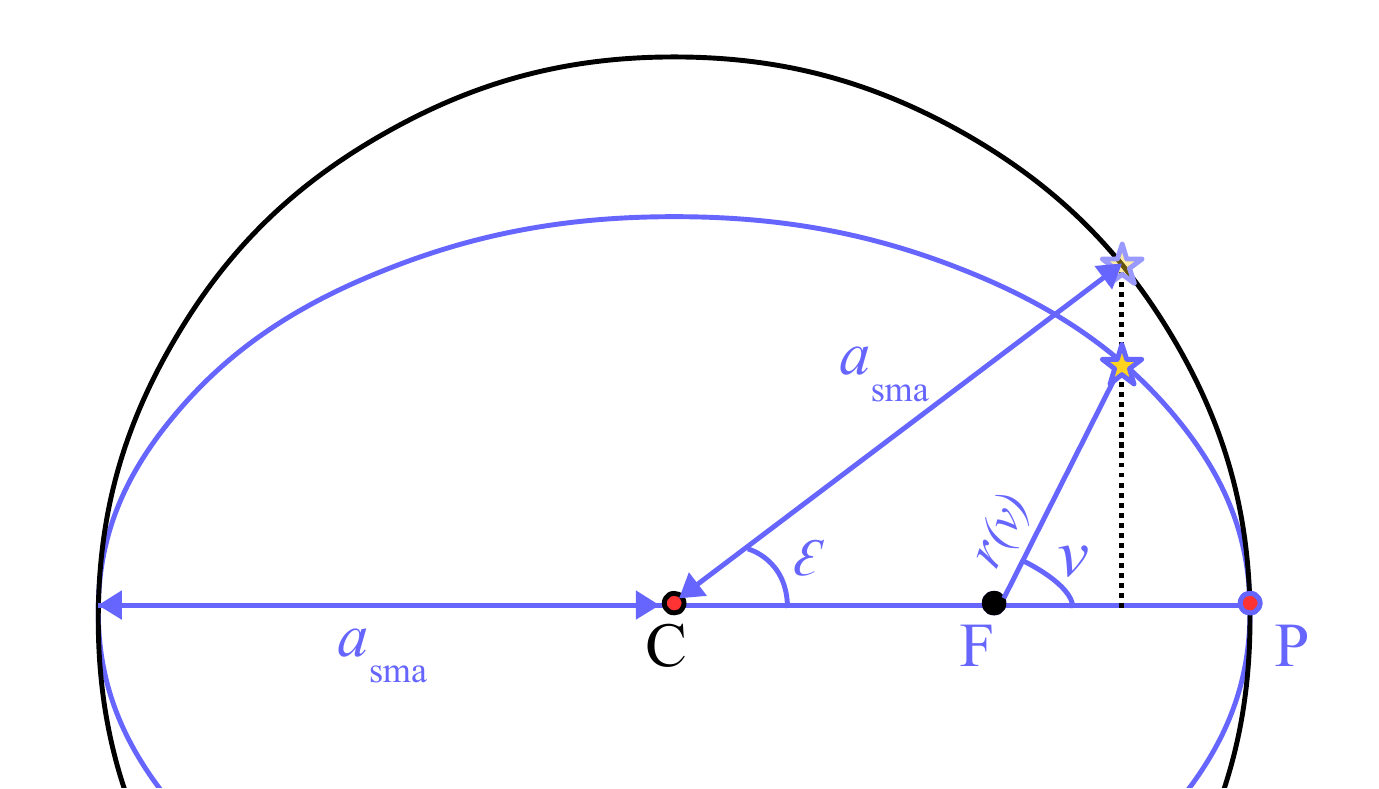}
   \caption{Illustration of the true $\nu$ and eccentric $\mathcal{E}$ anomalies. The black circle has a radius equal to the semi-major axis $a_{\mathrm{sma}}$. The point O is the origin of the circle, F is the focus of the orbit which corresponds to the black hole in our case, and P is the pericenter. The position $r(\nu)$ of the star is also present.}
        \label{EA}
\end{figure}

We define a new frame $(x',y',z')$ related to the observer frame defined in Sect.~\ref{obs} as: $x'=\delta$, $y'=\alpha$, and $z'=-z_{\mathrm{obs}}$ (see Fig.~\ref{frames_app}). Knowing the orbital parameters of the star, it is possible to get its position and velocity in the frame $(x',y',z')$.
To do so, we use the usual trajectory equation of a star orbiting a central mass, and originating from the equation of motion in the two-body problem
\begin{equation}
\label{eq:radius}
r(\nu) = \frac{a_{\mathrm{sma}}(1-e^2)}{1+e\cos{\nu}}
,\end{equation}
where $\nu$ is the true anomaly corresponding to the angle between the pericenter position and the star (see Fig.~\ref{EA}). This angle is obtained by using the following formula
\begin{equation}
\tan{\frac{\nu}{2}} = \sqrt{\frac{1+e}{1-e}} \tan{\frac{\mathcal{E}}{2}}
,\end{equation}
where $\mathcal{E}$ is the eccentric anomaly (see Fig.~\ref{EA}). To evaluate this angle we need to solve the Kepler equation given by
\begin{equation}
\mathcal{E} - e \sin{\mathcal{E}} - \mathcal{M} = 0
,\end{equation}
where $\mathcal{M}$ is the averaged anomaly (see Fig.~\ref{EA}). This last angle depends on the period, the time of the pericenter passage and a date $t$ as
\begin{equation}
\mathcal{M} = \frac{2 \pi}{T}(t - t_p).
\end{equation}

The cartesian coordinates of the star are expressed in the orbital plane $(x_{\mathrm{orb}},y_{\mathrm{orb}})$, at a given $t$, as
\begin{eqnarray}
(x_{s, \mathrm{orb}},y_{s, \mathrm{orb}}) & = & \left( r(\nu) \cos{\nu},r(\nu) \sin{\nu}  \right), \nonumber \\
        & =  & \left( a_{\mathrm{sma}} \left( \cos{\mathcal{E}} - e \right), a_{\mathrm{sma}} \sqrt{1 - e^2} \sin{\mathcal{E}} \right), \\
(v_{x_{s, \mathrm{orb}}}, v_{y_{s, \mathrm{orb}}}) & = & \left(- \frac{2\pi}{T} \frac{a_{\mathrm{sma}}^2}{r} \sin{\mathcal{E}},  \frac{2\pi}{T} \frac{a_{\mathrm{sma}}^2}{r} \sqrt{1-e^2} \cos{\mathcal{E}} \right).
\end{eqnarray}
By using these coordinates and the Thiele-Innes formulas given by \citep{2010arXiv1008.3416C}
\begin{align*}
& A =  \cos{\omega} \cos{\Omega} -  \sin{\omega} \sin{\Omega} \cos{i},   \notag \\
& B = \cos{\omega} \sin{\Omega}  +  \sin{\omega} \cos{\Omega} \cos{i},   \notag \\
& C = - \sin{\omega} \sin{i},  \notag \\
& F = -\sin{\omega} \cos{\Omega}  -  \cos{\omega} \sin{\Omega} \cos{i},   \notag \\
& G = -\sin{\omega} \sin{\Omega}  +  \cos{\omega} \cos{\Omega} \cos{i},   \notag \\
& H = - \cos{\omega} \sin{i},
\end{align*}
we can obtain the coordinates of the star in the new frame $(x',y',z')$
\begin{align*}
x'_s = A x_{s, \mathrm{orb}} + F y_{s, \mathrm{orb}}, 
\qquad
v'_{x_s} = A v_{x_{s, \mathrm{orb}}} + F v_{y_{s, \mathrm{orb}}}, \notag \\
y'_s = B x_{s, \mathrm{orb}} + G y_{s, \mathrm{orb}},
\qquad
v'_{y_s} = B v_{x_{s, \mathrm{orb}}} + G v_{y_{s, \mathrm{orb}}},\notag   \\
z'_s = C x_{s, \mathrm{orb}} + H y_{s, \mathrm{orb}},
\qquad
v'_{z_s} = C v_{x_{s, \mathrm{orb}}} + H v_{y_{s, \mathrm{orb}}}.
\end{align*}
Finally, the coordinates of the star in the observer frame $(\alpha,\delta,z_{\mathrm{obs}})$ are given by
\begin{align}
&\alpha_s = y'_s, 
\qquad
v_{\alpha_{s}} = v'_{y_s}, \notag \\
&\delta_s = x'_s, 
\qquad
v_{\delta_{s}} = v'_{x_s}, \notag \\
&z_{s, \mathrm{obs}} = -z'_s,
\qquad
v_{z_{s, \mathrm{obs}}} = - v'_{z_s},
\end{align}
where $(\alpha_s, \delta_s)$ is the astrometric position of the star and $v_{z_{s, \mathrm{obs}}} = \mathrm{V}$ is its radial velocity.


\section{Resolution of the Roemer equation}
\label{app_C}

In order to simulate the Roemer effect we solve the equation~\eqref{roemer} by using a dichotomy method. It is resolved to within $\approx~10^{-4}$ second which corresponds to a negligible astrometric shift: $<~10^{-6}$~$\mu$as. It also implies an error inferior to $10^{-7}$~km/s for radial velocities.


\section{Radial velocity of the star computed in models C to F}
\label{App_D}

In this section, we consider $G=M=c=1$. As in Appendix~\ref{App_A2}, we consider an emitter with a four-velocity $\vec{u}_{\mathrm{em}}$, emitting photons with a four-momentum $\vec{p}_{\mathrm{em}}$. These photons are received by a static observer possessing a four-velocity $\vec{u}_{\mathrm{obs}}$. The four-momentum of the photons at reception is $\vec{p}_{\mathrm{obs}}$. As the observer is static we have $u_{\mathrm{obs}}^{\alpha} = (u_{\mathrm{obs}}^{t},0,0,0)$. Besides, it is far from the black hole thus $g_{tt}|_{\mathrm{obs}}\footnote{this notation means that the coefficient $g_{tt}$ is evaluated at reception.}\approx-1$, so we get
\begin{align}
\vec{u}_{\mathrm{obs}} \cdot \vec{u}_{\mathrm{obs}} &= g_{tt}|_{\mathrm{obs}} (u_{\mathrm{obs}}^t)^2, \notag \\
-1 &\approx -(u_{\mathrm{obs}}^t)^2,
\end{align}
which leads to $u_{\mathrm{obs}}^{\alpha}~\approx~(1,0,0,0)$. The energy of the photon as measured by this observer is thus equal to
\begin{align}
E_{\mathrm{obs}} &= -\vec{u}_{\mathrm{obs}} \cdot \vec{p}_{\mathrm{obs}} \notag \\
&\approx - \boldsymbol{\partial}_t|_{\mathrm{obs}} \cdot \vec{p}_{\mathrm{obs}}. 
\end{align}
We decide to decompose the quantities $\vec{u}_{\mathrm{em}}$ and $\vec{p}_{\mathrm{em}}$ in the $3+1$ formalism of GR. To do so, we consider another observer called ZAMO (for Zero Angular Momentum Observer) with a four-velocity $\vec{u}_{\mathrm{ZAMO}}$. The four-velocity of the emitter and four-momentum of the photon, at emission, can be decomposed into a part parallel to $\vec{u}_{\mathrm{ZAMO}}$ (the temporal part) and a part orthogonal to $\vec{u}_{\mathrm{ZAMO}}$ (the spatial part) as
\begin{align}
\label{eq:ZAMO}
& \vec{u}_{\mathrm{em}} = \Gamma_{\mathrm{ZAMO}} (\vec{u}_{\mathrm{ZAMO}} + \vec{V}_{\mathrm{ZAMO}}), \notag \\
& \vec{p}_{\mathrm{em}} = E_{\mathrm{ZAMO}} (\vec{u}_{\mathrm{ZAMO}} + \vec{n}_{\mathrm{ZAMO}})
\end{align}
where 
\begin{align}
\Gamma_{\mathrm{ZAMO}} &= - \vec{u}_{\mathrm{ZAMO}} \cdot \vec{u}_{\mathrm{em}}, 
\end{align}
is the Lorentz factor of the star as measured by the ZAMO. The unit four-vector $\vec{n}_{\mathrm{ZAMO}}$ is the direction of emission of the photon as seen by the ZAMO, and $\vec{V}_{\mathrm{ZAMO}}$ is the four-velocity of the emitter as observed by the ZAMO. Finally, $E_{\mathrm{ZAMO}}$ is the energy of the photon at emission given by
\begin{align}
E_{\mathrm{ZAMO}} &= - \vec{u}_{\mathrm{ZAMO}} \cdot \vec{p}_{\mathrm{em}}, \notag \\
& \approx - \frac{1}{\sqrt{-g_{tt}|_{\mathrm{em}}}} \boldsymbol{\partial}_t|_\mathrm{em} \cdot \vec{p}_{\mathrm{em}},
\end{align}
where the four-velocity of the ZAMO is approximated by $\vec{u}_{\mathrm{ZAMO}} \approx \boldsymbol{\partial}_t|_\mathrm{em}/\sqrt{-g_{{tt}_{\mathrm{em}}}}$ since it is far from the black hole (the norm of the angular momentum of the black hole $a$ tends towards zero). As seen in Appendix~\ref{App_A2}, the total redshift is expressed as
\begin{align}
\mathcal{Z} &= \frac{1}{g} - 1
\end{align}
where 
\begin{align}
g  &= \frac{\vec{u_{\mathrm{obs}}} \cdot \vec{p_{\mathrm{obs}}}}{\vec{u_{\mathrm{em}}} \cdot \vec{p_{\mathrm{em}}}}.
\end{align}
Using the different previous expressions and the fact that the quantity $\boldsymbol{\partial}_t \cdot \vec{p}$ is conserved along the null geodesic, $g$ becomes
\begin{align}
g &=  \frac{\sqrt{-g_{tt}|_{\mathrm{em}}}}{\Gamma_{\mathrm{ZAMO}}(1 - \vec{V}_{\mathrm{ZAMO}} \cdot \vec{n}_{\mathrm{ZAMO}})}.
\end{align}
Now we want to obtain an expression of the velocity $\vec{V}_{\mathrm{ZAMO}}$ of the emitter as observed by the ZAMO. For doing so, we use the first equation of~\eqref{eq:ZAMO} and the approximation of the four-velocity of the ZAMO $\vec{u}_{\mathrm{ZAMO}}$. We get
\begin{align}
\vec{V}_{\mathrm{ZAMO}} \approx \frac{\vec{u}_{\mathrm{em}}}{\Gamma_{\mathrm{ZAMO}}} - \frac{1}{\sqrt{-g_{tt}|_{\mathrm{em}}}}\boldsymbol{\partial}_t|_{\mathrm{em}}.
\end{align}
The Lorentz factor can be approximated as
\begin{align}
\Gamma_{\mathrm{ZAMO}} &\approx - \frac{1}{\sqrt{-g_{tt}|_{\mathrm{em}}}} \boldsymbol{\partial}_t|_{\mathrm{em}} \cdot \vec{u}_{\mathrm{em}}, \notag \\
&\approx \sqrt{-g_{tt}|_{\mathrm{em}}} u_{\mathrm{em}}^t.
\end{align}
The new components of $\vec{V}_{\mathrm{ZAMO}}$ thus reduce to
\begin{align}
\vec{V}_{\mathrm{ZAMO}} &\approx \left(0,  \frac{1}{\sqrt{-g_{tt}|_{\mathrm{em}}}} \frac{u_{\mathrm{em}}^i}{u_{\mathrm{em}}^t}  \right), \notag \\
&\approx \frac{1}{\sqrt{-g_{tt}|_{\mathrm{em}}}} \boldsymbol{\mathcal{V}}
\end{align}
where $\boldsymbol{\mathcal{V}}$ is the four-velocity of the emitter in the black-hole frame with a null time coordinates. For simplicity, we nominate this velocity a three-velocity in the rest of the paper. Using the fact that the emitter is far from the black hole ($a \rightarrow 0$), we can write the quantity $g_{tt}|_{\mathrm{em}}$ as
\begin{equation}
g_{tt}|_{\mathrm{em}} = - (1 - \epsilon)
,\end{equation}
where $\epsilon$ is small compared to 1 and is equal to $\epsilon \approx 2 / r_{\mathrm{em}}$, with $r_{\mathrm{em}}$ the radial coordinate of the emitter in the black-hole frame. Finally, the GR redshift can be expressed as
\begin{align}
\mathcal{Z} &\approx \frac{1}{\sqrt{1-\epsilon}} \times \frac{1 + \mathcal{V}_{\mathrm{proj}}/\sqrt{1-\epsilon}}{\sqrt{1 - \mathcal{V}^2/(1-\epsilon)}},
\end{align}
where $\mathcal{V}_{\mathrm{proj}}$ is the velocity of the emitter projected along the line-of-sight.


\section{Measuring the impact of the different effects on both astrometric and spectroscopic measurements}
\label{app_E}

This Appendix defines the various effects that can impact the S2 observations, and present the methods used for computing these effects.

\subsection{Definition of each effect}
\label{app_E_1}

The different effects are:
\begin{itemize}
\item The Roemer time delay affecting both the astrometry and the spectroscopy. It is the only non-relativistic effect of this list. It is due to the finite speed of light propagating in a Newtonian spacetime. Depending on where the star is located along its orbit, the time needed by photons to reach the observer (following Euclidian straight lines) will defer.
\item The pericenter advance affecting both the astrometry and the spectroscopy. It is due to the spacetime curvature on the star trajectory. The orbit precesses because of the gravitational field caused by the central mass (the black hole in this paper). Thus, pericenter and apocenter of the star are shifted from one period to another.
\item The Lense-Thirring effect affecting both the astrometry and the spectroscopy. It is due to the angular momentum of the black hole. This effect varies with respect to the norm and the direction of the angular momentum. The Lense-Thirring effect affects both the star and the photon trajectories.
\item The gravitational lensing effect only affecting the astrometry. It is due to the curvature of the photon geodesic that changes the apparent position of the star on the plane of sky.
\item The Shapiro time delay affecting both the astrometry and the spectroscopy. This is due to the slowdown of the proper time of the photon with respect to the proper time of the observer when the photon crosses a gravitational field.
\item The relativistic redshifts only affecting the spectroscopy. The transversal Doppler shift appears in special relativity and is due to the relative motion between the emitter and the observer. The gravitational redshift appears in GR and is due to the spacetime curvature.
\end{itemize}

\subsection{Methods used to evaluate the impact of each effect on observations}
\label{app_E_2}

The astrometric contribution of the Roemer effect is obtained by using the formula $\Delta \mathrm{A}_{\mathrm{Roemer}} =\sqrt{\Delta \mathrm{A}_{\alpha}^2+\Delta \mathrm{A}_{\delta}^2}$ where $\Delta \mathrm{A}_{\alpha}~=~\alpha_\mathrm{A} - \alpha_{\mathrm{B}}$ et $\Delta \mathrm{A}_{\delta} = \delta_{\mathrm{A}} - \delta_{\mathrm{B}}$, with $(\alpha_\mathrm{A}, \delta_\mathrm{A})$ and $(\alpha_\mathrm{B}, \delta_\mathrm{B})$ the apparent positions simulated by models A and B, respectively. The radial velocity contribution is obtained by using the formula $\Delta \mathrm{V}_{\mathrm{Roemer}}~=~\mathrm{V}_{\mathrm{B}}~-~\mathrm{V}_{\mathrm{A}}$ where $\mathrm{V}_{\mathrm{A}}$ and $\mathrm{V}_{\mathrm{B}}$ are radial velocities estimated with models A and B, respectively.\\

The astrometric and spectroscopic contributions of the pericenter advance presented in Figs.~\ref{fig:AAP} and~\ref{fig:VAP} are obtained by comparing models C and D. For the astrometry, we use the formula $\Delta \mathrm{A}_{\mathrm{PA}} =\sqrt{\Delta \mathrm{A}_{\alpha}^2+\Delta \mathrm{A}_{\delta}^2}$ where $\Delta \mathrm{A}_{\alpha}~=~\alpha_\mathrm{C} - \alpha_{\mathrm{D}}$ and $\Delta \mathrm{A}_{\delta} = \delta_{\mathrm{C}} - \delta_{\mathrm{D}}$, with $(\alpha_\mathrm{C}, \delta_\mathrm{C})$ and $(\alpha_\mathrm{D}, \delta_\mathrm{D})$ the apparent positions simulated by models C and D, respectively. For the spectroscopy, we compute the difference $\Delta \mathrm{V}_{\mathrm{PA}}~=~\mathrm{V}_{\mathrm{C}}~-~\mathrm{V}_{\mathrm{D}}$ where $\mathrm{V}_{\mathrm{C}}$ and $\mathrm{V}_{\mathrm{D}}$ are radial velocities estimated with models C and D, respectively. \\

The Lense-Thirring effect on astrometric positions is computed with the formula $\Delta \mathrm{A}_{\mathrm{LT}} = \sqrt{\Delta \mathrm{A}_{\alpha}^2 + \Delta \mathrm{A}_{\delta}^2}$ where $\Delta \mathrm{A}_{\alpha}~=\alpha_{\mathrm{G},a = 0} - \alpha_{\mathrm{G},a = 0.99}$ and $\Delta \mathrm{A}_{\delta} = \delta_{\mathrm{G},a = 0} - \delta_{\mathrm{G},a = 0.99}$. The index G,$a$ = 0 and G,$a$ = 0.99 denote positions computed considering a spin of 0 and 0.99 in Model G, respectively. The radial velocity contribution of this effect is obtained with the formula $\Delta \mathrm{V}_{\mathrm{LT}} = \mathrm{V}_{\mathrm{G},a=0.99} - \mathrm{V}_{\mathrm{G},a=0}$ where $\mathrm{V}_{\mathrm{G},a=0.99}$ and $\mathrm{V}_{\mathrm{G},a=0}$ are radial velocities estimated considering $a=0.99$ and $a=0$ in Model G, respectively.\\

The gravitational lensing effect is computed by using the following formula: $\Delta \mathrm{A}_{\mathrm{GL}}~=~\sqrt{\Delta \mathrm{A}_{\alpha}^2~+~\Delta \mathrm{A}_{\delta}^2}$ where $\Delta \mathrm{A}_{\alpha}~=~\alpha_{\mathrm{G,GL = 0}}~-~\alpha_{\mathrm{G,GL = 1}}$ and $\Delta \mathrm{A}_{\delta}~=~\delta_{\mathrm{G,GL = 0}}~-~\delta_{\mathrm{G,GL = 1}}$. The index G,GL = 0 and G,GL = 1 denote positions computed without and with gravitational lensing in Model G, respectively. Let us better explain how the astrometric impact of this effect is reestimated. First we consider the star position ($\alpha_{\mathrm{G,GL = 1}}, \delta_{\mathrm{G,GL = 1}}$) on the observer screen as computed by the full-GR model (i.e., our Model G, see Appendix~\ref{app_A1}). A photon is thus ray traced backward in time from the observer screen until it reaches the star at some spacetime position $\mathcal{P}$. Let us now consider the modified position on the observer screen ($\alpha_{\mathrm{G,GL = 0}}, \delta_{\mathrm{G,GL = 0}}$) the star would have if light was propagating along Euclidian straight lines from $\mathcal{P}$ to the observer screen. For doing so, we project the coordinates of the star at $\mathcal{P}$ in the plane of sky by using the Thiele-Innes formulas. The leading astrometric shift ($\alpha_{\mathrm{G,GL = 1}}-\alpha_{\mathrm{G,GL = 0}}, \delta_{\mathrm{G,GL = 1}}-\delta_{\mathrm{G,GL = 0}}$) is due to both the gravitational lensing and the Lense-Thirring effect on the photon path. However, given that this latter effect is negligible for S2, this procedure gives access to the pure astrometric impact of the gravitational lensing effect.\\

To explain how we recover the astrometric impact of the Shapiro time delay let us still consider the position ($\alpha_{\mathrm{G,GL = 0}}, \delta_{\mathrm{G,GL = 0}}$) as obtained following the procedure above. In this procedure, the photon reaches the observer at a time $t_\mathrm{obs}$ that takes into account the Shapiro time delay. Let us now consider the position of the star on the observer screen as computed by the modified GR model without considering GR effects on photon path (i.e. Model E). In this model, the photon reaches the observer at a time $t_\mathrm{obs}'$ slightly different from the previous $t_\mathrm{obs}$, because at this time the photon path is an Euclidian straight line not affected by the gravitational field generated by the black hole. As a consequence, the star position at $t_\mathrm{obs}$ in Model E, ($\alpha_\mathrm{E}, \delta_\mathrm{E})$, will differ from ($\alpha_{\mathrm{G,GL = 0}}, \delta_{\mathrm{G,GL = 0}}$). The astrometric shift ($\alpha_{\mathrm{G,GL = 0}} - \alpha_\mathrm{E}, \delta_{\mathrm{G,GL = 0}} - \delta_\mathrm{E}$) is due to the Shapiro effect. The formula used to recover this effect is thus given by $\Delta \mathrm{A}_{\mathrm{Shapiro}}~=~\sqrt{\Delta \mathrm{A}_{\alpha}^2~+~\Delta \mathrm{A}_{\delta}^2}$ where $\Delta \mathrm{A}_{\alpha}~=~\alpha_{\mathrm{G,GL = 0}}~-~\alpha_{\mathrm{E}}$ and $\Delta \mathrm{A}_{\delta}~=~\delta_{\mathrm{G,GL = 0}}~-~\delta_{\mathrm{E}}$. \\

The spectroscopic HOPC contributions, including the Shapiro time delay, are obtained computing the difference between radial velocities estimated with Model E and those found in Model G.\\

Finally, the transverse Doppler shift and gravitational redshift are obtained by using the formula $\Delta \mathrm{V}_{\mathrm{TD,Grav}}~=~\mathrm{V}_{\mathrm{E, (TD,Grav)=1}}~-~\mathrm{V}_{\mathrm{E, (TD,Grav)=0}}$ where $\mathrm{V}_{\mathrm{E, (TD,Grav)=1}}$ and $\mathrm{V}_{\mathrm{E, (TD,Grav)=0}}$ are radial velocities estimated with and without implementing the transverse Doppler shift and gravitational redshift in Model E, respectively.

\end{document}